\documentclass[11pt]{article}
\usepackage{epsf}
\usepackage{color}
\usepackage{amsmath,amsfonts,amsbsy,amssymb}
\usepackage{indentfirst}
\usepackage{mathtools}

\newcommand{\fsl}[1]{\ensuremath{\mathrlap{\not{\phantom{#1}}}#1}}

\newcommand{\nn}{\nonumber}

\def\be{\begin{equation}}
\def\ee{\end{equation}}
\def\bse{\begin{subequations}}
\def\ese{\end{subequations}}
\def\bal{\begin{align}}
\def\ealn{\end{align}}
\def\tr{\text{tr}}
\def\bs{\boldsymbol}

\begin{document}

\begin{titlepage}

\def\slash#1{{\rlap{$#1$} \thinspace/}}

\begin{flushright} 

\end{flushright} 

\vspace{0.1cm}

\begin{Large}
\begin{center}


{\bf   
Relativistic Landau Models and Generation of Fuzzy Spheres  
}

\end{center}
\end{Large}

\vspace{1cm}

\begin{center}
{\bf Kazuki Hasebe}   \\ 
\vspace{0.5cm} 
\it{
Sendai National College of Technology, 
Ayashi, Sendai, 989-3128, Japan} \\ 

\vspace{0.5cm} 
{\sf
khasebe@sendai-nct.ac.jp} 

\vspace{0.8cm} 

{\today} 

\end{center}

\vspace{1.5cm}

\begin{abstract}
\noindent

\baselineskip=18pt

Non-commutative geometry naturally emerges in low energy physics of Landau models as a consequence of level projection. In this work, we proactively utilize the level projection as an effective tool to generate fuzzy geometry. The level projection is specifically applied to the relativistic Landau models. In the first half of the paper, a detail analysis of the relativistic Landau problems on a sphere is presented, where a concise expression of the Dirac-Landau operator eigenstates is obtained based on algebraic methods. We establish $SU(2)$ ``gauge'' transformation between the relativistic Landau model and the Pauli-Schr\"odinger non-relativistic quantum mechanics. After the $SU(2)$ transformation, the Dirac operator and the angular momentum operastors are found to satisfy the $SO(3,1)$ algebra.  
In the second half, the fuzzy geometries generated from the relativistic Landau levels are elucidated, where unique properties of the relativistic fuzzy geometries are clarified. We consider mass deformation of the relativistic Landau models and  demonstrate its geometrical effects to fuzzy geometry. Super fuzzy geometry is also constructed from a supersymmetric quantum mechanics as the square of the Dirac-Landau operator. Finally, we apply the level projection method to real graphene system to generate valley fuzzy spheres.   

\end{abstract}

\end{titlepage}

\newpage 

\tableofcontents

\newpage 

\section{Introduction}

 Quantization of the space-time is one of the most fundamental problems in physics. 
 Non-commutative geometry  is a promising  mathematical framework for the description of quantized space-time \cite{Connes-1994}.  
While string theory or matrix theory also suggests  appearance of  non-commutative geometry \cite{Connes-Douglas-Schwarz-1997}, the natural energy scale of the non-commutative geometry is considered to be  the Planck scale.     
Interestingly, however, it is well recognized that in low energy physics of some real materials, non-commutative geometry  naturally emerges.  
A well known example is the lowest Landau level physics of the quantum Hall effect, where the electron coordinates effectively satisfy non-commutative algebra due to the presence of strong magnetic field [see \cite{Hasebe-2010}  and references therein]. More precisely,  non-commutative geometry appears in any of the  Landau levels as well as the lowest Landau level as a consequence of the level projection.   Recently, higher dimensional non-commutative geometry has begun to be applied to  studies of topological insulators \cite{Hasebe-2014-1,Hasebe-2014-2, LiWu2013,LiZhangWu2013, EstienneRB2012, NeupertSRChMRB2012, Shiozaki-Fujimoto-2012}.  

Usually, non-commutative geometry is imposed on theories of interest in the beginning, and  within the mathematical framework we develop physical theories such as non-commutative quantum field theory.   
On the other hand, in the set-up of Landau models, non-commutative geometry is not postulated a priori but 
 ``generated''  as a consequence of level projection.        
In the work, we 
proactively utilize the level projection as a tool to derive fuzzy geometries. 
The merits of this scheme  are the following.  
First, the level projection basically yields a consistent framework of  non-commutative geometry. Generally it is far from obvious whether  non-commutative geometry can be incorporated in any manifolds, for instance, to curved manifolds,  
  keeping 
mathematical  
consistency. However, in the level projection scheme, we have a consistent Hilbert space of the quantum mechanics, and the level projection is just a method to extract a specific subspace of the consistent Hilbert space. 
Since the whole Hilbert space is well defined, we  need not to bother with the mathematical inconsistency in
 introducing the subspace and the corresponding  non-commutative geometry as well.    
Second, the level projection is rather mechanical, and one can readily introduce  fuzzy geometry  by following simple instructions to construct effective matrix representation in the subspace.   
Last, since the level projection scheme is based on  physical ideas, 
 mathematics of non-commutative geometry can be understood from a physical point of view, as we shall see in this work.

 In the first half of this work, we investigate relativistic Landau models described by Dirac-Landau operator on a sphere. 
  (We shall refer to the Dirac operator in magnetic field  as  Dirac-Landau operator.) 
We thus exploit a relativistic counterpart of the Haldane's sphere \cite{Haldane1983}.    
  Apart from applications to non-commutative geometry, the relativistic Landau models have increasing importance in  recent developments of Dirac matter  such as graphene and  topological insulator [there are many excellent books and reviews: see \cite{Aoki-Dresselhaus-Book-2014, Katsnelson-book-2012, Shen-Book-2012,  Bernevig-Hughes-2013} for instance].  
   Theoretical works of Dirac matter with Landau levels on a spherical geometry can be found in  Refs.\cite{Gozalez-Guinea-Vozmediano-1992,Gonzalez-Guinea-Vozmediano-1993} for fullerene, Refs.\cite{Lee-2009, Imura-Y-T-F-2012} for the surface of topological insulator, and Refs.\cite{Hasebe-2014-1,Li-Intriligator-Yu-Wu-2012} for higher dimensional topological insulators.  
Though the Dirac-Landau equation in flat space  has already been intensively investigated in various physical and mathematical contexts \cite{Thaller-book, Shnir-book} and on a sphere as well \cite{DeguchiKitsukawa2005, Jayewardena-1988}, many studies on a sphere  are restricted to zero mode solutions.  We present a full analysis of the relativistic Landau model on a sphere including all relativistic Landau level eigenstates.    Our method is based on an algebraic method, which  provides a concise way to solve  the Dirac-Landau operator and highlights a transparent  $SU(2)$ rotational symmetry of the present geometry  [Sec.\ref{Schwingerappend}]\footnote{The readers may find an analytic method for  solving the Dirac-Landau equation  in  Ref.\cite{Kolesnikov-Osipov-2006}.}.   
We establish $SU(2)$ transformation between  the relativistic Landau model and the  Pauli-Schwinger non-relativistic quantum mechanics obtained by Kazama et al. almost forty years ago  \cite{Kazama-Yang-Gldhaber-1977} [Sec.\ref{sec:relationpaulischnon}].  After the $SU(2)$ transformation,  the transformed Dirac operator and the angular momentum operators  are shown to satisfy the  $SO(3,1)$ algebra, which is the ``hidden'' symmetry of the system. 
In the second half, we discuss  fuzzy  geometries generated by  the level projection in the relativistic Landau models.  
In correspondence to each of the relativistic Landau levels, a relativistic fuzzy sphere is derived.   
  We compare behaviors of the relativistic  and  non-relativistic fuzzy spheres with respect to magnetic field,  where particular properties of the relativistic fuzzy spheres are  observed [Sec.\ref{sect:noncommutativegeo}]. We also investigate properties of fuzzy spheres under mass deformation [Sec.\ref{sect:aclasstopolo}]. Interestingly, th relativistic fuzzy spheres for opposite sign Landau levels balance their sizes keeping the sum of their radii invariant.  
  As the square of the Dirac-Landau operator, a supersymmetric quantum mechanics is constructed, where we demonstrate appearance of super fuzzy spheres [Sec.\ref{sect:superfuzzy}].  
Finally we apply the results to a realistic Dirac material, graphene,  to investigate fuzzy geometries with valley degrees of freedom and  behaviors under the change of mass parameter [Sec.\ref{sec:valleyfuzzysphere}].  
  Sec.\ref{sect:monopoleharmonics} is a review about the non-relativistic Landau problem  and  
Sec.\ref{sec:summarydiscussion} is devoted to summary and discussions.


\section{Review of the Non-Relativistic Landau Problem}\label{sect:monopoleharmonics}

\subsection{Monopole harmonics}

As a preliminary, we give a rather detail review of non-relativistic quantum mechanics for a charge-monopole system  mainly based on Refs.\cite{Shnir-book, Felsager1998,  Wu-Yang-1976}.  
We use the standard spherical coordinates,  
\be
x=r\sin\theta\cos\phi, ~~~y=r\sin\theta\sin\phi, ~~~z=r\cos\theta, 
\ee
and adopt the Schwinger gauge \cite{Felsager1998}\footnote{We utilize terminology,  Schwinger $\it{gauge}$, instead of the Schwinger $\it{formalism}$ in Ref.\cite{Felsager1998}. } 
[see Appendix \ref{appnsecDiracform} for the Dirac gauge] in which the monopole gauge field is given by  
\be
A=g\epsilon_{ij 3} \frac{z}{r(x^2+y^2)}x_j dx_i=-g\cos\theta d\phi, 
\label{u1gauschfiel}
\ee
or 
\begin{align}
&A_x=g\frac{z}{r(x^2+y^2)} y =g\frac{1}{r}\cot\theta\cdot \sin\phi, \nn\\
&A_y=-g\frac{z}{r(x^2+y^2)}x=-g\frac{1}{r}\cot\theta\cdot\cos\phi, \nn\\
&A_z=0, 
\label{gaugecartesianfield}
\end{align}
where  $g$ denotes the monopole charge. In this paper, we consider the case $g \ge 0$. (It is not difficult to expand similar discussions for  $g < 0$.)  In the Schwinger gauge the gauge field exhibits an infinite line singularity on the $z$-axis, and the direction of the monopole gauge field on the north hemisphere  is opposite to that   on the south hemisphere (on the equator, the monopole gauge field vanishes)\footnote{In the Dirac gauge [see Appendix \ref{appnsecDiracform}], the singularity of the gauge field is  a semi-infinite string either on the positive $z$-axis or on the negative $z$-axis, and  the directions of the monopole gauge fields are same on both hemispheres.}.          
The corresponding field strength is given by  
\be
F=dA=g\sin\theta ~d\theta\wedge d\phi, 
\label{sphericaexfieldst}
\ee
or 
\be
F_{i}=\epsilon_{ijk}\partial_jA_k=g \frac{1}{r^3}x_i. 
\label{monopolegaugefieldnonrel}
\ee
The covariant derivative is constructed as 
\be
D_i=\partial_i-iA_i,   
\label{covderiveflat}
\ee
or 
\be
-iD_r=-i\partial_r,~~~~-iD_{\theta}=-i\partial_{\theta}, ~~~~~-iD_{\phi}=-i\partial_{\phi}+g\cos\theta, 
\label{effectofmonopolecov}
\ee 
and the covariant angular momentum is 
\be
\Lambda_i^{(g)}=-i\epsilon_{ijk}x_jD_k , 
\label{covangmodef}
\ee
or 
\begin{align}
&\Lambda_x^{(g)}=L_x^{(0)}- g\frac{z^2}{r(x^2+y^2)}x =L_x^{(0)}-g\frac{\cos^2\theta}{\sin\theta}\cos\phi, \nn\\
&\Lambda_y^{(g)}=L_y^{(0)}- g\frac{z^2}{r(x^2+y^2)}y=L_y^{(0)}-g\frac{\cos^2\theta}{\sin\theta}\sin\phi, \nn\\
&\Lambda_z^{(g)}=L_z^{(0)}+g\frac{1}{r}z =L_z^{(0)}+g\cos\theta. 
\label{covariantangulamomts}
\end{align}
Here, $L_i^{(0)} =-i\epsilon_{ijk}x_j\frac{\partial}{\partial x_k}$ represent the free orbital angular momentum operators: 
\begin{align}
&L_x^{(0)}= i  (\sin\phi\partial_{\theta} +\cot\theta\cos\phi\partial_{\phi}), \nn\\
&L_y^{(0)}= -i (\cos\phi\partial_{\theta}-\cot\theta\sin\phi\partial_{\phi}), \nn\\
&L_z^{(0)}=-i\partial_{\phi}. 
\label{freeangumomentquant}
\end{align}
The total $SU(2)$ angular momentum is constructed as the sum of the covariant and the field angular momenta: 
\be
\bs{L}^{(g)}=\bs{\Lambda}^{(g)}-r^2 \bs{F}=\bs{\Lambda}^{(g)}-g\frac{1}{r}\bs{x},  
\label{nonretotalangmo} 
\ee
or 
\begin{align}
&L_x^{(g)}=i(\sin\phi D_\theta +\cos\phi\cot\theta D_{\phi})-g\frac{x}{r},  \nn\\
&L_y^{(g)}=-i(\cos\phi D_\theta -\sin\phi\cot\theta D_{\phi})-g\frac{y}{r}, \nn\\
&L_z^{(g)}=-iD_{\phi}-g\frac{z}{r}. 
\label{schwingerexsu2ooperators}
\end{align}
With use of (\ref{freeangumomentquant}), they are expressed as 
\begin{align}
&L_x^{(g)}=L_x^{(0)} -g\frac{r}{x^2+y^2}x=L_x^{(0)}-g\frac{\cos\phi}{\sin\theta}, \nn\\
&L_y^{(g)}=L_y^{(0)}-g\frac{r}{x^2+y^2}y =L_y^{(0)}-g\frac{\sin\phi}{\sin\theta}, \nn\\
&L_z^{(g)}=L_z^{(0)}. 
\label{explicitschwingerexsu2ooperators}
\end{align}
The square of $\bs{L}^{(g)}$ can be represented as 
\begin{align}
{{\bs{L}}^{(g)}}^2&=-\frac{1}{\sin\theta} \frac{\partial}{\partial\theta} (\sin\theta\frac{\partial}{\partial\theta})-\frac{1}{\sin^2\theta}(\frac{\partial}{\partial\phi} +ig \cos\theta )^2+g^2
\nn\\
&= {\bs{L}^{(0)}}^2 -2ig \frac{\cos\theta}{\sin^2\theta} \frac{\partial}{\partial \phi}+g^2\frac{1}{\sin^2\theta}\nn\\
&={\bs{L}^{(0)}}^2 +2g \frac{r z}{x^2+y^2} L_z^{(0)}+g^2\frac{r^2}{x^2+y^2}, 
\end{align}
where 
\be
{\bs{L}^{(0)}}^2= -\frac{1}{\sin\theta}\partial_{\theta}(\sin\theta\partial_{\theta})-\frac{1}{\sin^2\theta}{\partial_{\phi}}^2.
\ee
The monopole harmonics are introduced as the simultaneous eigenstates of ${\bs{L}^{(g)}}^2$ and $L_z^{(g)}$: 
\begin{align}
&{\bs{L}^{(g)}}^2 Y^g_{l, m}(\theta, \phi)=l(l+1) Y^g_{l, m}(\theta, \phi), \nn\\
&{L_z^{(g)}}Y^g_{l, m}(\theta, \phi)=m Y^g_{l, m}(\theta, \phi), 
\end{align}
where  $l$ and $m$ take the following values \cite{Wu-Yang-1976}: 
\begin{subequations}
\begin{align}
&l=g+n~~~~~~~(n=0, 1, 2, \cdots),\label{rangeofg}\\
&m=-l, -l+1, \cdots, l-1, l. \label{condmandlandg}
\end{align}\label{rangeoflandm}
\end{subequations}
The ladder operators are given by  
\begin{align}
&L^{(g)}_{+} =L_x^{(g)}+ iL_y^{(g)}=e^{ i\phi}( \partial_{\theta} +i\cot\theta\partial_{\phi}-g\frac{1}{\sin\theta}), \nn\\
&L^{(g)}_{-} =L_x^{(g)}- iL_y^{(g)}=e^{-i\phi}(-\partial_{\theta} +i\cot\theta\partial_{\phi}-g\frac{1}{\sin\theta}),  
\label{ladderoppangularmagg}
\end{align}
which act to the monopole harmonics as 
\be
L_{\pm}^{(g)}Y^g_{l,m} =\sqrt{(l\mp m)(l\pm m+1)}Y^g_{l, m\pm 1}. 
\ee
The irreducible representation of the monopole harmonics can be obtained by applying the  $SU(2)$ ladder operators to the lowest or highest weight state.   
The monopole harmonics are explicitly given by \cite{Wu-Yang-1976, Felsager1998}
\begin{align}
{Y}^g_{~l,m}(\theta, \phi)&=2^m \sqrt{\frac{(2l+1)(l-m)!(l+m)!}{4\pi (l-g )!(l+g)!}} (1-x)^{-\frac{m+g}{2}} (1+x)^{-\frac{m-g}{2}}P_{l+m}^{(-m-g, -m+g)} (x) \cdot e^{im\phi}\nn\\
&= \sqrt{\frac{(2l+1)(l-m)!(l+m)!}{4\pi (l-g )!(l+g)!}} \biggl(\sin\frac{\theta}{2}\biggr)^{-(m+g)} \biggl(\cos\frac{\theta}{2}\biggr)^{-({m-g})}P_{l+m}^{(-m-g, -m+g)} (\cos\theta) \cdot e^{im\phi}, 
\label{monopoleharmonicsjacobischwinger}
\end{align}
where $P_{n}^{(\alpha, \beta)} (x)$ denote the Jacobi polynomials [Appendix \ref{sec:jacobi-poly}]. For  uniqueness of the wavefunction, the magnetic quantum number of the  azimuthal part of (\ref{monopoleharmonicsjacobischwinger}) has to take an  integer value, $m=0, \pm 1, \pm 2, \cdots$. Due to (\ref{condmandlandg}),  the monopole charge $g$ should be quantized as an  integer in the Schwinger gauge \cite{Felsager1998}. 
Expressing the Jacobi polynomials by the trigonometric function, (\ref{monopoleharmonicsjacobischwinger}) can be rewritten as \cite{Goldberg-et-al-1967} 
\begin{align}
Y^g_{l,m} (\theta, \phi)&=(-1)^{l+m}\sqrt{\frac{(2l+1)}{4\pi}\frac{(l+m)!~(l-m)!}{(l+g)!~(l-g)!}}  ~e^{im\phi}\nn\\
&\cdot \sum_{n}(-1)^n\begin{pmatrix}
l-g \\
n
\end{pmatrix}
\begin{pmatrix}
l+g \\
g-m+n
\end{pmatrix} 
\biggl(\sin\frac{\theta}{2}\biggr)^{2l-2n-g+m}\biggl(\cos\frac{\theta}{2}\biggr)^{2n+g-m}, 
\label{simpleexpmonopoleharmonics} 
\end{align}
or 
\begin{align}
Y^g_{l,m} (\theta, \phi)=&(-1)^{l+m}\sqrt{\frac{(2l+1)}{4\pi}\frac{(l+m)!~(l-m)!}{(l+g)!~(l-g)!}}  \nn\\ &\cdot   \sum_{n}\begin{pmatrix}
l-g \\
n
\end{pmatrix}
\begin{pmatrix}
l+g \\
g-m+n
\end{pmatrix} ~ (-1)^{n}~ u^{g-m+n}~v^{l-n+m}~{u^*}^{n}~{v^*}^{l-g-n}, 
\label{monopoehamonicseffechopf}
\end{align}
where $u$ and $v$ are the components of the Hopf spinor \cite{Hasebe-2010}: 
\be
u = \cos\frac{\theta}{2} e^{-i\frac{1}{2}\phi},~~~
v= \sin\frac{\theta}{2} e^{i\frac{1}{2}\phi}, 
\label{hopfspinorcompo}
\ee
and ${{u}^*}$ and ${{v}^*}$ are their complex conjugates.    
For instance, in the case $g=1$ and $l=2$, we have  
\begin{align}
&Y^1_{2,2}=  -\sqrt{\frac{5}{\pi}}\sin^3\frac{\theta}{2}\cos\frac{\theta}{2} e^{2i\phi} ,~~Y^1_{2,1}= \frac{1}{2}\sqrt{\frac{5}{\pi}}   (1+2\cos\theta)\sin^2\frac{\theta}{2} e^{i\phi}  ,\nn\\
&Y^1_{2,0}= -\frac{1}{2}\sqrt{\frac{15}{2\pi}}\sin\theta  \cos\theta   ,~~Y^1_{2,-1}=  \frac{1}{2}\sqrt{\frac{5}{\pi}}   (-1+2\cos\theta)\cos^2\frac{\theta}{2} e^{-i\phi}   ,\nn\\
&Y^1_{2,-2}=  \sqrt{\frac{5}{\pi}}\sin\frac{\theta}{2}\cos^3\frac{\theta}{2} e^{-2i\phi} . 
\end{align}

The non-relativistic Landau Hamiltonian  in a monopole background is given by \cite{Haldane1983} 
\begin{align}
H&=-\frac{1}{2M}\sum_{i=1}^3 {D_i}^2\nn\\
&=-\frac{1}{2M}\biggl(\frac{1}{r^2}D_r(r^2D_r)
+\frac{1}{r^2\sin\theta}D_{\theta}(\sin\theta D_{\theta})+\frac{1}{r^2\sin^2\theta}{D_{\phi}}^2\biggr)\nn\\
\nn\\
&=
-\frac{1}{2M}\frac{\partial^2}{\partial r^2} -\frac{1}{Mr}\frac{\partial}{\partial r}+\frac{1}{2M r^2}{\bs{\Lambda}^{(g)}}^2,  
\end{align}
which, on a sphere $r=1$, reduces  to  
\begin{align}
H^{(g)}=-\frac{1}{2M}\biggl(\frac{1}{\sin\theta}D_{\theta}(\sin\theta D_{\theta})
+\frac{1}{\sin^2\theta}{D_{\phi}}^2\biggr)=
\frac{1}{2M}{\bs{\Lambda}^{(g)}}^2=
\frac{1}{2M}({\bs{L}^{(g)}}^2-g^2). 
\label{hamiltonianmonopolenonrela}
\end{align}
In the following, we take $r=1$. (We sometimes recover $r$ to indicate the dimensions of quantities of  interest.)   
Since we have already solved the eigenvalue problem of ${\bs{L}^{(g)}}^2$, the eigenvalues of (\ref{hamiltonianmonopolenonrela}) can readily be obtained as 
\be
E_n^{(g)}=\frac{1}{2M}(n(n+1)+g(2n+1)),  
\label{non-relaeigenvalues}
\ee
where we used (\ref{rangeofg}),   
and the degenerate eigenstates of the $n$th Landau level are the monopole harmonics (\ref{monopoleharmonicsjacobischwinger}) with degeneracy, 
\be
2l+1=2g+1+2n.
\ee
In the lowest Landau level  $n=0$ $(l=g)$\footnote{For $g<0$,  the monopole harmonics in the lowest Landau level ($l=|g|$) are given by 
\be
Y^g_{~~|g|, m}(\theta, \phi)= \sqrt{\frac{(2|g|+1)!}{4\pi (|g|+m)!(|g|-m)!}} (u^*)^{|g|+m}
(v^*)^{|g|-m}. 
\ee
}, the monopole harmonics are represented as 
\begin{align}
Y^g_{g, m}(\theta, \phi)&=(-1)^{m+g} \sqrt{\frac{(2g+1)!}{4\pi (g+m)!(g-m)!}} \biggl(\sin\frac{\theta}{2}\biggr)^{m+g}\biggl(\cos\frac{\theta}{2}\biggr)^{-m+g} e^{im\phi}\nn\\
&=(-1)^{m+g} \sqrt{\frac{(2g+1)!}{4\pi (g+m)!(g-m)!}} u^{g-m}
v^{g+m}. \label{monopoleharmolll}
\end{align}
The lowest Landau level eigenstates are  homogeneous holomorphic polynomials of the Hopf spinor.  

\subsection{Edth operators}

The monopole harmonics carry  two $SU(2)$ spin indices, $m$ and $g$. (With fixed $l$, both $m$ and $g$ range from $-l$ to $l$.\footnote{This is the basic observation about the equivalence between the monopole harmonics and  spin-weighted spherical harmonics \cite{Dray1985, Dray1986}.})   
One may expect that ladder operators for $g$ may exist just like the ladder operators, $L_{\pm}^{(g)}$, for $m$. Such operators are known as the edth differential operators  \cite{Newman-Penrose-1966}\footnote{$\eth_{+}$ and $\eth_{-}$ respectively correspond to $\eth$ and $\bar{\eth}$ in Refs.\cite{Newman-Penrose-1966, Dray1985, Dray1986}}: 
\begin{align}
&\eth_+^{(g)}\equiv (\sin\theta)^g ({\partial_\theta} +i\frac{1}{\sin\theta}{\partial_\phi})(\sin\theta)^{-g}=\partial_{\theta}-g\cot\theta +i\frac{1}{\sin\theta}\partial_{\phi}
\nn\\
&\eth_-^{(g)}\equiv (\sin\theta)^{-g} ({\partial_\theta} -i\frac{1}{\sin\theta}{\partial_\phi})(\sin\theta)^{g} 
=\partial_{\theta}+g\cot\theta -i\frac{1}{\sin\theta}\partial_{\phi}, 
\label{expliladdercharge}
\end{align}
or 
\begin{align}
&\eth_+^{(g)}=D_{\theta}+i\frac{1}{\sin\theta}D_{\phi}, \nn\\
&\eth_-^{(g)}=D_{\theta}-i\frac{1}{\sin\theta}D_{\phi}, 
\label{edthcovderi}
\end{align}
where $D_{\theta}=\partial_{\theta}$ and $D_{\phi}=\partial_{\phi}+ig\cos\theta$ are the covariant derivatives  (\ref{effectofmonopolecov}). 
The edth operators indeed act to the monopole harmonics as \cite{Dray1985, Dray1986}
\footnote{ In the Cartesian coordinates, the edth operators are represented as 
\be
\eth_{\pm}^{(g)}=\frac{z}{\sqrt{x^2+y^2}}(x\partial_x+y\partial_y)-\sqrt{x^2+y^2}\partial_z \pm i\frac{r}{\sqrt{x^2+y^2}}(x\partial_y-y\partial_x)\mp g \frac{z}{\sqrt{x^2+y^2}} 
\ee
or, with use of the angular momentum operators (\ref{freeangumomentquant}) and (\ref{covariantangulamomts}),  
\begin{align}
\eth_{\pm}^{(g)}&=i\frac{1}{\sqrt{x^2+y^2}}( xL_y^{(0)}-y L_x^{(0)}) \mp g \frac{z}{\sqrt{x^2+y^2}} \mp \frac{r}{\sqrt{x^2+y^2}}L_z^{(0)} \nn\\
&=i\frac{1}{\sqrt{x^2+y^2}}( x\Lambda_y^{(g)}-y \Lambda_x^{(g)} \mp r \Lambda_z^{(g)}).  
\label{edthcobarang}
\end{align}
}
\begin{align}
&\eth_+^{(g)} Y^g_{l, m}(\theta, \phi)= \sqrt{(l-g)(l+g+1)} ~Y^{g+1}_{l,m}(\theta, \phi), \nn\\
&\eth_-^{(g)} Y^g_{l, m}(\theta, \phi)= -\sqrt{(l+g)(l-g+1)} ~Y^{g-1}_{l,m}(\theta, \phi). 
\label{ladderschargemonohar}
\end{align}
Notice that, while $\eth_+^{(g)}$ and $\eth_-^{(g)}$ respectively increases and decreases the monopole charge by $1$, they are inert with  the $SU(2)$ index $l$ (and the magnetic quantum number $m$). Therefore, in the language of  Landau level $n=l-g$, the edth operators act as the ladder operators of the Landau levels. In more detail, since $\eth_+^{(g)}$/$\eth_-^{(g)}$ acts as the raising/lowering operator for the monopole charge, as for the Landau levels, $\eth_+^{(g)}$/$\eth_-^{(g)}$ plays the opposite;  lowering/raising operator for the Landau level. This implies  that the edth operators are the covariant derivatives on a sphere in  monopole magnetic field.    

From (\ref{expliladdercharge}),  we obtain 
\begin{subequations}
\begin{align}
\eth_+^{(g-1)}\eth_-^{(g)}-\eth_-^{(g+1)}\eth_+^{(g)}=-2g, \label{commutationcovsphere}
\end{align}
and 
\begin{align}
\eth_+^{(g-1)}\eth_-^{(g)}+\eth_-^{(g+1)}\eth_+^{(g)}= -2({\bs{L}^{(g)}}^2- g^2).    
\label{anticommutationcovsphere}
\end{align}\label{commutationcovspheretot}
\end{subequations}
These relations are essentially the same as of the ladder operators (in the $L_z$ diagonalized basis) with replacement of $m$ with  $g$:
\begin{subequations}
\begin{align}
L_+^{(g)}L_-^{(g)}-L_-^{(g)}L_+^{(g)}=2m, 
\end{align}
and 
\begin{align}
L_+^{(g)}L_-^{(g)}+L_-^{(g)}L_+^{(g)}=2({\bs{L}^{(g)}}^2- m^2). 
\end{align}
\end{subequations}
From the  point of  view of three-sphere, the analogies between  the edth operators and the angular momentum operators are clearly understood  [Appendix \ref{append:secmaurer}].    
The edth and angular momentum operators are mutually commutative: 
\be
\bs{L}^{(g+1)}\eth_+^{(g)}-\eth_+^{(g)}\bs{L}^{(g)}=0,~~~~~\bs{L}^{(g-1)}\eth_-^{(g)}-\eth_-^{(g)}\bs{L}^{(g)}=0. 
\label{commutationrelaethls}
\ee
In other words, the edth operators are singlet under the $SU(2)$ angular momentum transformations.  
Due to the relation 
(\ref{anticommutationcovsphere}),  
the Landau Hamiltonian  (\ref{hamiltonianmonopolenonrela}) can be expressed 
as\footnote{
Alternatively using (\ref{edthcovderi}), one may explicitly verify 
\be
-\frac{1}{2M} \eth_-^{(g+1)}\eth_+^{(g)}+\frac{g}{2M}=-\frac{1}{2M}\biggl(\frac{1}{\sin\theta}D_{\theta}(\sin\theta D_{\theta})+\frac{1}{\sin^2\theta} {D_{\phi}}^2\biggr)=H^{(g)}. 
\ee
}  
\begin{align}
H^{(g)}&=-\frac{1}{4M}(\eth_+^{(g-1)}\eth_-^{(g)}+\eth_-^{(g+1)}\eth_+^{(g)})\nn\\
&=-\frac{1}{2M} \eth_-^{(g+1)}\eth_+^{(g)}+\frac{g}{2M}. 
\label{Hamilcovs2}
\end{align}
Eq.(\ref{commutationrelaethls}) implies that the Hamiltonian (\ref{Hamilcovs2}) is invariant under the $SU(2)$ rotations: 
\be
[H^{(g)}, \bs{L}^{(g)}]=0. 
\ee
It is straightforward to confirm that $Y^g_{l, m}(\theta, \phi)$  is the eigenstate of  the Hamiltonian (\ref{Hamilcovs2}) with the eigenvalues (\ref{non-relaeigenvalues}) with use of (\ref{ladderschargemonohar}). 
One may find analogies between (\ref{Hamilcovs2}) and the Landau  Hamiltonian on a plane,  $H_{\text{plane}}=-\frac{1}{2M} ({D_x}^2+{D_y}^2)$ with $[D_x,D_y]=-iB$:  
\begin{align}
H_{\text{plane}}&=-\frac{1}{4M}(D\bar{D}+\bar{D}D)\nn\\
&=-\frac{1}{2M}\bar{D}D+\frac{B}{2M},  
\end{align}
where 
$D=D_x+iD_y,$ and $\bar{D}=D_x-iD_y$.  
The covariant derivatives  satisfy 
\be
[D,\bar{D}]=-2i[D_x, D_y]=-2B, 
\ee
which corresponds to (\ref{commutationcovsphere}). Also from these relations,  the edth operators turn out  to play the covariant derivatives of the Landau model on the sphere.   
Furthermore, 
 the center-of-mass coordinates,  $X=x-i\frac{1}{B}D_y$ and $Y=y+i\frac{1}{B}D_x$, or the magnetic translation operators which commute with the covariant derivatives  correspond to  the angular momentum operator  $\bs{L}^{(g)}$  on the sphere.    
Then the correspondences between the plane and sphere cases are summarized as    
\begin{align}
&{D},\bar{D}~~\longleftrightarrow~~\eth_+, \eth_-, \nn\\ 
&X, Y~~\longleftrightarrow~~{L}_x, L_y, L_z. 
\end{align}

\section{Relativistic Landau Problem on a Sphere}\label{Schwingerappend}

\subsection{Spin connection and  the $SU(2)$ angular momentum operator}

From the metric on a two-sphere 
\be
ds^2=d\theta^2+\sin^2\theta d\phi^2, 
\ee
zweibein  can be adopted as [see Appendix \ref{sec:genforms2geo} for details]  
\be
e^1=d\theta,~~~e^2=\sin\theta d\phi. 
\ee
The torsion free condition, $de^a+\omega_{ab}e^b=0$, determines the spin connection: 
\be
\omega_{12}(=-\omega_{21})=-\cos\theta d\phi. 
\label{spinconnectioncomp}
\ee
We choose the $SO(2)$ gamma matrices and generator as 
\begin{align}
&\gamma^1=\sigma_x, ~~~\gamma^2=\sigma_y, \nn\\ 
&\sigma^{12}=-\sigma^{21}=-i\frac{1}{4}[\gamma^a, \gamma^b]=\frac{1}{2}\sigma_z,  
\end{align}
to have matrix valued spin connection 
\be
\omega =\frac{1}{2}\omega_{ ab}\sigma^{ab}= -\frac{1}{2} \sigma_z \cos\theta~d\phi . 
\label{spinconneomega}
\ee
Notice that (\ref{spinconneomega}) coincides with the monopole gauge field (\ref{u1gauschfiel}) 
 with 
 $g=-\frac{1}{2}\sigma_z$. This is because that the $SO(2)$ holonomy of the base-manifold $S^2$  is  isomorphic to the $U(1)$  gauge group of the monopole. 
Consequently, the spin connection effectively modifies the monopole charge by  $\mp \frac{1}{2}$   
depending on  up and down-components of the spinor.    
The components of the Dirac-Landau operator are given by  
\be
-i\mathcal{D}_{\mu}
=-i\partial_{\mu}+\omega_{\mu}\otimes 1-1\otimes A_{\mu}
=-i\partial_{\mu}-\mathcal{A}_{\mu},  
\label{absderivcov}
\ee
where $\mathcal{A}$  denotes a matrix valued gauge field: 
\be
\mathcal{A}=-g_s\cos\theta d\phi, 
\ee
with 
\be
g_{\text{s}}\equiv 1\otimes g - \frac{1}{2}\sigma_z\otimes 1=g-\frac{1}{2}\sigma_z. 
\label{replacemonopolecharge}
\ee
(\ref{absderivcov}) is thus obtained as 
\be
-i\mathcal{D}_{\theta}=-i\partial_{\theta}, ~~~~-i\mathcal{D}_{\phi}=-i\partial_{\phi}+g_s\cos\theta . 
\label{spherecovderi}
\ee 
It is straightforward to expand similar discussions to Section \ref{sect:monopoleharmonics} with replacement:  
\be
g~~\rightarrow~~g_s. 
\ee
The field strength for  $\mathcal{A}$ is derived as 
\be
\mathcal{F}_{\theta\phi}=-i[\mathcal{D}_{\theta}, \mathcal{D}_{\phi}]=\partial_{\theta}\mathcal{A}_{\phi}- \partial_{\phi}\mathcal{A}_{\theta}=g_s \sin\theta, 
\ee
or 
\be
\mathcal{F}_i=g_s\frac{1}{r^3}x_i.  
\ee
The total angular momentum operator is 
\be
\bs{J}=\bs{L}^{(g_s)}\equiv \begin{pmatrix}
\bs{L}^{(g-\frac{1}{2})} & 0 \\
0 & \bs{L}^{(g+\frac{1}{2})}
\end{pmatrix}, 
\label{JfromLs}
\ee
or 
\begin{align}
&J_x=i(\sin\phi \mathcal{D}_\theta +\cos\phi\cot\theta \mathcal{D}_{\phi})+g_{\text{s}}\frac{1}{r}x,  \nn\\
&J_y=-i(\cos\phi \mathcal{D}_\theta -\sin\phi\cot\theta \mathcal{D}_{\phi})+g_{\text{s}}\frac{1}{r}y, \nn\\
&J_z=-i\mathcal{D}_{\phi}+g_{\text{s}}\frac{1}{r}z, 
\label{schwingerexsu2ooperatorsnew}
\end{align}
which satisfy the $SU(2)$ algebra: 
\be
[J_i, J_j]=i\epsilon_{ijk}J_k. 
\ee
$\bs{J}$ can be represented as   
\be
J_x
=L_x^{(0)}-g_{\text{s}}\frac{\cos\phi}{\sin\theta},~~~~~~~~
J_y
=L_y^{(0)}-g_{\text{s}}\frac{\sin\phi}{\sin\theta},~~~~~~~~
J_z
=L_z^{(0)}, 
\label{totalanwithmono}
\ee
and the $SU(2)$ Casimir operator is   
\begin{align}
\bs{J}^2&=  {\bs{L}^{(0)}}^2 -2ig_s \frac{\cos\theta}{\sin^2\theta} \frac{\partial}{\partial \phi}+{g_s}^2\frac{1}{\sin^2\theta}\nn\\
&={\bs{L}^{(0)}}^2
-2ig\frac{\cos\theta}{\sin^2\theta}\partial_{\phi}+\frac{1}{4\sin^2\theta}(1+4g^2) +i\frac{1}{\sin^2\theta} \sigma_z({\cos\theta}\partial_{\phi}+ig). 
\label{jsquareschwinger}
\end{align}
Since $\bs{J}$  commutes with the chiral matrix $\sigma_z$: 
\be
[\sigma_z, {J}_i]=0,  
\ee
 we can diagonalize 
 $\bs{J}^2$  in each chiral sector.  
The  eigenvalues of $\bs{J}^2$  are given by 
\be
j(j+1), 
\ee
where\footnote{Strictly speaking, Eq.(\ref{rangeofj}) holds for non-zero $g$. 
 For $g=0$, we have  $j=\frac{1}{2}+n$~~$(n=0, 1, 2, \cdots)$. } 
\be
j=g- \frac{1}{2}+n ~~~~~(n=0, 1, 2, \cdots). 
\label{rangeofj}
\ee
For $j=g-\frac{1}{2}$, the corresponding eigenstates are 
\be
{\Upsilon'}^{g}_{j=g-\frac{1}{2}, m}=
\begin{pmatrix}
{Y}^{g-\frac{1}{2}}_{j=g-\frac{1}{2} ,m}(\theta, \phi) \\
0
\end{pmatrix},
\label{minimjstates}
\ee
with degeneracy $2j+1|_{j=g-\frac{1}{2}}=2g$, while for $j=g-\frac{1}{2}+n$ $(n = 1, 2, \cdots)$, the corresponding eigenstates are 
\be
{\Upsilon'}^{g}_{j=g-\frac{1}{2}+n, m}=
\begin{pmatrix}
{Y}^{g-\frac{1}{2}}_{j=g-\frac{1}{2}+n ,m}(\theta, \phi) \\
0
\end{pmatrix},~~~~~\Upsilon^{g}_{j=g-\frac{1}{2}+n,m}=\begin{pmatrix}
0 \\
{Y}^{g+\frac{1}{2}}_{j=g-\frac{1}{2}+n ,m} (\theta, \phi)
\end{pmatrix}, 
\label{higherjstates}
\ee
with degeneracy $2\cdot (2j+1)|_{j=g-\frac{1}{2}+n}=4(g+n)$.   

\subsection{Dirac-Landau operator and eigenvalue problem}\label{sec:eigenstatediracop}

Using (\ref{spherecovderi}), we construct 
the Dirac-Landau operator,  
 $-i\fsl{\mathcal{D}}=-ie_{m}^{~~\mu}\gamma^{m}\mathcal{D}_{\mu}$, 
as 
\begin{align}
-i\fsl{\mathcal{D}}&=-i\sigma_x \partial_{\theta}-i\frac{1}{\sin\theta}\sigma_y (\partial_{\phi}+ig_s\cos\theta)\nn\\
&=-i\sigma_x (\partial_{\theta}+\frac{1}{2}\cot\theta)-i\sigma_y \frac{1}{\sin\theta}(\partial_{\phi}+ig\cos\theta)\nn\\
&=\begin{pmatrix}
0 & -i\partial_{\theta}-\frac{1}{\sin\theta}(\partial_{\phi}+i(g+\frac{1}{2})\cos\theta) \\
  -i\partial_{\theta}+\frac{1}{\sin\theta}(\partial_{\phi}+i(g-\frac{1}{2})\cos\theta)  & 0  
\end{pmatrix}. 
\label{explicidiragaugesch}
\end{align}
With  the edth operators (\ref{expliladdercharge}) or\footnote{The edth operators are generally given by $\eth^{(g)}_m=e_m^{~~\mu}D_{\mu}$ $(m=x,y)$. See Appendix \ref{appnsecDiracform} also.}  
\begin{align}
&\eth_x^{(g)}\equiv \frac{1}{2}(\eth_+^{(g)}+\eth_-^{(g)})=\partial_{\theta}=D_{\theta},\nn\\ 
&\eth_y^{(g)}\equiv -i\frac{1}{2}(\eth_+^{(g)}-\eth_-^{(g)})=\frac{1}{\sin\theta}(\partial_{\phi}+ig \cos\theta)=\frac{1}{\sin\theta}D_{\phi}, 
\end{align}
the Dirac-Landau operator is concisely expressed as
\begin{align}
-i\fsl{\mathcal{D}}
=-i\sigma_x \eth_{x}^{(g_s)}-i\sigma_y \eth_{y}^{(g_s)}
=\begin{pmatrix}
0 & -i\eth_-^{(g+\frac{1}{2})} \\
-i\eth_+^{(g-\frac{1}{2})} & 0 
\end{pmatrix}. \label{diracopfrometh}
\end{align}
Note $\eth_{x}^{(g_s)}=\mathcal{D}_{\theta}$ and  $\eth_{y}^{(g_s)}=\frac{1}{\sin\theta}\mathcal{D}_{\phi}$. 
The spin connection term\footnote{The spin connection term yields the non-hermitian term, $-i\frac{1}{2}\cot\theta$, in (\ref{explicidiragaugesch}). It is well known that on 2D manifolds, the spin connection term vanishes when we modify the Dirac operator to be hermitian [see \cite{Nakahara-book} or \cite{Bertlmann-book} for instance]. Though the present Dirac operator contains the non-hermitian term, its  eigenvalues are real numbers.     
  } induces a difference between monopole charges by 1 in the off-diagonal components, and such ``discrepancy'' is crucial in the following  discussions.

It is not difficult to derive the eigenvalues of the Dirac-Landau operator on a sphere \cite{Hasebe-2014-1, Dolan2003}. 
The square of the Dirac-Landau operator gives the $SU(2)$ Casimir of the angular momentum $\bs{J}$: 
\begin{align}
(-i\fsl{\mathcal{D}})^2
&=-
\begin{pmatrix}
\eth_-^{(g+\frac{1}{2})}\eth_+^{(g-\frac{1}{2})} & 0 \\
0 & \eth_+^{(g-\frac{1}{2})}\eth_-^{(g+\frac{1}{2})}
\end{pmatrix}=\begin{pmatrix}
{\bs{L}^{(g-\frac{1}{2})}}^2 +\frac{1}{4}-g^2 & 0 \\
0 & {\bs{L}^{(g+\frac{1}{2})}}^2  +\frac{1}{4}-g^2
\end{pmatrix}\nn\\
&=\bs{J}^2 +\frac{1}{4}-g^2, 
\label{squarediracangular}
\end{align}
where we used (\ref{commutationcovspheretot}). Eq.(\ref{squarediracangular}) is consistent with the general formula \cite{Dolan2003, Hasebe-2014-1}:  
\be
(-i\fsl{\mathcal{D}})^2=\boldsymbol{J}^2-g^2+\frac{{R}}{8},   
\label{generalformdsquarelsquare}
\ee
with scalar curvature  ${R}=2$ for two-sphere.  
 Therefore the eigenvalues of $(-i\fsl{\mathcal{D}})^2$ are obtained as 
\be
(-i\fsl{\mathcal{D}})^2 
= (j+\frac{1}{2}-g)(j+\frac{1}{2}+g)=n(2g+n), 
\ee
and those of the Dirac-Landau operator are  
\be
\pm \lambda_n=\pm\sqrt{n(2g+n)}~~~~~~(n=0, 1, 2, \cdots).
\ee
The eigenstates of the square of the Dirac-Landau operator are exactly same as of the $SU(2)$ Casimir $\bs{J}^2$. For $n=0$, the eigenstates of $(-i\fsl{\mathcal{D}})^2$ are ${\Upsilon'}^{g}_{j=g-\frac{1}{2}, m}$ (\ref{minimjstates}) with degeneracy $2g$, and for $n=1, 2, \cdots$ the eigenstates are ${\Upsilon'}^{g}_{j=g-\frac{1}{2}+n, m}$ and $\Upsilon^{g}_{j=g-\frac{1}{2}+n, m}$ (\ref{higherjstates}) with degeneracy $4(g+n)$.

From Eqs.(\ref{JfromLs}) and (\ref{diracopfrometh}), we can verify that  the Dirac-Landau operator itself is invariant  under the $SU(2)$ rotations,
\be
[\bs{J}, \fsl{\mathcal{D}}]= \begin{pmatrix}
0 &  -\bs{L}^{(g-\frac{1}{2})} \eth_-^{(g+\frac{1}{2})} +\eth_-^{(g+\frac{1}{2})} \bs{L}^{(g+\frac{1}{2})}\\
 \bs{L}^{(g+\frac{1}{2})} \eth_+^{(g-\frac{1}{2})} -\eth_+^{(g-\frac{1}{2})} \bs{L}^{(g-\frac{1}{2})}  &  0 
\end{pmatrix}=0,  
\ee
where  (\ref{commutationrelaethls}) was  used. 
Since the Dirac operator is invariant under the $SU(2)$ transformation, the relativistic Landau levels have the $SU(2)$ degeneracy and  the eigenstates of the Dirac-Landau operator may be constructed by some linear combination of the eigenstates of $(-i\fsl{\mathcal{D}})^2$, $i.e.$, 
${\Upsilon'}^{g}_{j, m}$ and $\Upsilon^{g}_{j, m}$. 
The Dirac-Landau operator also respects the chiral ``symmetry'':   
\be
\{ -i\fsl{\mathcal{D}}, \sigma_z\}=0, 
\label{diracchiralsymm}
\ee
and  the eigenstates for opposite sign eigenvalues are related by the chiral transformation\footnote{The Dirac operator does not commute with the chiral matrix,  
\be
[-i\fsl{\mathcal{D}}, \sigma_z] \neq 0,  
\ee
and hence there do not exist simultaneous eigenstates of the Dirac-Landau operator and the chiral matrix except for the zero modes   (\ref{zero-modeeigendirac})} except for the zero modes.  

\subsubsection{Zero modes ($n=0$)}

For  $n=0$, the relativistic Landau level and the $SU(2)$ index are respectively given by    
\be
\lambda_{n=0}=0, ~~~~j=g-\frac{1}{2},  
\ee
and the corresponding zero modes are\footnote{For $g<0$, the zero modes are given by 
\be
\begin{pmatrix}
0\\
{Y}^{-|g|+\frac{1}{2}}_{|g|-\frac{1}{2} ,m}(\theta, \phi)  
\end{pmatrix}. 
\ee
} 
\be
\Psi_{\lambda_0=0, m}^g(\theta, \phi)=\begin{pmatrix}
{Y}^{g-\frac{1}{2}}_{g-\frac{1}{2} ,m}(\theta, \phi) \\
0
\end{pmatrix} ~~(m=-g+\frac{1}{2}, -g+\frac{3}{2}, \cdots, g-\frac{1}{2}),  \label{zero-modeeigendirac}
\ee
where 
\begin{align}
{Y}^{g-\frac{1}{2}}_{g-\frac{1}{2}, m}(\theta, \phi)  
&=(-1)^{g+m-\frac{1}{2}} 
 \sqrt{\frac{(2g)!}{4\pi (g+m-\frac{1}{2})! (g-m-\frac{1}{2})!}}(\sin\frac{\theta}{2})^{(m+g-\frac{1}{2})}(\cos\frac{\theta}{2})^{(-m+g-\frac{1}{2})} \cdot e^{im\phi}\nn\\
&=(-1)^{g+m-\frac{1}{2}} 
 \sqrt{\frac{(2g)!}{4\pi (g+m-\frac{1}{2})! (g-m-\frac{1}{2})!}}u^{-m+g-\frac{1}{2}}v^{m+g-\frac{1}{2}}.  
\end{align}
The zero modes are equal to the lowest Landau level monopole harmonics (\ref{monopoleharmolll}) with  
the reduced monopole charge from $g$ to  $g-\frac{1}{2}$. 
The degeneracy is  
\be
2(g-\frac{1}{2})+1=2g. 
\label{degezeromodes}
\ee
It is easy to see  that  $\Psi_{\lambda_0=0, m}^g$ (\ref{zero-modeeigendirac}) are the Dirac operator zero modes  with the formula (\ref{ladderschargemonohar}) and  (\ref{diracopfrometh}).  
For $g=3/2$, we have three fold degenerate zero modes: 
\begin{align}
&\Psi^{3/2}_{0,1}= \frac{1}{2}\sqrt{\frac{3}{\pi}}\sin^2\frac{\theta}{2} e^{i\phi} \begin{pmatrix}
1 \\
0
\end{pmatrix},~
&\Psi^{3/2}_{0,0}= -\frac{1}{2}\sqrt{\frac{3}{2\pi}}\sin\theta\begin{pmatrix}
1 \\
0
\end{pmatrix},~
&\Psi^{3/2}_{0,-1}=  \frac{1}{2}\sqrt{\frac{3}{\pi}}\cos^2\frac{\theta}{2}e^{-i\phi}  \begin{pmatrix}
1 \\
0
\end{pmatrix}, 
\end{align}
which are in accordance with the results of  Ref.\cite{Gozalez-Guinea-Vozmediano-1992}. 
The degeneracy of zero modes is expected from  the index theorem \cite{Hasebe-2014-1,Dolan2003};  the 1st Chern number of the monopole gauge field configuration (\ref{sphericaexfieldst}) is given by     
\be
c_1=\frac{1}{2\pi}\int_{S^2} F= 2g,   
\ee
which is equal to (\ref{degezeromodes}). 

\subsubsection{Non-zero modes ($n=1, 2, \cdots$)}

We take a linear combination of ${\Upsilon'}^{g}_{j, m}(\theta, \phi)$ and $\Upsilon^{g}_{j, m}(\theta, \phi)$ so that it becomes the eigenstate of $-i\fsl{\mathcal{D}}$  with non-zero eigenvalue:  
\be
\pm \lambda_n =\pm \sqrt{n(n+2g)} ~~~~~(n=1,2, \cdots).
\label{eigenvaluesofdirac}
\ee
 With the aid of  (\ref{ladderschargemonohar}), the  linear combination is readily obtained by taking a linear combination of ${\Upsilon'}^{g}_{j, m}(\theta, \phi)$ and $\Upsilon^{g}_{j, m}(\theta, \phi)$ with  same weights:   
\be
\Psi_{\pm \lambda_n, m}^g=
\frac{1}{\sqrt{2}} ({\Upsilon'}^{g}_{j, m}(\theta, \phi)\mp i\Upsilon^{g}_{j, m}(\theta, \phi) )
\label{eigennonzerogdiraceig}
\ee
or 
\be
\Psi_{\pm \lambda_n, m}^g 
=\frac{1}{\sqrt{2}}\begin{pmatrix}
Y^{g-\frac{1}{2}}_{j=g-\frac{1}{2}+n, m}(\theta, \phi) \\
\mp iY^{g+\frac{1}{2}}_{j=g+\frac{1}{2}+(n-1), m}(\theta, \phi)
\end{pmatrix},
\label{eigennonzerogdiraceig}
\ee
where 
\begin{subequations}
\begin{align}
&j=g-\frac{1}{2}+n,\\ 
&m=-j, -j+1, \cdots, j-1, j. 
\end{align}
\end{subequations}
One may directly check that $\Psi_{\pm \lambda_n, m}^g$ (\ref{eigennonzerogdiraceig}) are indeed the eigenstates of the Dirac operator (\ref{diracopfrometh}) of the eigenvalues (\ref{eigenvaluesofdirac}) using the formula (\ref{ladderschargemonohar}).  
Notice that, when $g$ is an integer  (half-integer), $j$ and $m$ should be half-integers (integers)\footnote{In the non-relativistic case (\ref{ladderoppangularmagg}), when $g$ is an integer (half-integer), $j$ and $m$ should  be integers  (half-integers).}.  
Both $\Psi_{+ \lambda_n, m}^g$ and $\Psi_{- \lambda_n, m}^g$  are  $SU(2)$ irreducible representations with the $SU(2)$ index $j$, and the relativistic Landau levels,  $+\lambda_n$ and  $-\lambda_n$, respectively have  the following degeneracy:   
\be
2j+1=2(g+n). 
\ee
For $g=1/2$,  three fold degenerate eigenstates at $+\lambda_{n=1}=\sqrt{2}$ are given by  
\begin{align}
&\Psi^{1/2}_{ \sqrt{2}, 1}=-\sqrt{\frac{3}{8\pi}}
\begin{pmatrix}
\sqrt{2}\sin\frac{\theta}{2}\cos\frac{\theta}{2} \\
i \sin^2\frac{\theta}{2}
\end{pmatrix}e^{i\phi}, ~~~~~~\Psi^{1/2}_{\pm \sqrt{2}, 0}=\frac{1}{4}\sqrt{\frac{3}{\pi}}
\begin{pmatrix}
\sqrt{2}\cos\theta \\
 i \sin\theta
\end{pmatrix},\nn\\
&\Psi^{1/2}_{\pm \sqrt{2}, -1}=\sqrt{\frac{3}{8\pi}}
\begin{pmatrix}
\sqrt{2}\sin\frac{\theta}{2}\cos\frac{\theta}{2} \\
- i \sin^2\frac{\theta}{2}
\end{pmatrix}e^{-i\phi}. 
\end{align}

\begin{figure}[tbph]\center
\includegraphics*[width=100mm]{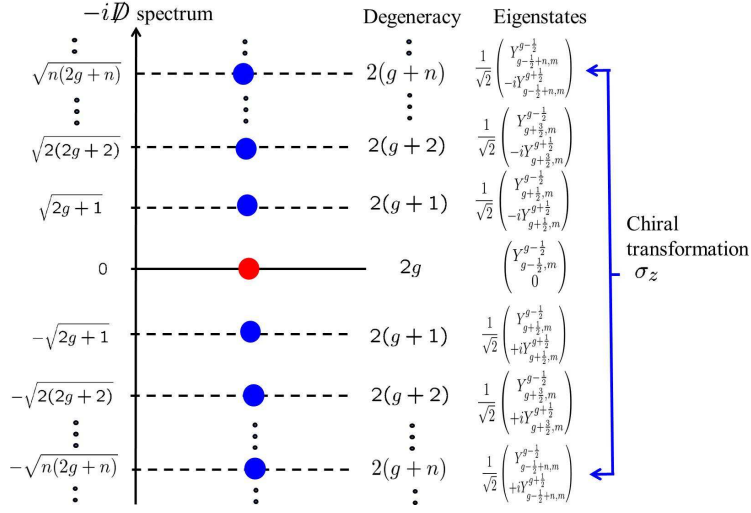}
\caption{ The Dirac-Landau operator eigenvalues,  eigenstates and   degeneracy. Eigenstates with opposite  sign eigenvalues  are related by the chiral transformation.    }
\label{RelationsFig}
\end{figure}

We add several comments here.  
First,  (\ref{eigennonzerogdiraceig}) 
consists of the monopole harmonics of the $n$th non-relativistic Landau level  for monopole charge $g-\frac{1}{2}$ (upper component) and the monopole harmonics of the $(n-1)$th non-relativistic Landau level  for monopole charge $g+\frac{1}{2}$ (lower component). This reminds the eigenstates of the Dirac-Landau Hamiltonian on a plane (see Refs.\cite{Shon-Ando-1998,Zheng-Ando-2002} for instance): 
\be
\frac{1}{\sqrt{2}}
\begin{pmatrix}
|n\rangle \\
|n-1\rangle
\end{pmatrix}. 
\ee
In the limit $g>\!\!> n$, the relativistic Landau levels on a plane, $\pm \sqrt{2B \cdot n}$, are reproduced from  (\ref{eigenvaluesofdirac}) with $B=g$\footnote{More precisely, to reproduce the plane result, we recover the  sphere radius $r$ as $\pm \lambda_n=\pm \frac{1}{r}\sqrt{n(n+2g)}$ and take the thermodynamic limit, $r,~g~\rightarrow~\infty$ with fixing $B=g/r^2$ finite.}.  
Second, for $g=0$, (\ref{eigennonzerogdiraceig}) reduces to the free Dirac operator eigenstates with  eigenvalues  $\pm \lambda_n=\pm n$ $(n=1, 2,   \cdots)$: 
\be
\Psi^{g=0}_{\pm n, m}(\theta, \phi)= 
\frac{1}{\sqrt{2}}\begin{pmatrix}
{Y}^{-\frac{1}{2}}_{n-\frac{1}{2} ,m}(\theta, \phi) \\
\mp i{Y}^{\frac{1}{2}}_{n-\frac{1}{2} ,m} (\theta, \phi)
\end{pmatrix}. 
\ee
This is a concise representation of the Abrikosov's result \cite{AbrikosovJr2002, AbrikosovJr2001}.  
   Third, $\Psi^{g}_{+\lambda_n, m}$ and $\Psi^{g}_{-\lambda_n, m}$ are related by the chiral transformation as expected from (\ref{diracchiralsymm}): 
\be
\Psi^{g}_{\mp\lambda_n, m} =\sigma_z \Psi^{g}_{\pm\lambda_n, m}.
\ee
The relativistic Landau levels and corresponding eigenstates are summarized  in Fig.\ref{RelationsFig}. 

\section{Relations to the Pauli-Schr\"odinger Non-Relativistic System}\label{sec:relationpaulischnon}

We have discussed the relativistic Landau problem on a sphere. 
In non-relativistic quantum mechanics, the Landau problem with spin degrees of freedom is described by the Pauli-Schr\"odinger Hamiltonian in a monopole background.  
The eigenvalue problem of the Pauli-Schr\"odinger Hamiltonian was solved by   Kazama et al \cite{Kazama-Yang-Gldhaber-1977}, in which the eigenvalues of the parity operator that constitutes the Pauli-Schr\"odinger Hamiltonian turned out to be  
\be
\bs{\Lambda}^{(g)}\cdot \bs{\sigma}=\pm \lambda_n -1.  
\ee
Here, $\pm \lambda_n= \pm \sqrt{n(2g+n)}$ are exactly the eigenvalues of the Dirac-Landau operator. This implies a close relation between the relativistic Landau model and the Pauli-Schr\"odinger system.    
 In this section, we demonstrate that these two systems are indeed related by a simple $SU(2)$ ``gauge'' transformation. For this goal, 
we generalize the work of   Abrikosov about free Dirac operator \cite{AbrikosovJr2002, AbrikosovJr2001} to include monopole gauge field.

\subsection{The $SU(2)$ ``gauge'' transformation and $SO(3, 1)$ algebra}

Abrikosov showed that the free Dirac operator eigenstates and the spinor spherical harmonics are related by the $SU(2)$ transformation\cite{AbrikosovJr2002, AbrikosovJr2001}\footnote{
With use of  $D$ functions [Appendix \ref{append:secmaurer}], $V(\theta, \phi)$ and $V(\theta, \phi)^{\dagger}$ are  represented as 
\begin{subequations}
\begin{align}
&V(\theta,\phi)=
\begin{pmatrix}
{D}_{\frac{1}{2}, \frac{1}{2}, \frac{1}{2}}(\phi, \theta, 0) & {D}_{\frac{1}{2}, \frac{1}{2}, -\frac{1}{2}}(\phi, \theta, 0) \\
{D}_{\frac{1}{2}, -\frac{1}{2}, \frac{1}{2}}(\phi, \theta, 0) & {D}_{\frac{1}{2}, -\frac{1}{2}, -\frac{1}{2}}(\phi, \theta, 0)
\end{pmatrix}
=
\begin{pmatrix}
u & -v^* \\
v & u^*
\end{pmatrix},  \\
&V(\theta, \phi)^{\dagger}=\begin{pmatrix}
D_{\frac{1}{2}, \frac{1}{2}, \frac{1}{2}}(0, -\theta, -\phi) &   D_{\frac{1}{2}, \frac{1}{2}, -\frac{1}{2}}(0, -\theta, -\phi) \\
 D_{\frac{1}{2}, -\frac{1}{2}, \frac{1}{2}}(0, -\theta, -\phi)  & D_{\frac{1}{2}, -\frac{1}{2}, -\frac{1}{2}}(0, -\theta, -\phi)
\end{pmatrix}=
\begin{pmatrix}
u^* & v^* \\
-v & u
\end{pmatrix}, \label{Vdaggerdfuncs}
\end{align}
\end{subequations}
where $u$ and $v$ are the components of the  Hopf spinor (\ref{hopfspinorcompo}).    
}:  
\be
V(\theta,\phi)\equiv 
e^{-i\frac{1}{2}\sigma_z \phi} e^{-i\frac{1}{2}\sigma_y \theta} = 
\begin{pmatrix}
\cos\frac{\theta}{2} e^{-i\frac{1}{2}\phi} & -\sin\frac{\theta}{2} e^{-i\frac{1}{2}\phi} \\
\sin\frac{\theta}{2} e^{i\frac{1}{2}\phi} & \cos\frac{\theta}{2} e^{i\frac{1}{2}\phi} 
\end{pmatrix}. 
\label{expresionVexp}
\ee
$V(\theta, \phi)$ is the  $SU(2)$ matrix that induces a spacial rotation of the Pauli matrices: 
\be
V(\theta,\phi)^{\dagger}\sigma_i V(\theta,\phi)=\sigma_j R_{ji}(\theta,\phi), 
\ee
where 
\be
R_{ij}(\theta,\phi)=\begin{pmatrix}
\cos\theta\cos\phi & \cos\theta\sin\phi & -\sin\theta \\ 
-\sin\phi & \cos\phi & 0 \\
\sin\theta\cos\phi & \sin\theta\sin\phi & \cos\theta
\end{pmatrix}. 
\ee
Notice that $V$ also generates a $SU(2)$ pure gauge field  $(dW+iW^2=0)$ as  
\be
W 
=
-iV^{\dagger}dV 
=
\frac{1}{2}
\begin{pmatrix}
-\cos\theta d\phi & id\theta +\sin\theta d\phi \\
-id\theta +\sin\theta d\phi & \cos\theta d\phi
\end{pmatrix},   
\ee
and the diagonal part of $W$ gives the $U(1)$ monopole gauge field (\ref{u1gauschfiel}):    
\be
A =-ig~\!\tr (\sigma_z V^{\dagger}dV). 
\ee
Thus interestingly, the role of $V(\theta, \phi)$ is two-fold: One is 
the  $SO(3)$ spacial rotation of the Pauli matrices, and the other is the  $SU(2)$ gauge transformation whose $U(1)$ part corresponds to the monopole. In the former case, the Pauli matrices of $V$ are interpreted as the 
spacial rotation generators,  while in the latter they are the gauge group generators. 

While both $\bs{J}$ and $-i\fsl{\mathcal{D}}$ are (Pauli) matrix valued differential operators,  under  the $V$ transformation they are completely decoupled to a differential operator part and Pauli matrix part:  
\begin{subequations}
\begin{align}
&V(\theta, \phi) ~\bs{J}~ V(\theta, \phi)^{\dagger}= \bs{L}^{(g)}+\frac{1}{2}\bs{\sigma}, \\
&V(\theta, \phi)~(-i\fsl{\mathcal{D}})~V(\theta, \phi)^{\dagger}=\bs{K}^{(g)}\cdot \bs{\sigma}. 
\end{align}\label{trasnrelatononre}
\end{subequations}
Here, $\bs{L}^{(g)}$ is the non-relativistic angular momentum operator 
(\ref{nonretotalangmo}) while $\bs{K}^{(g)}$ represents ``boost'' operator given by 
\begin{align}
&K_x^{(g)}\equiv -i\cos\theta\cos\phi \frac{\partial}{\partial \theta} +i\frac{1}{\sin\theta}\sin\phi\frac{\partial}{\partial\phi} -g \cot\theta \sin\phi  +i\sin\theta\cos\phi, \nn\\
&K_y^{(g)}\equiv -i\cos\theta\sin\phi \frac{\partial}{\partial \theta} -i\frac{1}{\sin\theta}\cos\phi\frac{\partial}{\partial\phi}+g \cot{\theta} \cos\phi  +i\sin\theta\sin\phi , \nn\\
&K_z^{(g)}\equiv i\sin\theta\frac{\partial}{\partial\theta}+i\cos\theta. 
\label{compboostop}
\end{align}
The Dirac-Landau operator is transformed to the ``helicity operator'',   $\bs{K}^{(g)}\cdot \bs{\sigma}$. 
Unlike  $\mathcal{D}_{\mu}$ (\ref{spherecovderi}), $K_i^{(g)}$ are simple differential operators (not matrix valued).   
The role of $V$ becomes even transparent in the inverse transformation of (\ref{trasnrelatononre}):   
\begin{subequations}
\begin{align}
&V^{\dagger} ~\bs{L}^{(g)}~V+V^{\dagger}~\frac{1}{2}\bs{\sigma} ~ V = \bs{J} , \\
&V^{\dagger}~ \bs{K}^{(g)}~V\cdot V^{\dagger}~\bs{\sigma} V=-i\fsl{\mathcal{D}}. 
\end{align}\label{trasnnonrelatore}
\end{subequations}
In (\ref{trasnnonrelatore}), 
$V$ acts as $SU(2)$  gauge transformation for $\bs{K}^{(g)}$ and $\bs{L}^{(g)}$, while 
acts as $SO(3)$ spacial rotation for $\bs{\sigma}$, as mentioned above.

$\bs{K}^{(g)}$ is  concisely represented as 
\be
\bs{K}^{(g)}=-i\bs{D}|_{r=1}+i\frac{1}{r}{\bs{x}}, 
\label{concisereprKs}
\ee
where $\bs{D}$ represents the Cartesian covariant derivatives in  $\it{flat}$ 3D space\footnote{
For comparison,  we represent the Dirac operator in flat 3D space by spherical coordinates:    
\begin{align}
-i\sum_{i=1}^3\sigma_i\frac{\partial}{\partial x_i}&=-i\sigma_x\frac{\partial}{\partial x} -i\sigma_y\frac{\partial}{\partial y} -i\sigma_z\frac{\partial}{\partial z} \nn\\
&=-i\sigma_x ( \frac{\cos\theta\cos\phi}{r} \frac{\partial}{\partial\theta}-\frac{\sin\phi}{r\sin\theta} \frac{\partial}{\partial \phi}  +\sin\theta\cos\phi\frac{\partial}{\partial r} )\nn\\
&~~-i\sigma_y (\frac{\cos\theta\sin\phi} {r}\frac{\partial}{\partial \theta}    +\frac{\cos\phi}{r\sin\theta}\frac{\partial}{\partial\phi} +\sin\theta \sin\phi\frac{\partial}{\partial r} )-i\sigma_z (-\frac{\sin\theta}{r}\frac{\partial}{\partial \theta} +\cos\theta \frac{\partial}{\partial r} ). 
\end{align}
}:  
\be
D_i=\partial_i -iA_i, ~~~(i=x,y,z)
\ee
with the gauge field (\ref{gaugecartesianfield}). 
Notice that $i\frac{1}{r}\bs{x}$ of (\ref{concisereprKs}) is non-hermitian and comes from the spin-connection term of the original Dirac-Landau operator. 
With the explicit form of $\bs{K}^{(g)}$ (\ref{compboostop}) and $\bs{L}^{(g)}$ (\ref{explicitschwingerexsu2ooperators}), $\bs{K}^{(g)}$ and $\bs{L}^{(g)}$ satisfy the $SO(3,1)$ algebra: 
\begin{align}
&[K_i^{(g)}, K^{(g)}_j]=-i\epsilon_{ijk}L^{(g)}_k,  \nn\\
&[{L}_i^{(g)}, K^{(g)}_j]=i\epsilon_{ijk} {K}^{(g)}_k, \nn\\
&[{L}_i^{(g)}, L^{(g)}_j]=i\epsilon_{ijk} {L}^{(g)}_k,  
\label{so31kandl}
\end{align}
and hence we refer to $\bs{K}^{(g)}$ as ``boost operator''. 
Eq.(\ref{so31kandl}) holds even if the non-hermitian term $i\frac{1}{r}\bs{x}$ was not present in (\ref{concisereprKs}). 
The square of ${\bs{K}}^{(g)}$ is explicitly represented as 
\be
{\bs{K}^{(g)}}^2=-\frac{1}{\sin\theta}\partial_{\theta}(\sin\theta\partial_{\theta})-\frac{1}{\sin^2\theta}{\partial_{\phi}}^2-2ig\frac{\cos\theta}{\sin^2\theta}\partial_{\phi}+g^2\frac{\cos^2\theta}{\sin^2\theta}+1, \label{squareKequal}
\ee
which is\footnote{The last term $1$ of (\ref{squareKequal}) comes from the non-hermitian term (\ref{concisereprKs}).}  
\be
{\bs{K}^{(g)}}^2={\bs{L}^{(g)}}^2-g^2+1={\bs{\Lambda}^{(g)}}^2+1. 
\label{ksquareislamsqare}
\ee
${\bs{K}^{(g)}}^2$ is essentially the non-relativistic Landau Hamiltonian (\ref{hamiltonianmonopolenonrela}): 
\be
H=\frac{1}{2M}{\bs{\Lambda}^{(g)}}^2=\frac{1}{2M}({\bs{K}^{(g)}}^2-1). 
\label{nonrelahammono}
\ee

\subsection{Relations to spinor monopole harmonics}

 Here, we give a detail discussion on the helicity operator, $\bs{K}^{(g)}\cdot \bs{\sigma}$. 
From the algebra (\ref{so31kandl}), it is verified that the square of the helicity operator yields a non-relativistic Hamiltonian, 
\be
H'=\frac{1}{2M}((\bs{K}^{(g)}\cdot \bs{\sigma})^2-1)=\frac{1}{2M}({\bs{K}^{(g)}}^2+\bs{L}^{(g)}\cdot \bs{\sigma}-1). 
\ee
With use of (\ref{squareKequal}), we have 
\be
H'=\frac{1}{2M}({\bs{\Lambda}^{(g)}}^2+\bs{L}^{(g)}\cdot \bs{\sigma})= \frac{1}{2M}({\bs{\Lambda}^{(g)}}^2 +  \bs{\Lambda}^{(g)}\cdot \bs{\sigma} -\bs{F}\cdot \bs{\sigma}). 
\label{Ksigmasquare}
\ee
Here, $\frac{1}{2M}{\bs{\Lambda}^{(g)}}^2$ denotes the non-relativistic Landau Hamiltonian (\ref{nonrelahammono}), $\bs{\Lambda}^{(g)}\cdot\bs{\sigma}$ represents the spin-orbit coupling term known as the Parity operator, and $\bs{F}\cdot\bs{\sigma}$ stands for the Zeeman coupling. $H'$  is  a supersymmetric quantum mechanical Hamiltonian, since it is $SU(2)$ gauge equivalent to 
$(-i\fsl{\mathcal{D}})^2$ [see Sec.\ref{subsection:susyqm} for details] up to a constant. 
From (\ref{so31kandl}) and (\ref{ksquareislamsqare}), we have 
\begin{subequations}
\begin{align}
&(\bs{K}^{(g)}\cdot \bs{\sigma})^2={\bs{L}^{(g)}}^2+{\bs{L}^{(g)}}\cdot \bs{\sigma} -(g+1)(g-1)=(\bs{L}^{(g)}+\frac{1}{2}\bs{\sigma})^2-g^2+\frac{1}{4}, \label{ksigmasquare}\\
&[\bs{K}^{(g)}\cdot \bs{\sigma}, ~\bs{L}^{(g)}+\frac{1}{2}\bs{\sigma}]=[\bs{K}^{(g)}, \bs{L}^{(g)}]\cdot \bs{\sigma}+\frac{1}{2}\bs{K}^{(g)}\cdot [\bs{\sigma}, \bs{\sigma}]=i(\bs{K}^{(g)}\times \bs{\sigma}-\bs{K}^{(g)}\times \bs{\sigma})=0, 
\end{align}
\end{subequations}
which correspond to 
\begin{subequations}
\begin{align}
&(-i\fsl{\mathcal{D}})^2=\bs{J}^2-g^2+\frac{1}{4}, \\
&[-i\fsl{\mathcal{D}}, \bs{J}]=0. 
\end{align}
\end{subequations}
The $SU(2)$ Casimir  eigenvalues for  $\bs{L}^{(g)}+\frac{1}{2}\bs{\sigma}$ are 
\be
(\bs{L}^{(g)}+\frac{1}{2}\bs{\sigma})^2=j(j+1), 
\ee
with $j=g-\frac{1}{2}+n$ ($n=0, 1, 2, \cdots$),  
and then from (\ref{ksigmasquare}) the eigenvalues of $(\bs{K}\cdot \bs{\sigma})^2$ are  
\be
(j+\frac{1}{2})^2-g^2=n(n+2g), 
\ee
and hence 
\be
\bs{K}^{(g)}\cdot \bs{\sigma}=\pm \sqrt{n(n+2g)}, 
\ee
which are identical to the relativistic Landau level (\ref{eigenvaluesofdirac}) as expected.  
In a similar manner to Sec.\ref{sec:eigenstatediracop}, we can derive the eigenstates of the helicity operator $\bs{K}^{(g)}\cdot \bs{\sigma}$.   
The eigenstates of  the $SU(2)$ Casimir for $\bs{L}+\frac{1}{2}\bs{\sigma}$, 
\begin{subequations}
\begin{align}
&(\bs{L}^{(g)}+\frac{1}{2}\bs{\sigma})^2 \Omega_{j,m}=j(j+1)\Omega_{j,m}, \\
&(L_z^{(g)}+\frac{1}{2}\sigma_z)\Omega_{j,m}=m\Omega_{j,m}, ~~~~~~~~~(m=-j, -j+1, \cdots, j)
\end{align}
\end{subequations}
are given by the spinor monopole harmonics: 
\be\Omega_{j, m}^g=\frac{1}{\sqrt{2j}} 
\begin{pmatrix}
\sqrt{j+m} ~Y^g_{j-\frac{1}{2} , m-\frac{1}{2}} \\
\sqrt{j-m} ~Y^g_{j-\frac{1}{2}, m+\frac{1}{2}}
\end{pmatrix}, ~~{\Omega'}_{j, m}^g=\frac{1}{\sqrt{2j+2}} 
\begin{pmatrix}
-\sqrt{j-m+1} ~Y^g_{j+\frac{1}{2} , m-\frac{1}{2}} \\
\sqrt{j+m+1} ~Y^g_{j+\frac{1}{2}, m+\frac{1}{2}}
\end{pmatrix}. 
\label{omegasasaeigensu2}
\ee 
The eigenstates of the helicity operator $\bs{K}^{(g)}\cdot \bs{\sigma}$ with $\pm \lambda_n=\pm\sqrt{n(n+2g)}$ $(n=1,2, \cdots)$ are constructed by their linear combinations: 
\begin{align}
\Phi_{\pm \lambda_n, m}^g&\equiv \frac{1}{\sqrt{2}}(  {\Omega'}_{j, m}^g(\theta, \phi)  \pm i\Omega_{j, m}^g(\theta, \phi))\nn\\
&=\frac{1}{2}
\begin{pmatrix}
-\sqrt{\frac{j-m+1}{j+1}} ~Y^g_{j+\frac{1}{2} , m-\frac{1}{2}} \pm i\sqrt{\frac{j+m}{j}} ~Y^g_{j-\frac{1}{2} , m-\frac{1}{2}}
\\
\sqrt{\frac{j+m+1}{j+1}} ~Y^g_{j+\frac{1}{2}, m+\frac{1}{2}}\pm i\sqrt{\frac{j-m}{j}} ~Y^g_{j-\frac{1}{2}, m+\frac{1}{2}}
\end{pmatrix}. ~~~(j=g-\frac{1}{2}+n)
\label{lincomphiomega}
\end{align}    
The zero modes $\lambda_{n=0}=0$ are  
\be
\Phi^{g}_{\lambda_0=0, m}={\Omega'}^g_{g-\frac{1}{2}, m}=\frac{1}{\sqrt{2g+1}}
\begin{pmatrix}
-\sqrt{g+\frac{1}{2}-m} ~Y_{g, m-\frac{1}{2}}^g \\
\sqrt{g+\frac{1}{2}+m} ~Y_{g, m+\frac{1}{2}}^g
\end{pmatrix} ~~~~~(m=-g+\frac{1}{2}, -g+\frac{3}{2}, \cdots , g-\frac{1}{2}).
\ee
A bit of calculation\footnote{
The monopole harmonics are equivalent to the $D$ functions (\ref{reldfuncmonopoeharm}) with decomposition formula: 
\be
{D}_{l, m_1, m_2}\otimes {D}_{l', m'_1, m'_2} 
= \sum_{L, M_1, M_2}  C_{l, m_2;~ l', m_2'}^{L, M_2} ~{D}_{L, M_1, M_2}~ C^{L, M_1}_{l, m_1;~ l', m'_1}, 
\label{productsofDmatrix}
\ee
where 
$C_{l, m; ~l', m'}^{L, M}=\langle L, M|l, m;~ l', m'\rangle =\langle l, m;~ l', m'|L, M\rangle$    
are  the Clebsch-Gordan coefficients. 
Since ${D}_{l, m_1, m_2}$ have  two $SU(2)$ indices, $m_1$ and $m_2$, the $SU(2)$ angular momentum decomposition is respectively applied to two pairs, $(m_1, m'_1)$ and $(m_2, m'_2)$. 
To derive (\ref{upsilonomegas}), we used 
\be
V(\theta, \phi)^{\dagger}~{\Omega'}^g_{j, m}=\frac{1}{\sqrt{2j+1}}\begin{pmatrix}
-\sqrt{j+g+\frac{1}{2}}~Y^{g-\frac{1}{2}}_{j, m} \\
\sqrt{j-g+\frac{1}{2}}~Y^{g+\frac{1}{2}}_{j, m}
\end{pmatrix},  ~~~
V(\theta, \phi)^{\dagger}~\Omega_{j, m}^g=  \frac{1}{\sqrt{2j+1}}\begin{pmatrix}
\sqrt{j-g+\frac{1}{2}}~Y^{g-\frac{1}{2}}_{j, m} \\
\sqrt{j+g+\frac{1}{2}}~Y^{g+\frac{1}{2}}_{j, m}
\end{pmatrix},  
\label{relationsupsilonomega}
\ee
which is verified by  Eq.(\ref{Vdaggerdfuncs}) and  (\ref{productsofDmatrix}) with the following Clebsch-Gordan coefficients: 
\begin{align}
&C_{1/2, 1/2;~l, m}^{L, M}= 
\delta_{L, l+\frac{1}{2}} \delta_{M, m+\frac{1}{2}}\sqrt{\frac{l+m+1}{2l+1}} +\delta_{L, l-\frac{1}{2}} \delta_{M, m+\frac{1}{2}}\sqrt{\frac{l-m}{2l+1}}, \nn\\ 
&C_{1/2, -1/2;~l, m}^{L, M}= 
\delta_{L, l+\frac{1}{2}} \delta_{M, m-\frac{1}{2}}\sqrt{\frac{l-m+1}{2l+1}} -\delta_{L, l-\frac{1}{2}} \delta_{M, m-\frac{1}{2}}\sqrt{\frac{l+m}{2l+1}}. 
\end{align}
} shows that 
linear combinations of $\Omega_{j, m}^g$ and ${\Omega'}_{j, m}^g$ (\ref{omegasasaeigensu2}) are related to $\Upsilon^{g}_{j,m}$ and ${\Upsilon'}^{g}_{j,m}$ (\ref{higherjstates}) by the $SU(2)$ transformation: 
\begin{subequations}
\begin{align}
&{\Upsilon'}^{g}_{j, m}=V(\theta, \phi)^{\dagger} ~(-\cos\alpha \cdot {\Omega'}_{j, m}^g +\sin \alpha \cdot {\Omega}_{j, m}^g) ,\\
&\Upsilon^{g}_{j, m}=V(\theta, \phi)^{\dagger} ~(\sin\alpha \cdot {\Omega'}_{j, m}^g +\cos \alpha \cdot {\Omega}_{j, m}^g),  
\end{align}\label{upsilonomegas}
\end{subequations}
where  
\be
\tan \alpha = \sqrt{\frac{j-g+\frac{1}{2}}{j+g+\frac{1}{2}}}. 
\ee
Consequently we have  
\be
\Psi^g_{\pm \lambda_n, m}=\frac{1}{\sqrt{2}}({\Upsilon'}^{g}_{j, m}\mp i\Upsilon_{j, m}^{g})=-e^{\pm i\alpha}~ V(\theta, \phi)^{\dagger}~\Phi^g_{\pm\lambda_n, m}. 
\label{nonzerodirachelicity}
\ee
Thus up to the irrelevant phase factor,  $\Phi^g_{\pm\lambda, m}$ is transformed to  $\Psi^g_{\pm\lambda, m}$ by the $SU(2)$ matrix $V$. For zero modes,  
\begin{align}
\Psi_{\lambda_0=0, m}^g 
=-V(\theta, \phi)^{\dagger}~ {\Phi}^g_{\lambda_0=0, m}. 
\label{zerodirachelicity}
\end{align}

\subsection{Relations to the Pauli-Sch\"odinger eigenstates}

\begin{figure}[tbph]\center
\includegraphics*[width=130mm]{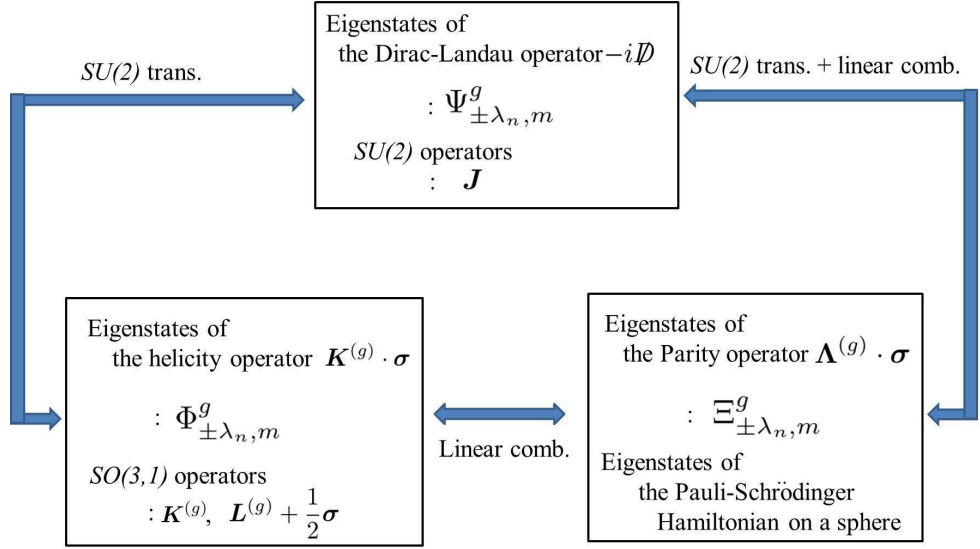}
\caption{ 
The eigenstates of the Dirac-Landau operator are related to those of the helicity operator by the $SU(2)$ transformations, (\ref{nonzerodirachelicity}) and (\ref{zerodirachelicity}). The linear combinations of the eigenstates  of the helicity operator give  the Parity operator eigenstates,   (\ref{nonzerohelicityparity}) and (\ref{zerohelicityparity}).    
The Dirac-Landau operator eigenstates are transformed to the Parity operator eigenstates by the $SU(2)$ transformations and the linear combinations, (\ref{nonzerodiracparity}) and (\ref{zerodiracparity}).   }
\label{MutRel.fig}
\end{figure}

Next, we establish relations between the relativistic Landau model and the Pauli-Sch\"odinger non-relativistic system.  
The Pauli-Sch\"odinger  Hamiltonian is given by 
\be
H_{\text{P-S}}=-\frac{1}{2M}\sum_{i=1}^3(\sigma_i D_i)^2=-\frac{1}{2M}\sum_{i=1}^3 D_i -\frac{1}{2M}\bs{F}\cdot \bs{\sigma}, 
\ee
where $D_i$ denote the covariant derivative in 3D $\it{flat}$ space  (\ref{covderiveflat}) and $\bs{F}$ represents an external magnetic field, in the present case, the monopole field strength (\ref{monopolegaugefieldnonrel}). In the spherical coordinates, $H_{\text{P-S}}$ is expressed as\footnote{Interestingly, the  Pauli-Sch\"odinger Lagrangian  enjoys the  $OSp(1|2)$ super-conformal symmetry and  $(\bs{\Lambda}^{(g)}\cdot\bs{\sigma}+1)$ plays the role of $OSp(1|2)$ Scasimir operator \cite{DHoker-Vinet-1984}. }     
\be
H_{\text{P-S}}
=  -\frac{1}{2M r^2}\frac{\partial}{\partial r} r^2\frac{\partial}{\partial r} +\frac{1}{2M r^2} (\bs{\Lambda}^{(g)}\cdot \bs{\sigma})(\bs{\Lambda}^{(g)}\cdot \bs{\sigma}+1).  \label{sqarediracandpauliham}
\ee 
On a sphere, we have  
\be
H_{\text{P-S}}|_{r=1}=\frac{1}{2M}(\bs{\Lambda}^{(g)}\cdot \bs{\sigma})(\bs{\Lambda}^{(g)}\cdot \bs{\sigma}+1).  
\label{squareofdiracopsymm}
\ee
Since the Pauli-Sch\"odinger Hamiltonian consists of the Parity operator $\bs{\Lambda}^{(g)}\cdot \bs{\sigma}$, the Parity operator eigenstates are automatically the eigenstates of the Pauli-Sch\"odinger Hamiltonian (\ref{squareofdiracopsymm}). 
 The eigenvalues of  $(\bs{\Lambda}^{(g)}\cdot \bs{\sigma}+1)$  are  exactly same as those of the helicity operator $\bs{K}^{(g)}\cdot \bs{\sigma}$, $\pm \lambda_n=\pm\sqrt{n(2g+n)}$, 
and the corresponding eigenstates of  $(\bs{\Lambda}^{(g)}\cdot \bs{\sigma}+1)$  are 
\cite{Kazama-Yang-Gldhaber-1977}  
\be
\Xi^{ g}_{\pm \lambda_n,  m} =\frac{1}{2}\biggl(\sqrt{1+\frac{g}{j+\frac{1}{2}}}   \pm  \sqrt{1-\frac{g}{j+\frac{1}{2}}}\biggr) \Omega^g_{j m } +   \frac{1}{2}\biggl(\mp \sqrt{1+\frac{g}{j+\frac{1}{2}}}   +\sqrt{1-\frac{g}{j+\frac{1}{2}}}\biggr) {\Omega'}^g_{j m } ~~~(n=1,2, \cdots),
\label{xiandeomegarel}
\ee
where $\Omega^g_{j m }$ and $ {\Omega'}^g_{j m }$ are the spinor monopole harmonics (\ref{omegasasaeigensu2}).  
The ``coincidence'' between the eigenvalues  of the parity operator $(\bs{\Lambda}^{(g)}\cdot \bs{\sigma} +1)$ and the helicity operator $\bs{K}^{(g)}\cdot\bs{\sigma}$ is understood by noticing that the relations between the Patiry operator and helicity operator: 
\be
(\bs{\Lambda}^{(g)}\cdot \bs{\sigma}+1)^2= {\bs{\Lambda}^{(g)}}^2 +\bs{L}^{(g)}\cdot \bs{\sigma}+1=(\bs{K}^{(g)}\cdot \bs{\sigma})^2, 
\ee
where we used the commutation relations of $\bs{\Lambda}^{(g)}$(\ref{covangmodef}): 
\be
[\Lambda_i^{(g)}, \Lambda_j^{(g)}]=-i\epsilon_{ijk}(L_k^{(g)}-2\Lambda_k^{(g)}). 
\ee
Therefore, the eigenvalues of $\bs{K}^{(g)}\cdot \bs{\sigma}$ and those of $\bs{\Lambda}^{(g)}\cdot\bs{\sigma}+1$ 
 are exactly  the same. 
From (\ref{lincomphiomega}) and (\ref{xiandeomegarel}), we can relate $\Xi^{g}_{\pm\lambda_n, m}$ and $\Phi^g_{\pm\lambda_n, m }$ as 
\begin{align}
&\Xi_{\lambda_n, m}^{g} =-\frac{1}{\sqrt{2}}(e^{- i \beta} \cdot \Phi_{ \lambda_n, m}^g + e^{ i \beta} \cdot \Phi_{ -\lambda_n, m}^g) , \nn\\
&\Xi_{-\lambda_n, m}^{g} =-i\frac{1}{\sqrt{2}}(e^{- i \beta} \cdot \Phi_{ \lambda_n, m}^g - e^{ i \beta} \cdot \Phi_{ -\lambda_n, m}^g), 
\label{nonzerohelicityparity}
\end{align}
where 
\be
\tan\beta =-\frac{j+\frac{1}{2}}{g} \biggl(1+\sqrt{1-\biggl(\frac{g}{j+\frac{1}{2}}\biggr)^2}\biggr). 
\ee
Consequently, relations between the eigenstates of $-i\fsl{\mathcal{D}}$ and $H_{\text{P-S}}$ are given by   
\be
\Psi_{\pm\lambda_n, m}^g=\frac{1}{\sqrt{2}}e^{\pm i\gamma} \cdot V(\theta, \phi)^{\dagger}~(\Xi_{\lambda_n, m}^g\mp i \Xi_{-\lambda_n, m}^g)~~~~~(n= 1, 2, \cdots),
\label{nonzerodiracparity}
\ee
where $\gamma~\equiv ~\alpha+\beta$, or 
\be
\tan \gamma = -\frac{j(j+\frac{1}{2})\biggl(\sqrt{j+g+\frac{1}{2}}+\sqrt{j-g+\frac{1}{2}}\biggl)}{j(j+\frac{1}{2})\sqrt{j-g+\frac{1}{2}}+g(g-\frac{1}{2})\sqrt{j+g+\frac{1}{2}}}. 
\ee
Similarly, the zero modes $(\lambda_{0}=0)$ are given by 
\be
\Xi_{\lambda_0=0, m}^g=-{\Omega'}^g_{g-\frac{1}{2}, m}=-{\Phi}^g_{\lambda_0=0, m}, 
\label{zerohelicityparity}
\ee
and then 
\be 
\Psi_{\lambda_0=0, m}^g 
=V(\theta, \phi)^{\dagger}~ {\Xi}^g_{\lambda_0=0, m}. 
\label{zerodiracparity}
\ee
Fig.\ref{MutRel.fig} summarizes the mutual relations discussed in this section.  

\section{Non-Commutative Geometry in Relativistic Landau Levels}\label{sect:noncommutativegeo}

\subsection{Landau level projection and non-commutative geometry}

By diagonalizing the Landau Hamiltonian, we obtain an infinite dimensional Hilbert space spanned by the monopole harmonics. The  Hilbert space consists of finite dimensional subspaces labeled by the Landau level index $n$. Sandwiching an operator of interest with the monopole harmonics, 
we have a matrix representation of the operator. 
In general, the matrix representation is given by an infinite dimensional matrix made of block matrices.  For instance,  matrix representation of  Cartesian coordinates is given by  
\be
x_i=\begin{pmatrix}
* & * & 0 & 0 & 0 & 0 & 0 \\
* & * & * & 0 & 0 & 0 & 0  \\
0 & * & * & X_i(n-1,n) & 0 & 0 & 0   \\
0 & 0 & X_i(n,n-1) & X_i(n,n) & X_i(n,n+1) & 0 & 0  \\
0 & 0 & 0 & X_i(n+1,n) & * & * & 0 \\
0 & 0 & 0 & 0 & * & * & *  \\
0 & 0 & 0 & 0 & 0 & * & *  
\end{pmatrix}, \label{matrixformxs}
\ee
where $X_i(n_1, n_2)$ denotes $(2g+2n_1+1)\times (2g+2n_2+1)$ 
block matrix  between $n_1$ and $n_2$th Landau levels.  (In the case of $x_i$, only the  matrix elements of adjacent and intra Landau levels take non-zero values.) 
The original coordinates $x_i$ are commutative: 
\be
x_ix_j=x_jx_i. 
\label{commurelaxs}
\ee
Let us concentrate the $n$th intra Landau level block of (\ref{commurelaxs}); from (\ref{matrixformxs}) the left-hand side gives 
\be
x_ix_j(n,n)=X_i(n, n-1)X_j(n-1, n)+X_i(n,n)X_j(n,n)+X_i(n, n+1)X_j(n+1, n), 
\label{lhsxixj}
\ee
while the right-hand side of (\ref{commurelaxs}) yields 
\be
x_jx_i(n,n)=X_j(n, n-1)X_i(n-1, n)+X_j(n,n)X_i(n,n)+X_j(n, n+1)X_i(n+1, n). 
\label{lhsxjxi}
\ee
Since (\ref{lhsxixj}) and (\ref{lhsxjxi}) are equal, we have 
\be
[X_i(n, n), X_j(n, n)]=-[X_i(n, n-1),X_j(n-1, n)]-[X_i(n+1, n),X_j(n, n+1)].   
\label{noncommutativealgebragene}
\ee
Though 
 each of the commutators on the right-hand side of (\ref{noncommutativealgebragene})  gives both inter and intra Landau level block matrices, the sum of  the commutators amounts to be an intra Landau level block matrix only:  
\be
-[X_i(n, n-1), X_j(n-1, n)]-[X_i(n+1, n), X_j(n, n+1)]=-i\alpha_{n}^{(g)}\epsilon_{ijk}X_k(n,n).  
\label{commuoffblock}
\ee
(Here, $\alpha_n^{(g)}$ denotes a proportional coefficient which will be identified as (\ref{defnoncompara})). It may be a good exercise for readers to check (\ref{commuoffblock}) in low dimensional matrices.     
Consequently, (\ref{noncommutativealgebragene}) can be rewritten as   
\be
[X_i(n, n), X_j(n, n)]=-i\alpha_{n}^{(g)}\epsilon_{ijk}X_k(n,n).   
\label{fuzzyspherealgebradef}
\ee
(\ref{fuzzyspherealgebradef}) is exactly the algebra of the fuzzy sphere \cite{Hoppe1982, Hoppe1990, madore1992}. 
 As demonstrated above, the off-diagonal blocks are the seed of the non-commutative geometry.  
  Though the coordinates are commutative in the whole Hilbert space, restricted to a  subspace, the coordinates (expressed by intra Landau level matrix elements) are no longer commutative due to the existence of the matrix elements between inter Landau levels.   The level projection is the heart of non-commutativity.

\subsection{Projection to the non-relativistic Landau levels}

\begin{figure}[tbph]\center
\includegraphics*[width=120mm]{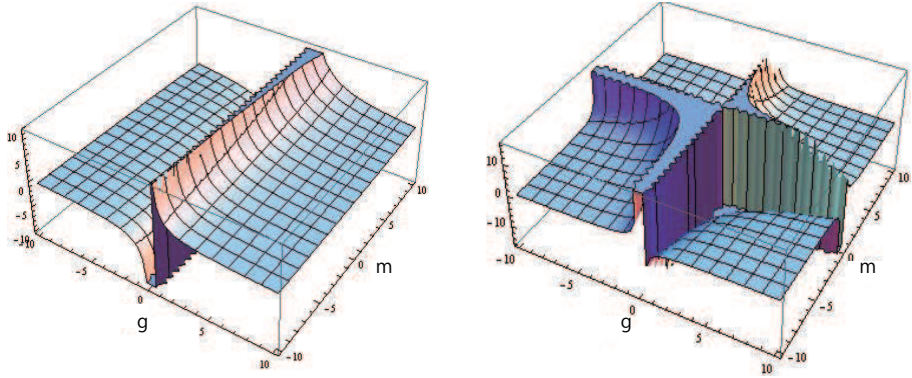}
\caption{ The left-figure shows 
$\frac{(X_1+iX_2)(n+1, n)}{(X_1+iX_2)(n,n)}$ $(n=5)$ with respect to the monopole charge $g$ and the magnetic quantum number $m$, while the right figure   shows  $\frac{ X_3(n+1, n )}{X_3(n,n)}$ $(n=5)$. In the limit $g\rightarrow \infty$, the ratios approach zero, implying that the inter-Landau level components (numerator) are negligible compared to the intra-Landau level components (denominator). (In the right figure, there exists a singularity around $m=0$  coming from the small intra-Landau level  components of $X_3$ around  $m=0$.)}
\label{ratiointerintra.fig}
\end{figure}

We expand more detail discussions about the appearance of the fuzzy geometry. 
The matrix elements of the coordinates (\ref{matrixformxs}) are  explicitly given by  
\begin{subequations}
\begin{align}
\langle Y^{g}_{l', m'}|\frac{1}{r}(x\pm iy)|Y^{g}_{l, m}\rangle &=\frac{g}{l(l+1)}\sqrt{(l\mp m)(l\pm m+1)}\delta_{l',l}\delta_{m', m\pm 1}\nn\\
&\pm \frac{1}{l+1}\sqrt{\frac{((l+1)^2-g^2)(l\pm m+2)(l\pm m+1)}{(2l+1)(2l+3)}}\delta_{l',l+1}\delta_{m', m\pm 1}\nn\\
&\mp \frac{1}{l}\sqrt{\frac{(l^2-g^2)(l\mp m)(l\mp m-1)}{(2l-1)(2l+1)}}\delta_{l',l-1}\delta_{m',m\pm 1}, \\
\langle Y^{g}_{l', m'}|\frac{1}{r}z|Y^{g}_{l, m}\rangle &=
\frac{g}{l(l+1)}m\delta_{l',l}\delta_{m', m}\nn\\
&-\frac{1}{l+1}\sqrt{\frac{((l+1)^2-g^2)((l+1)^2-m^2)}{(2l+1)(2l+3)}}\delta_{l',l+1}\delta_{m', m}\nn\\
&+\frac{1}{l}\sqrt{\frac{(l^2-g^2)(l^2-m^2)}{(2l-1)(2l+1)}}\delta_{l',l-1}\delta_{m',m},
\end{align}\label{matrixelecoordexp}
\end{subequations}
where the $SU(2)$ indices, $l$ and $l'$, are related to the Landau level indicies, $n$ and $n'$, as $l=g+n$ and $l'=g+n'$.   
The first components of the right-hand sides of (\ref{matrixelecoordexp}) represent the matrix elements of intra Landau level, $X_i(n, n)$, while the second and third terms stand for those of the adjacent inter Landau levels, $X_i(n,n')$ with $|n-n'|=1$.  
In the limit 
\be
g ~>\!>~n, 
\ee
which we refer to as the non-commutative limit,  the diagonal blocks $X_i(n,n)$ behave as   $O(1)$, while  
the off-diagonal blocks $X_i(n, n')$ $(|n-n'|=1)$  as  
$O(\sqrt{\frac{n}{g}})$.  
Thus in the non-commutative limit, the  intra Landau level block matrices become dominant compared to  inter Landau level block matrices [Fig.\ref{ratiointerintra.fig}].  
The intra Landau level matrix elements can be expressed as 
\be
\bs{X}(n, n)_{m', m}=\langle Y^g_{l, m'}|\bs{x}| Y^g_{l, m}\rangle=-r\frac{g}{l(l+1)} \langle Y^g_{l, m'}| \bs{L}^{(g)}| Y^g_{l, m}\rangle,  ~~~(l=g+n)
\ee
where  $ \langle Y^g_{l, m'}| \bs{L}^{(g)}| Y^g_{l, m}\rangle$ represents the ordinary $SU(2)$ matrices with  spin magnitude $l$: 
\begin{align}
&\langle Y_{l,m'}^g |L_{\pm}^{(g)}|Y^g_{l,m}\rangle = \sqrt{(l\mp m)(l\pm m+1)}\delta_{m', m\pm1}, \nn\\
&\langle Y_{l,m'}^g |L_z^{(g)}|Y^g_{l,m}\rangle = m\delta_{m', m}, 
\end{align}
and then  $\bs{X}(n ,n )$ is simply represented as    
\be
\bs{X}(n,n)= -\alpha^{(g)}_n  {\bs{S}}_{s=n+g}, 
\label{identxands}
\ee
where $\bs{S}_{s=n+g}$ represents the ordinary $(2s+1)\times (2s+1)$ $SU(2)$ matrices with spin magnitude $s=g+n$\footnote{For instance, $\bs{S}_{s=\frac{1}{2}} =\frac{1}{2}\bs{\sigma}$.}, and 
\be
\alpha^{(g)}_n=r\frac{g}{(g+n)(g+n+1)}.  
\label{defnoncompara}
\ee
The square of the radius of fuzzy sphere  is obtained as 
\be
\bs{X}\cdot \bs{X}={\alpha_n^{(g)}}^2 (g+n)(g+n+1)\equiv {R^{(g)}_n}^2,  
\ee
where 
\begin{align}
R_n^{(g)}&=\alpha_n^{(g)}\sqrt{(g+n)(g+n+1)}\nn\\
&=r~\frac{g}{\sqrt{(g+n)(g+n+1)}}~~~~(n=0,1,2,\cdots).
\end{align}
(Hereinafter, we abbreviate the Landau level index $n$ of $\bs{X}(n, n)$ for notational brevity.) 
One may find that the radius of fuzzy sphere  depends on the Landau level index $n$.   

\begin{figure}[tbph]\center
\includegraphics*[width=70mm]{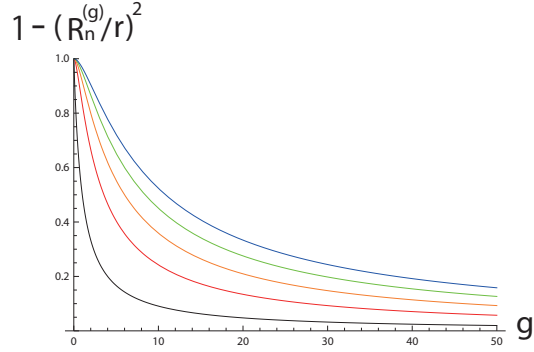}
\caption{ 
Behaviors of the  ratio  with respect to the monopole charge $g$. The black, red, orange, green and blue curves respectively correspond to  the Landau levels with $n=0, 1, 2, 3, 4$.  }
\label{MatrixElements.fig}
\end{figure}

Also based on 
(\ref{nonretotalangmo}), one can understand  the appearance of fuzzy sphere. 
In the $n$th Landau level, the matrix elements of the covariant angular momentum are derived as 
\be
\langle Y_{l,m'}^g |\boldsymbol{\Lambda}^{(g)}|Y^g_{l,m}\rangle =\biggl(1-\biggl(\frac{R^{(g)}_n}{r}\biggr)^2\biggr)
\langle Y_{l,m'}^g |\boldsymbol{L}^{(g)}|Y^g_{l,m}\rangle. \label{matrixcompLambda} 
\ee
Notice that
 the proportional factor on the right-hand side of (\ref{matrixcompLambda}) 
does not depend on magnetic angular momenta, $m$ and $m'$, as expected by the Wigner-Eckart theorem, so the proportional factor is solely determined by the  Landau level index $n$.    
Though matrix elements of $\bs{\Lambda}^{(g)}$ take non-zero values in each Landau level (\ref{matrixcompLambda}), the matrix elements become negligible compared to those of $\bs{L}^{(g)}$ in the non-commutative limit, ${R^{(g)}_n}/{r}\overset{{g}/{n}\rightarrow \infty}\longrightarrow 1$.  
Indeed, the factor, 
$1-(\frac{R^{(g)}_n}{r})^2=1-\frac{g^2}{(n+g)(n+g+1)}$, 
 monotonically decreases as  $g$ increases [Fig.\ref{MatrixElements.fig}]. 
Thus in the non-commutative limit, the covariant angular momentum no longer contributes to the total angular momentum in (\ref{nonretotalangmo}) and hence $\bs{x}$ can be identified with the operator $-\frac{r}{g}\bs{L}^{(g)}$ that satisfy the $SU(2)$ algebra of fuzzy sphere (\ref{fuzzyspherealgebradef}).

\subsection{Projection to the relativistic Landau levels}\label{sec:projrelLLs}

\begin{figure}[tbph]\center
\includegraphics*[width=60mm]{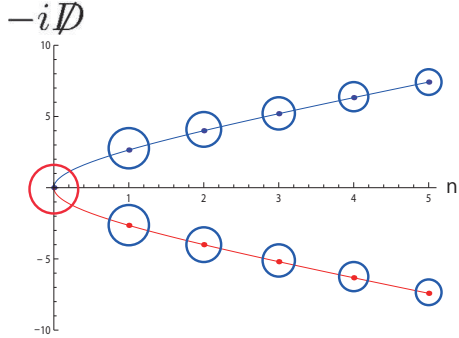}
\caption{The circles schematically represent the fuzzy spheres on the corresponding relativistic Landau levels. 
The sizes of two fuzzy spheres on the levels,  $+\lambda_n$  and $-\lambda_n$, are identical. The size monotonically decreases as $n$ increases. ($g=3$ and $M=2$ are adopted in the figure.) }
\label{radiirelnonrel1.fig}
\end{figure}

With the matrix elements by the monopole harmonics (\ref{identxands}), 
it is easy to derive matrix elements for the relativistic case.  
The eigenstates of the Dirac-Landau operator are respectively given by 
\begin{subequations}
\begin{align}
&n=0~~~~~~~~~:~~\Psi^g_{\lambda_0=0, m} =\begin{pmatrix}
Y^{g-\frac{1}{2}}_{g-\frac{1}{2}, m} \\
0 
\end{pmatrix}, \\
&n=1,2,\cdots~:~~\Psi_{\pm\lambda_n, m}^g
 =\frac{1}{\sqrt{2}}
\begin{pmatrix}
Y^{g-\frac{1}{2}}_{j=g-\frac{1}{2}+n ,m } \\
\mp iY^{g+\frac{1}{2}}_{j=g-\frac{1}{2}+n ,m }
\end{pmatrix}, 
\end{align}
\end{subequations}
and the matrix elements of $\bs{x}$ are derived as\footnote{(\ref{innerproxrel}) should be interpreted as the abbreviation form of $\bs{X}_{m, m'}
\equiv \langle \Psi^g_{\pm \lambda_n, m} |\bs{x}|\Psi^g_{\pm \lambda_n, m'}\rangle$. } 
\be
\bs{X}
\equiv \langle \Psi^g_{\pm \lambda_n} |\bs{x}|\Psi^g_{\pm \lambda_n}\rangle 
=-{\alpha'}_{n}^{(g)}\bs{S}_{s=n+g-\frac{1}{2}}, \label{innerproxrel}
\ee
where 
\begin{subequations}
\begin{align}
&n=0~~~~~~~~~:~~{\alpha'_0}^{(g)}= \alpha_0^{(g-\frac{1}{2})}=r\frac{1}{g+\frac{1}{2}} , \label{noncoscal0mode}\\
&n=1,2,\cdots~:~~{\alpha'}_{n}^{(g)}\equiv \frac{1}{2}(\alpha_{n}^{(g-\frac{1}{2})}+\alpha_{n-1}^{(g+\frac{1}{2})})=r\frac{g}{(g+n-\frac{1}{2})(g+n+\frac{1}{2})}. 
\end{align}
\end{subequations}
Notice that the matrix elements $\bs{X}$ are completely identical for positive and negative eigenvalues $\pm\lambda_n$. 
$X_i$ satisfy the  fuzzy sphere algebra:   
\be
[X_i, X_j]=-i{\alpha'}_{n}^{(g)}\epsilon_{ijk}X_k, 
\ee
and the squares of their radii are  given by 
\be
\bs{X}\cdot \bs{X}={{\alpha'}_{n}^{(g)}}^2 (n+g-\frac{1}{2})(n+g+\frac{1}{2})\equiv {{R'}_n^{(g)}}^2,  
\ee
where 
\begin{subequations}
\begin{align}
&n=0~~~~~~~~~:~~{R'}_{0}^{(g)}=\alpha_0^{(g-\frac{1}{2})} \sqrt{(g-\frac{1}{2})(g+\frac{1}{2})}=r\sqrt{\frac{g-\frac{1}{2}}{g+\frac{1}{2}}}, \\
&n=1,2,\cdots~:~~{R'}_{n}^{(g)}={\alpha'}_{n}^{(g)}\sqrt{(n+g-\frac{1}{2})(n+g+\frac{1}{2})}=r\frac{g}{\sqrt{(g+n-\frac{1}{2})(g+n+\frac{1}{2})}}.
\end{align}\label{relfuzzyradiiexp}
\end{subequations}
The sizes of the fuzzy spheres are ordered as [Figs.\ref{radiirelnonrel1.fig}]
\be
R'^{(g)}_{0} ~>R'^{(g)}_{1} ~>R'^{(g)}_{2} ~>~ \cdots. 
\ee

\begin{figure}[tbph]\center
\includegraphics*[width=95mm]{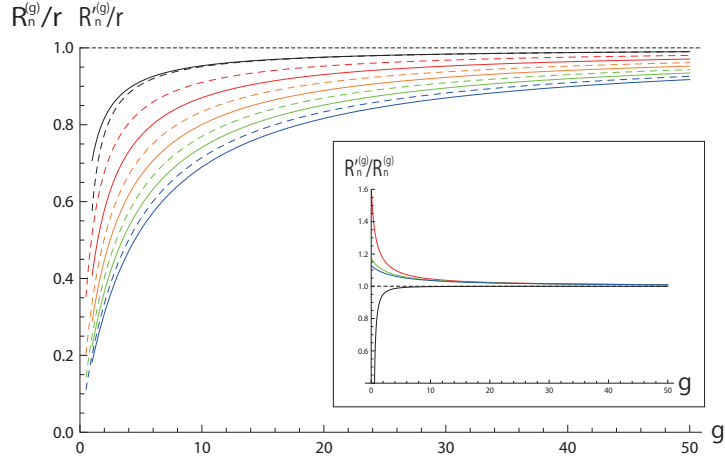}
\caption{Plot of the radii (\ref{relfuzzyradiiexp}) with respect to $g$. The solid and dashed curves are respectively for the non-relativistic and relativistic cases, $R_n^{(g)}/r$ and ${R'}_n^{(g)}/r$,  and  black, red, green, blue for  $n=0, 1, 2, 3$.  Inset depicts the ratios of (\ref{radiiratiosnonrelrelagaex})  with same color assignment for $n$.  }
\label{radiirelnonrel2.fig}
\end{figure}

Here, we compare  the sizes of the relativistic and non-relativistic fuzzy spheres. 
The ratios between the radii are given by 
\begin{align}
&n=0~~~~~~~~~:~~\frac{{R'}^{(g)}_0}{R_0^{(g)}}=\frac{\alpha_0^{(g-\frac{1}{2})}}{\alpha_0^{(g)}}~\sqrt{\frac{(g-\frac{1}{2})(g+\frac{1}{2})}{g(g+1)}}=\sqrt{\frac{(g-\frac{1}{2})(g+1)}{g(g+\frac{1}{2})}} ~< ~1, \nn\\
&n=1,2,\cdots~:~~\frac{{R'}^{(g)}_{n}}{R_{n}^{(g)}}=\frac{{\alpha'}_n^{(g)}}{\alpha_n^{(g)}}~\sqrt{\frac{(g+n-\frac{1}{2})(g+n+\frac{1}{2})}{(g+n)(g+n+1)}}=\sqrt{\frac{(g+n)(g+n+1)}{(g+n-\frac{1}{2})(g+n+\frac{1}{2})}} ~> ~1.  \label{radiiratiosnonrelrelagaex}
\end{align} 
Thus, the radius of the fuzzy sphere for $n=0$ reduces,  
 while those for $n=1,2,\cdots$ enhance. 
From (\ref{radiiratiosnonrelrelagaex}), the ratios are ordered as [Fig.\ref{radiirelnonrel2.fig}] 
\be
\frac{{R'}^{(g)}_{1}}{R_{1}^{(g)}} > \frac{{R'}^{(g)}_{2}}{R_{2}^{(g)}}> \frac{{R'}^{(g)}_{3}}{R_{3}^{(g)}} > \cdots > 1 > \frac{{R'}^{(g)}_{0}}{R_{0}^{(g)}}. 
\ee

\section{Mass Deformation and Balanced  Fuzzy Spheres}\label{sect:aclasstopolo}

We consider mass deformation of the relativistic Landau model. 
In real Dirac matter, mass term is physically induced by  Zeeman effect on the surface of topological insulator \cite{QiHZ2008} and  sublattice asymmetry between  A and B sites in graphene \cite{Hatsugai-M-K-H-A-2013}.

\subsection{Mass deformation}

Mass term is  added to the Dirac-Landau operator as 
\be
-i\fsl{\mathcal{D}}+\sigma_z M =\begin{pmatrix}
M & -i\eth_-^{(g+\frac{1}{2})} \\
 -i\eth_+^{(g-\frac{1}{2})}    & -M 
\end{pmatrix}. 
\label{mass2dhamildirac}
\ee
The $SU(2)$ rotational symmetry is still kept exact under the mass deformation 
\be
[\sigma_z M, \bs{J}]=0, 
\ee
but the chiral symmetry is broken:  
\be
\{-i\fsl{\mathcal{D}}+\sigma_z M, \sigma_z\}=2M \neq 0.
\ee

\begin{figure}[tbph]\center
\includegraphics*[width=150mm]{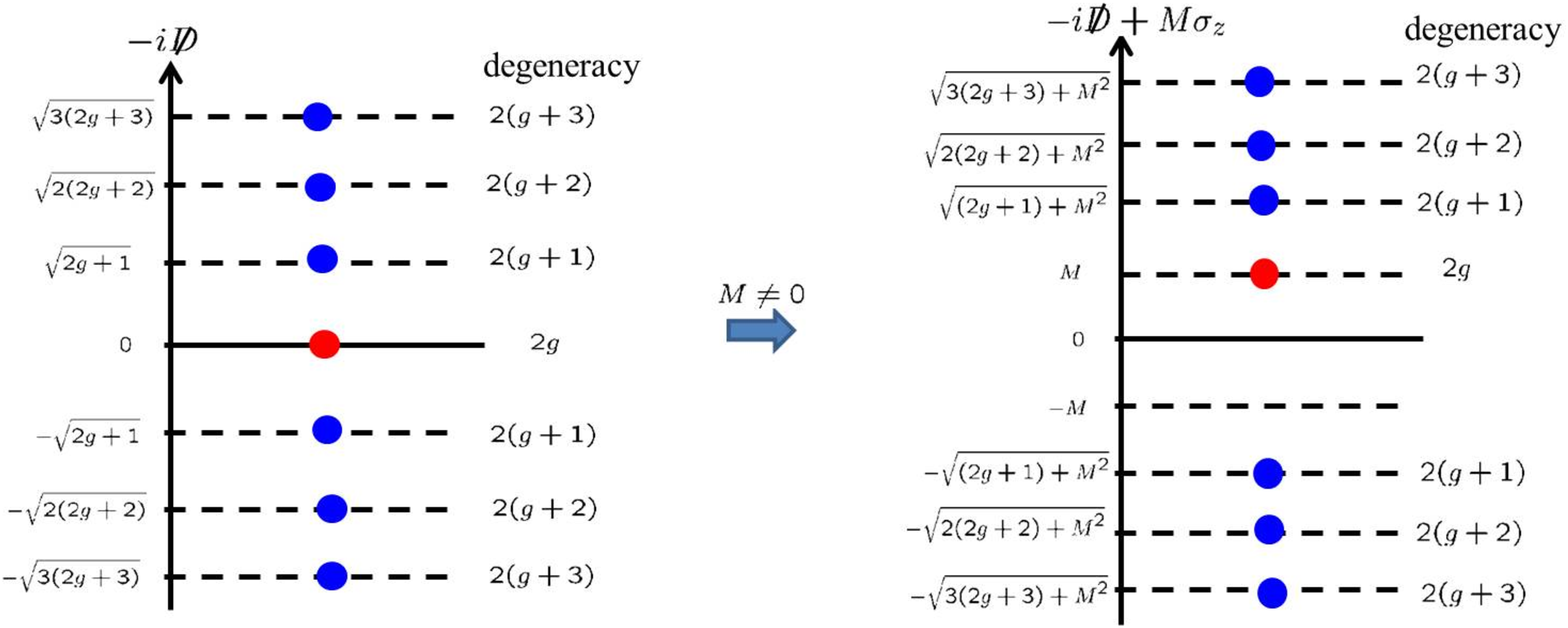}
\caption{  
The eigenvalues of the Dirac-Landau operator change from the left to the right by the mass 
deformation $(M>0)$. 
The massive Dirac-Landau operator has the eigenvalue,  $+M$,  but  not $-M$,  which is known as   the ``parity anomaly''.     
}
\label{Diracwithmass.fig}
\end{figure}

The kinetic term $-i\fsl{\mathcal{D}}$ and the mass term $M\sigma_z$ do not commute and hence their simultaneous eigenstates do not exist in general except for the zero modes. 
Square of the massive Dirac-Landau operator is given by 
\be
(-i\fsl{\mathcal{D}}+\sigma_z M)^2=(-i\fsl{\mathcal{D}})^2+M^2, 
\ee
where we used the chiral symmetry of the Dirac-Landau operator, 
 $\{-i\fsl{\mathcal{D}}, \sigma_z\}=0$. 
Therefore, the eigenvalues of $(-i\fsl{\mathcal{D}}+M\sigma_z)^2$ are given by 
\be
{\Lambda_n}^2~\equiv~ {\lambda_n}^2+M^2=n(n+2g)+M^2.  
\ee
The eigenvalues of the mass deformed Dirac-Landau operator are\footnote{For $g<0$, we have  $\Lambda_{n=0}=-M$ instead of (\ref{lambdazeromass}).}  
\begin{subequations}
\begin{align}
&n=0~~~~~~~~~:~~~\Lambda_{n=0}=+M,~~~~~~~~~~~~~~~~~~~~~~~~~~~~~~~~~~~~~~~~~~~~~~ \label{lambdazeromass} \\
&n=1,2, \cdots~:~~\pm \Lambda_n = \pm \sqrt{{\lambda_n}^2+M^2}= \pm \sqrt{n(n+2g)+M^2}.
\end{align}
\end{subequations}
Notice  the absence of $-M$ in the eigenvalues. The zero modes of the (massless) Dirac-Landau operator correspond to those of the  massive Dirac-Landau operator with the eigenvalue $+M$. Explicitly, the corresponding eigenstates are given by  
\begin{subequations}
\begin{align}
&n=0~~~~~~~~~:~~~\Psi^{g}_{\Lambda_{0}=M, m}=\Psi^{g}_{\lambda_{0}=0, m}=\begin{pmatrix}
Y_{g-\frac{1}{2}, m}^{g-\frac{1}{2}} \\
0
\end{pmatrix},~~~~~~~~~~~~~~~~~~~~~~~~~~~~~~~ \label{lambdazeromass}\\
&n=1,2, \cdots~:~~\Psi^{g}_{\pm \Lambda_n, m}= \sqrt{\frac{\Lambda_n+ \lambda_n}{2\Lambda_n}}( \Psi^{g}_{\pm \lambda_n, m} \pm \frac{M}{\Lambda_n+\lambda_n} \Psi^{g}_{\mp \lambda_n, m})\nn\\
&~~~~~~~~~~~~~~~~~~~~~~~~~~~~~~~~=\frac{1}{2}\sqrt{\frac{\Lambda_n+\lambda_n}{\Lambda_n}} \begin{pmatrix}
(1\pm \frac{M}{\Lambda_n +\lambda_n})Y_{j=g-\frac{1}{2}+n, m}^{g-\frac{1}{2}} \\ 
\mp i(1\mp \frac{M}{\Lambda_n +\lambda_n})Y_{j=g+\frac{1}{2}+(n-1), m}^{g+\frac{1}{2}}
\end{pmatrix}. 
\label{eigenvaluemassivedirac}
\end{align}\label{massdefeigen}
\end{subequations}
Eq.(\ref{massdefeigen}) can be chosen as the simultaneous eigenstates of the  $SU(2)$ Casimir $\bs{J}^2$ due to the existence of the $SU(2)$ symmetry.
 Eq.(\ref{eigenvaluemassivedirac}) shows that the mass term mixes the massless  eigenstates with opposite sign eigenvalues of same magnitude. For  $\Psi_{+\Lambda_n, m}^g$, the mass term enhances/reduces the weight of the upper/lower component, while for $\Psi_{-\Lambda_n, m}^g$, the opposite.    
The mass deformed Dirac-Landau operator  exhibits the symmetric spectra with respect to the zero energy except for $\Lambda_0=+M$ [Figs.\ref{Diracwithmass.fig}]. 
 The Landau level degeneracies do not change under the mass deformation: 
\be
(2j+1)|_{j=n+g-\frac{1}{2}}=2(n+g). 
\ee
It is easy to see that, in the massless limit $M\rightarrow 0$, (\ref{massdefeigen}) are reduced to (\ref{eigennonzerogdiraceig}): 
\be
\Psi^{g}_{\pm \Lambda_n, m} ~\rightarrow~\Psi^{g}_{\pm \lambda_n, m} ~~(n= 1, 2, \cdots).
\ee
Also  in the limit  $M ~\rightarrow~\infty$, we have 
\begin{subequations}
\begin{align}
&\Lambda_n -M ~\simeq ~   \frac{{\lambda_n}^2}{2M}=\frac{1}{2M}n(n+2g), \\
&\Psi^{g}_{ +\Lambda_n, m}~\rightarrow ~\frac{1}{\sqrt{2}}(\Psi_{\lambda_n, m}^{g}+\Psi_{-\lambda_n, m}^{g} )=\begin{pmatrix}
Y^{g-\frac{1}{2}}_{l=n+g-\frac{1}{2}, m} \\
0
\end{pmatrix},\\
&\Psi^{g}_{ -\Lambda_n, m}~\rightarrow ~\frac{1}{\sqrt{2}} 
(\Psi_{-\lambda_n, m}^g -\Psi^g_{\lambda_n, m})=i\begin{pmatrix}
0 \\
Y^{g+\frac{1}{2}}_{j=g+\frac{1}{2}+(n-1), m}
\end{pmatrix},   
\end{align}
\end{subequations}
which reproduce  the non-relativistic results, (\ref{non-relaeigenvalues}) and (\ref{monopoleharmonicsjacobischwinger}) with replacement of  $(n, g)$ by $(n, g-\frac{1}{2})$ or $(n-1, g+\frac{1}{2})$ up to  constant.   
  
Though the massive Dirac-Landau operator does not respect the original chiral symmetry, the spectrum structure suggests the existence of some generalized chiral  operator that anti-commutates with the mass deformed Dirac-Landau operator. 
Such a chiral operator is given by  
\be
\mathcal{R} =-i\sigma_z\fsl{\mathcal{D}}=\frac{1}{2}[\sigma_z, -i\fsl{\mathcal{D}}], 
\label{geralizechiral}
\ee
or  
\be
\mathcal{R} 
=(\partial_{\theta}+\frac{1}{2}\cot\theta)\sigma_y
-\frac{1}{\sin\theta}(\partial_{\phi}+ig\cos\theta)\sigma_x. 
\ee
It is straightforward to demonstrate 
\be
\{\mathcal{R}, -i\fsl{\mathcal{D}}+M\sigma_z\}=\frac{1}{2}
\{[\sigma_z, -i\fsl{\mathcal{D}}+M\sigma_z], -i\fsl{\mathcal{D}}+M\sigma_z\}=\frac{1}{2}[\sigma_z, 
(-i\fsl{\mathcal{D}}+M\sigma_z)^2]=
\frac{1}{2}[\sigma_z, 
-{\fsl{\mathcal{D}}}^2+M^2]=0, 
\ee
and the eigenstates for $+\Lambda_n$ and $-\Lambda_n$ are related by $\mathcal{R}$: 
\be
\mathcal{R}\Psi_{\pm \Lambda_n, m}^{g}=\pm \lambda_n \Psi_{\mp \Lambda_n, m}^{g}. 
\ee
Since $-i\fsl{\mathcal{D}} \rightarrow \pm \lambda_n$ in the massless limit, 
$\mathcal{R}$ (\ref{geralizechiral}) is reduced to the original chiral matrix $\sigma_z$ (times $\pm\lambda_n$).

\subsection{Balanced fuzzy spheres}

\begin{figure}[tbph]\center
\includegraphics*[width=60mm]{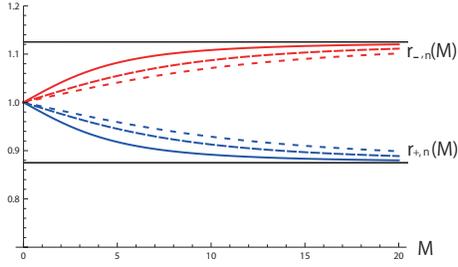}
\caption{Mass dependence of the ratios (\ref{ratiosmassdep}). 
 The blue curves represent $r_{+, n}$ for $n=3, 7, 11$ (solid, dashed and dotted), while  the red  curves denote $r_{-, n}$. ($g=4$ is adopted in the figure.  The blue curves and red curves approach to  $1\pm \frac{1}{8}$ respectively as $M$ goes infinity.)} 
\label{ratiossphereradiiwithmass.fig}
\end{figure}

\begin{figure}[tbph]\center
\includegraphics*[width=130mm]{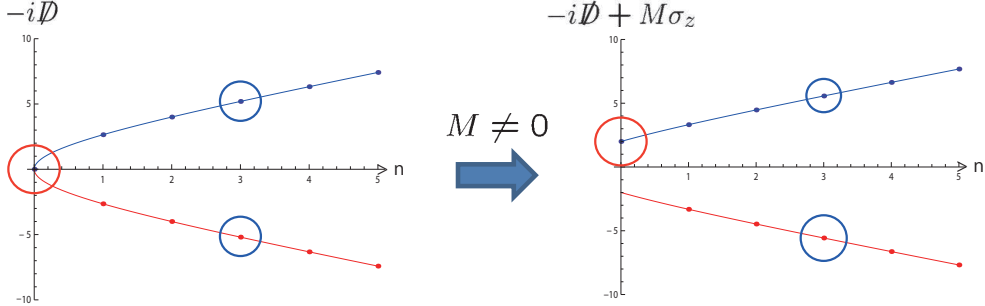}
\caption{Size change of fuzzy spheres under the mass deformation.  
 The red circle represents the fuzzy sphere for $n=0$,  while the blue circles indicate the sizes of the fuzzy spheres for $\pm\lambda_{n=3}$. ($g=3$ and $M=2$ are adopted in the figure.) } 
\label{2dfuzzyspheremassdirac.fig}
\end{figure}

Mass deformed Dirac-Landau model introduces  fuzzy spheres as    
\begin{subequations}
\begin{align}
&n=0~~~~~~~~~:~~\bs{X}_{\Lambda_0=+M}=\langle \Psi_{\Lambda_0=M}^{g}|\bs{x}|\Psi_{\Lambda_0=M}^{g}\rangle ={\alpha'}^{(g)}_{0}\cdot \bs{S}_{s=g-\frac{1}{2}},~~~~~~~~~~~~~~\\
&n=1,2,\cdots~:~~\bs{X}_{\pm \Lambda_n}=\langle \Psi_{\pm \Lambda_n}^{g}|\bs{x}|\Psi_{\pm \Lambda_n}^{g}\rangle ={\alpha'}^{(g)}_{\pm \Lambda_n}(M)\cdot \bs{S}_{s=n+g-\frac{1}{2}}, ~~~ \label{matelexshigherlrel}
\end{align}
\end{subequations}
where 
\begin{subequations}
\begin{align}
&n=0~~~~~~~~~:~~{\alpha'}^{(g)}_{0}=r\frac{1}{g+\frac{1}{2}}, \\ 
&n=1,2,\cdots~:~~{\alpha'}^{(g)}_{\pm \Lambda_n}(M)\equiv  
(1\mp \frac{1}{2g} \frac{M}{\Lambda_n}){\alpha'}^{(g)}_{n}. 
\label{noncomscalemassdepM}
\end{align}
\end{subequations}
To derive (\ref{matelexshigherlrel}), we used (\ref{innerproxrel}) and $\langle \Psi^g_{\pm \lambda_n} |\bs{x}|\Psi^g_{\mp \lambda_n}\rangle 
=\frac{1}{2g}{\alpha'}_{n}^{(g)}\bs{S}_{s=n+g-\frac{1}{2}}$. 
For $\Lambda_0=+M$, everything is same as of the fuzzy sphere of the zero modes of the Dirac-Landau operator.  
(\ref{noncomscalemassdepM}) suggests that the mass parameter unevenly affects the non-commutative length scales, ${\alpha'}^{(g)}_{\pm\Lambda_n}$ ($n=1, 2, \cdots$),  
which have the following properties: 
\begin{subequations}
\begin{align} 
&{\alpha'}^{(g)}_{\pm \Lambda_n}(-M)={\alpha'}^{(g)}_{\mp \Lambda_n}(M) 
\label{propertyofalphaM}, \\
&{\alpha'}^{(g)}_{ +\Lambda_n}(M)+{\alpha'}^{(g)}_{ -\Lambda_n}(M)=2{\alpha'}^{(g)}_n.  
\end{align} 
\end{subequations}
The radii of the fuzzy spheres are 
\be
\bs{X}_{\pm \Lambda_n} \cdot \bs{X}_{\pm \Lambda_n}  = {{R'}_{\pm \Lambda_n}^{(g)}(M)}^2, 
\ee
where 
\be
{R'}_{\pm \Lambda_n}^{(g)}(M)\equiv {\alpha'}^{(g)}_{\pm \Lambda_n}(M)\cdot \sqrt{(n+g-\frac{1}{2})(n+g+\frac{1}{2})}. 
\label{massdepradii}
\ee
Sum of the radii of the fuzzy spheres for $+\Lambda_n$ and $-\Lambda_n$ is immune to the mass deformation and  same as in the massless case:  
\be
{R'}_{+\Lambda_n}^{(g)}(M)+{R'}_{- \Lambda_n}^{(g)}(M)=  2{\alpha'}^{(g)}_{n}\cdot \sqrt{(n+g-\frac{1}{2})(n+g+\frac{1}{2})}=2 {R'}^{(g)}_{n}. 
\ee
To investigate behaviors of the radii under the mass deformation, we define 
\be
r_{\pm, n}(M)\equiv \frac{{R'}_{\pm \Lambda_n}^{(g)}(M)}{{R'}_{n}^{(g)}}=\frac{{\alpha'}^{(g)}_{\pm \Lambda_n}(M)}{{\alpha'}^{(g)}_{n}}=1\mp \frac{1}{2g}\frac{M}{\Lambda_n}. 
\label{ratiosmassdep}
\ee
$r_{\pm, n}(M)$ denote the ratios of ${R'}_{\pm \Lambda_n}^{(g)}(M)$ with respect to their massless limit, and 
 are depicted  in Fig.\ref{ratiossphereradiiwithmass.fig}.  
When $M=0$, there exist two identical fuzzy spheres for $+\lambda_n$ and $-\lambda_n$: 
\be
r_{+, n}(M)|_{M=0}=r_{-, n}(M)|_{M=0}= 1. \label{mzerolimratios}
\ee
As the mass parameter is turned, these two fuzzy spheres begin to ``correlate''  and  their radii monotonically change until their sizes  reach  $1\pm\frac{1}{2g}$ of their original sizes, which are the radii of the non-relativistic fuzzy spheres of $(n-\frac{1}{2}\pm\frac{1}{2}, g\mp\frac{1}{2})$: 
\begin{align}
&{R'}^{(g)}_{+\Lambda_n}(M)~\overset{M\rightarrow \infty}\longrightarrow~~ R_n^{(g-\frac{1}{2})}~(<~{R'}_{n}^{(g)}~), \nn\\
&{R'}^{(g)}_{-\Lambda_n}(M)~\overset{M\rightarrow \infty}\longrightarrow~~ R_{n-1}^{(g+\frac{1}{2})}~(>~{R'}_{n}^{(g)}~). 
 \label{minfinilimratios}
\end{align}
It may be visualized as if the fuzzy sphere of $+\lambda_n$ is ``absorbed'' in the fuzzy sphere of  $-\lambda_n$ as $M$ increases [Fig.\ref{2dfuzzyspheremassdirac.fig}]. 
Thus, we can tune the sizes of the fuzzy spheres (with their radii sum fixed) by changing the mass parameter. 

\section{Supersymmetric Landau Model and Super Fuzzy Spheres}\label{sect:superfuzzy}

A close connection is well known between Dirac operator and supersymmetric quantum mechanics [see Ref.\cite{Thaller-book} for instance]. 
Here, we construct supersymmetric quantum mechanical Hamiltonian from the Dirac-Landau operator,  and construct super fuzzy spheres by the level projection to supersymmetric Landau models.

\subsection{Square of the Dirac-Landau operator}\label{subsection:susyqm}

\begin{figure}[tbph]\center
\includegraphics*[width=60mm]{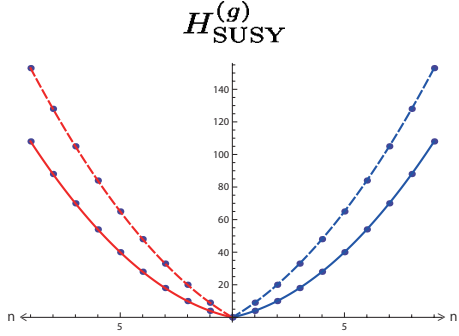}
\caption{The spectra of the supersymmetric Landau Hamiltonian.  The solid and dashed curves respectively correspond to (\ref{spectrasusyhamil}) for $g=3$ and $8$.    }
\label{spectrasusyQM.fig}
\end{figure}

Square of the Dirac operator yields a supersymmetric quantum Hamiltonian\footnote{
In the thermodynamic limit $g\rightarrow \infty$ with $g/r^2$ fixed,  $H_{\text{SUSY}}$ is reduced to the supersymmetric Pauli Hamiltonian on a plane \cite{Crombrugghe-Rittenberg-1983}:  
\be
H=-\frac{1}{2M}\sum_{i=1,2}{D_i}^2-\frac{g}{2M} \sigma_z, 
\ee
which is diagonalized as 
\be
\frac{g}{M}
\begin{pmatrix}
n & 0 \\
0 & n+1
\end{pmatrix}. ~~~(n=0,1,2,\cdots)
\ee
}, 
\be
H^{(g)}_{\text{SUSY}}=\frac{1}{2M}(-i\fsl{\mathcal{D}})^2=H^{(g_s)} -\frac{1}{2M} g_s\sigma_z, \label{defsusyhamil}
\ee
or 
\be
H_{\text{SUSY}}^{(g)} 
= \begin{pmatrix}
H^{(g-\frac{1}{2})} -\frac{1}{2M}(g-\frac{1}{2}) & 0 \\
0 & H^{(g+\frac{1}{2})} +\frac{1}{2M}(g+\frac{1}{2})
\end{pmatrix}. 
\ee
Here, $H^{(g_s)}$ is given by 
\be
H^{(g_s)}\equiv \begin{pmatrix}
H^{(g-\frac{1}{2})} & 0 \\
0 & H^{(g+\frac{1}{2})}
\end{pmatrix}=\frac{1}{2M}
\begin{pmatrix}
{\bs{\Lambda}^{(g-\frac{1}{2})}}^2 & 0 \\
0 &{\bs{\Lambda}^{(g+\frac{1}{2})}}^2  
\end{pmatrix},  
\label{diaghamilspindep}
\ee
with  $H^{(g)}$ (\ref{hamiltonianmonopolenonrela}). The second term of the right-hand side of (\ref{defsusyhamil}) represents the Zeeman term.  
As partially discussed in Sec.\ref{Schwingerappend},  the square of the Dirac-Landau operator enjoys both $SU(2)$ and chiral symmetries: 
\begin{subequations}
\begin{align}
&[H^{(g)}_{\text{SUSY}}, \bs{J}]=0, \label{symmesusyham1} \\
&[H^{(g)}_{\text{SUSY}}, \sigma_z]=0. \label{symmesusyham2}
\end{align}
\end{subequations}
One may readily verify (\ref{symmesusyham1}) and (\ref{symmesusyham2}) from 
$[-i\fsl{\mathcal{D}}, \bs{J}]=0$  and $\{ -i\fsl{\mathcal{D}}, \sigma_z\}=0$ using identities $[A^2, B]= \{A,[A,B]\}$ and $[A^2, B]=[A, \{A,B\}]$ respectively.  The energy eigenvalues  of the supersymmetric Landau Hamiltonian (\ref{diaghamilspindep}) are given by [Fig.\ref{spectrasusyQM.fig}]
\be
E_n=\frac{1}{2M}(n(n+2g))~~~~(n=0, 1, 2, \cdots),
\label{spectrasusyhamil}
\ee
with degeneracy 
\begin{subequations}
\begin{align}
&n=0~~~~~~~~~~:~~2g,\\
&n=1,2, \cdots~~:~~4(g+n).  
\end{align}
\end{subequations}
The corresponding energy eigenstates with definite chiralities are given by  
\begin{subequations}
\begin{align}
&n=0~~~~~~~~~~:~~~{\Upsilon'}_{j=g-\frac{1}{2}, m}^{g}=\begin{pmatrix}
Y^{g-\frac{1}{2}}_{j=g-\frac{1}{2}, m}(\theta, \phi)\\
0
\end{pmatrix}, ~~~~~~\\
&n=1, 2, \cdots~~:~~~{\Upsilon'}^{g}_{j=g-\frac{1}{2}+n, m}=\begin{pmatrix}
Y^{g-\frac{1}{2}}_{j=g-\frac{1}{2}+n, m}(\theta, \phi)\\
0
\end{pmatrix},~~~~{\Upsilon}^{g}_{j=g-\frac{1}{2}+n, m}=\begin{pmatrix}
0 \\
Y^{g+\frac{1}{2}}_{j=g+\frac{1}{2}+(n-1), m}(\theta, \phi)
\end{pmatrix}.  
\end{align}\label{supereigenstates}
\end{subequations}

The supersymmetric structure becomes obvious when we express  
$H_{\text{SUSY}}^{(g)}$ as 
\be
H_{\text{SUSY}}^{(g)}=-\frac{1}{2M}
\begin{pmatrix}
\eth_-^{(g+\frac{1}{2})} \eth_+^{(g-\frac{1}{2})} & 0 \\
0 & \eth_+^{(g-\frac{1}{2})} \eth_-^{(g+\frac{1}{2})} 
\end{pmatrix}=\{ Q^{(g)}, \bar{Q}^{(g)}\}, 
\ee
where $Q^{(g)}$ and $\bar{Q}^{(g)}$ are   nilpotent 
 super-charges:  
\be
(Q^{(g)})^2=(\bar{Q}^{(g)})^2=0,  
\label{squareq0}
\ee
as given by 
\begin{align}
&{Q}^{(g)}=-\frac{1}{\sqrt{2M}}\sigma_+ \eth_-^{(g+\frac{1}{2})}=-\frac{1}{\sqrt{2M}} \begin{pmatrix}
0 & \eth_-^{(g+\frac{1}{2})}  \\
0 & 0 \
\end{pmatrix}, \\ 
&\bar{Q}^{(g)}=\frac{1}{\sqrt{2M}}\sigma_- \eth_+^{(g-\frac{1}{2})}= \frac{1}{\sqrt{2M}} \begin{pmatrix}
0 & 0\\ 
\eth_+^{(g-\frac{1}{2})} & 0 \
\end{pmatrix}. 
\end{align}
From the nilpotency of the supercharges (\ref{squareq0}), it is obvious that the supersymmetric Landau Hamiltonian respects the supersymmetry:   
\be
[H^{(g)}_{\text{SUSY}}, Q^{(g)}]=[H_{\text{SUSY}}^{(g)}, \bar{Q}^{(g)}]=0.  
\ee
The supercharges are also $SU(2)$ singlet operators, 
\be
[\bs{J}, Q^{(g)}]=[\bs{J}, \bar{Q}^{(g)}]=0,  
\ee
which anticommute with the chirality matrix:  
\be
\{Q^{(g)}, \sigma_z\}=\{\bar{Q}^{(g)}, \sigma_z\}=0.
\ee
 $Q^{(g)}$ and $\bar{Q}^{(g)}$ act to   the opposite chirality eigenstates of the $n$th Landau level as   
\begin{align}
&{Q}^{(g)} \Upsilon^{g}_{j=g+n+\frac{1}{2}, m}=\sqrt{\frac{(n+2g+1)(n+1)}{2M}}~ {\Upsilon'}^{g}_{j=g+n+\frac{1}{2}, m}, ~~~~~~\bar{Q}^{(g)} \Upsilon_{j=g+n+\frac{1}{2}, m}^{g}=0, \nn\\
&\bar{Q}^{(g)}{\Upsilon'}_{j=g+n-\frac{1}{2}, m}^{g} =\sqrt{\frac{(n+2g)n}{2M}}~ {\Upsilon}_{j=g+n-\frac{1}{2}, m}^{g},~~~~~~~~~~~~~~~~~~~{Q}^{(g)}{\Upsilon'}_{j=g+n-\frac{1}{2}, m}^{g}=0.  
\label{susytransupsilrel}
\end{align}

\subsection{Super fuzzy spheres}

\begin{figure}[tbph]\center
\includegraphics*[width=90mm]{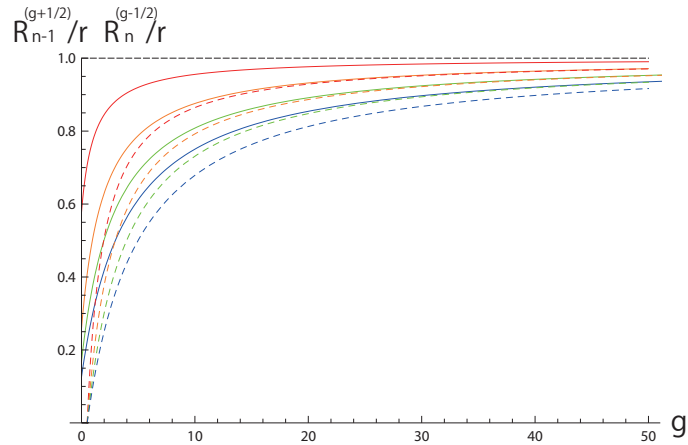} 
\caption{  ${R_{n-1}^{(g+\frac{1}{2})}}/r$ ($n=1, 2, 3, 4$) correspond to the solid curves (red, orange, green, blue), while ${R_{n}^{(g-\frac{1}{2})}}/r$ the dashed curves.  }
\label{superfuzzy.fig}
\end{figure}

For each supersymmetric Landau level of $n\neq 0$, we introduce two fuzzy spheres from the opposite chirality states, ${\Upsilon'}^g_{j=g+n+\frac{1}{2}, m}$ and ${\Upsilon}^g_{j=g+n+\frac{1}{2}, m}$:    
\begin{subequations}
\begin{align}
&\bs{X}^{(-)}\equiv \langle  {\Upsilon'}^g_{j=g+n+\frac{1}{2}} | \bs{x} |{\Upsilon'}^g_{j=g+n+\frac{1}{2}}\rangle=  -\alpha_n^{(g-\frac{1}{2})}\bs{S}_{s=n+g-\frac{1}{2}}, \\
&\bs{X}^{(+)}\equiv \langle  \Upsilon^g_{j=g+n+\frac{1}{2}, m} | \bs{x} |\Upsilon^g_{j=g+n+\frac{1}{2}, m} \rangle=  -\alpha_{n-1}^{(g+\frac{1}{2})}\bs{S}_{s=n+g-\frac{1}{2}}. 
\end{align}\label{superfuzzysphereintro}
\end{subequations}
(Eigenstates of the supersymmetric $n=0$ Landau level are  same as of the zero modes of the relativistic Landau model as  discussed in Sec.\ref{sec:projrelLLs}.)   
These two fuzzy spheres may be considered as super partners, since   ${\Upsilon'}^{g}_{j, m}$  and $\Upsilon_{j, m}^g$ are related by the supersymmetric transformation (\ref{susytransupsilrel}).  We shall refer to these two fuzzy spheres as super fuzzy spheres\footnote{We adopt a terminology,  super fuzzy sphere,  instead of  fuzzy supersphere since  fuzzy supersphere usually means  a fuzzy sphere made of  graded Lie algebra [see Ref.\cite{Hasebe2011NPB} for instance.] Fuzzy superspheres appear in the Landau levels of the $UOSp(1|2)$ invariant Landau model \cite{Hasebe-Kimura-2004,Hasebe-2005}.}.  
The radii of the super fuzzy spheres (\ref{superfuzzysphereintro}) slightly differ as 
\begin{subequations}
\begin{align}
&{R_n^{(g-\frac{1}{2})}}=\alpha_n^{(g-\frac{1}{2})}\sqrt{j(j+1)}|_{j=g+n-\frac{1}{2}}=r\frac{g-\frac{1}{2}}{\sqrt{(g+n-\frac{1}{2})(g+n+\frac{1}{2})}},\\
&{R_{n-1}^{(g+\frac{1}{2})}}=\alpha_{n-1}^{(g+\frac{1}{2})}\sqrt{j(j+1)}|_{j=g+n-\frac{1}{2}}=r\frac{g+\frac{1}{2}}{\sqrt{(g+n-\frac{1}{2})(g+n+\frac{1}{2})}}. 
\end{align}
\end{subequations}
Their behaviors with respect to $g$  are plotted in Fig.\ref{superfuzzy.fig}.  As $g$ increases,  $R_n^{(g-\frac{1}{2})}$ and $R_{n-1}^{(g+\frac{1}{2})}$ asymptotically approach to  same value, $r\frac{g}{g+n}$. 
The radius of the relativistic fuzzy sphere (\ref{relfuzzyradiiexp}) is the average of the radii of the super fuzzy spheres: 
\be
{R'}_n^{(g)}=\frac{1}{2}(R_{n-1}^{(g+\frac{1}{2})} +R_{n}^{(g-\frac{1}{2})}). 
\ee
The mass deformation just brings  a constant shift to the supersymmetric Landau Hamiltonian:  
\be
\frac{1}{2M}(-i\fsl{\mathcal{D}}+M \sigma_z )^2=\frac{1}{2M}(-i\fsl{\mathcal{D}})^2+\frac{1}{2}M=H_{\text{SUSY}}^{(g)}+\frac{1}{2}M, 
\ee
and does not affect the supersymmetric eigenstates (\ref{supereigenstates}) and so the super fuzzy spheres either.

\section{Valley Fuzzy Spheres from Graphene}\label{sec:valleyfuzzysphere}

In this section, we apply the above analysis to the realistic graphene system. 

\subsection{Graphene spectrum}

In graphene,  the spinor components of the Dirac operator indicate A and B sub-lattice degrees of freedom. 
In addition to the sub-lattice degrees of freedom, graphene accommodates the valley degrees of freedom of  $K$ and $K'$ points, in which low energy physics is described by 
\be
-i\fsl{\mathcal{D}}_{K\oplus K'}\equiv \begin{pmatrix}
-i\fsl{\mathcal{D}}_K & 0 \\
0 & -i\fsl{\mathcal{D}}_{K'}
\end{pmatrix},   
\label{grapheneHamiltonianonepart}
\ee 
where   
\be
-i\fsl{\mathcal{D}}_K\equiv -i\sigma_x \mathcal{D}_{\theta}-i\frac{1}{\sin\theta}\sigma_y \mathcal{D}_{\phi}, ~~~~-i\fsl{\mathcal{D}}_{K'}\equiv -i\sigma_x \mathcal{D}_{\theta}+i\frac{1}{\sin\theta}\sigma_y \mathcal{D}_{\phi}, 
\ee
with $\mathcal{D}_{\theta}$  and $\mathcal{D}_{\phi}$ (\ref{spherecovderi}). 
These are related as 
\be
-i\fsl{\mathcal{D}}_K =\sigma_x (-i\fsl{\mathcal{D}}_{K'})\sigma_x.    
\ee
The $SU(2)$ operator that commutes with $-i\fsl{\mathcal{D}}_{K\oplus K'}$ is given by 
\be
\bs{J}=\begin{pmatrix}
\bs{L}^{(g_s)} & 0 \\
0 & \bs{L}^{(\bar{g_s})}
\end{pmatrix}=
\begin{pmatrix}
\bs{L}^{(g-\frac{1}{2})} & 0 & 0 & 0 \\
0 & \bs{L}^{(g+\frac{1}{2})} & 0 & 0 \\
0 & 0 &  \bs{L}^{(g+\frac{1}{2})}  & 0 \\
0 & 0 & 0 & \bs{L}^{(g-\frac{1}{2})}  
\end{pmatrix}, 
\ee
where $g_s=g-\frac{1}{2}\sigma_z$ and $\bar{g_s}=g+\frac{1}{2}\sigma_z$. $\bs{J}$ satisfies 
\be
[J_i, J_j]=i\epsilon_{ijk}J_k, 
\ee
and 
\be
\bs{J}^2=\begin{pmatrix}
j_K (j_K+1)\bs{1}_{2j_K+1} & 0  \\
0 & j_{K'} (j_{K'}+1)\bs{1}_{2j_{K'}+1}
\end{pmatrix}, 
\ee
where $\bs{1}_{2j+1}$ denotes $(2j+1)\times (2j+1)$ unit matrix and 
\be
j_K=g-\frac{1}{2}+n_K,~~~~
j_{K'}=g-\frac{1}{2}+n_{K'}.~~~~(n_K,~n_{K'}=0,1,2,\cdots)
\ee
Square of the graphene Hamiltonian (\ref{grapheneHamiltonianonepart})  is given by 
\be
(-i\fsl{\mathcal{D}}_{K\oplus K'})^2 
=\bs{J}^2-g^2+\frac{1}{4}. 
\ee
$-i\fsl{\mathcal{D}}_{K}$ and $-i\fsl{\mathcal{D}}_{K'}$ have the same spectrum, and so the spectrum of  $-i\fsl{\mathcal{D}}_{K\oplus K'}$ is equally given by  
\be
\pm \lambda_n
= \pm \sqrt{n(2g+n)} ~~~~(n=0, 1, 2, \cdots), 
\ee
and the corresponding  degeneracy for each of $+\lambda_n$ and $-\lambda_n$ is   
\be
2\times (2j+1)=4(g+n) ~~(n=0, 1, 2, \cdots). 
\ee
Obviously, $2\times$ comes from the valley degrees of freedom. 
The eigenstates are denoted as 
\begin{subequations}
\begin{align}
&n=0~~~~~~~~~~:~~
\Psi^g_{\lambda_0=0, m;~K}=
\begin{pmatrix}
{Y}^{g-\frac{1}{2}}_{j=g-\frac{1}{2} ,m} \\
0
\end{pmatrix},~~~
\Psi^g_{\lambda_0=0, m;~K'} = 
\begin{pmatrix}
0 \\
{Y}^{g-\frac{1}{2}}_{j=g-\frac{1}{2} ,m} 
\end{pmatrix},  
\label{masslessgrapheneeigens}\\
&n=1,2, \cdots~~:~~\Psi_{\pm \lambda_n, m;~ K}^g=
\begin{pmatrix}
{Y}^{g-\frac{1}{2}}_{j ,m} \\
\mp i{Y}^{g+\frac{1}{2}}_{j ,m} 
\end{pmatrix},~~~~~~
\Psi_{\pm \lambda_n, m; ~K'}^g 
=
\begin{pmatrix}
\mp i{Y}^{g+\frac{1}{2}}_{j ,m} \\
{Y}^{g-\frac{1}{2}}_{j ,m}
\end{pmatrix}, 
 \label{Kdashpointmonopoleharmonics}
 \end{align}
\end{subequations} 
which are related as 
\be
\Psi^g_{\pm \lambda_n, m;~K}=\sigma_x\Psi^g_{\pm \lambda_n, m;~K'}.  
\ee

\subsection{Mass deformation  and valley fuzzy spheres}

\begin{figure}[tbph]\center
\includegraphics*[width=130mm]{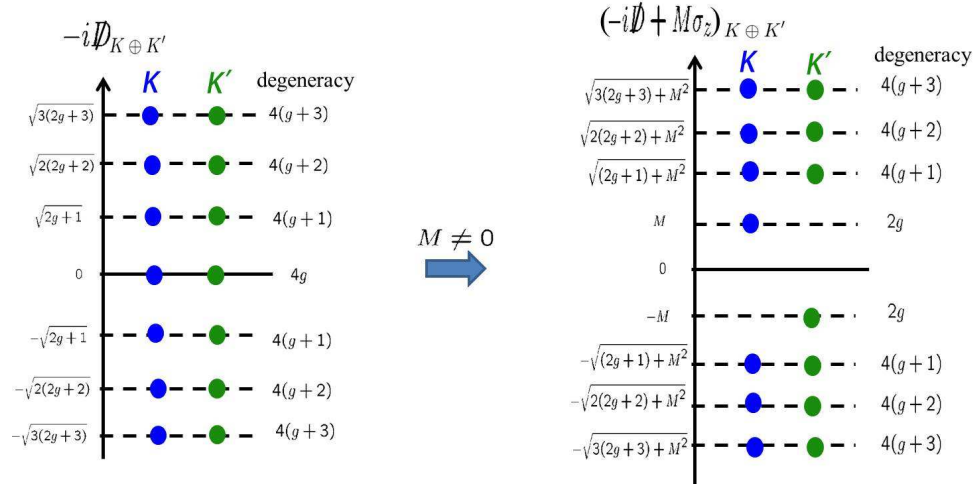}
\caption{The blue and green blobs respectively correspond to the eigenstates of the Dirac-Landau operators at $K$ and $K'$ points.  The degeneracy of  zero modes is lifted to  $M$ and $-M$ when the mass term is added.    }
\label{graphenefig}
\end{figure}

We consider mass deformation of the Dirac-Landau operators at $K$ and $K'$ points:   
\be 
-i \fsl{\mathcal{D}}_K +M \sigma_z, 
~~~~-i\fsl{\mathcal{D}}_{K'} +M \sigma_z, 
\ee
to have 
\be
(-i\fsl{\mathcal{D}}+M\sigma_z)_{K\oplus K'}\equiv 
\begin{pmatrix}
-i\fsl{\mathcal{D}}_K+M\sigma_z & 0 \\
0 & -i\fsl{\mathcal{D}}_{K'}+M\sigma_z
\end{pmatrix}. 
\ee
In each valley, the mass deformed Dirac-Landau operator is readily diagonalized:  
\begin{subequations}
\begin{align}
&K~:~~+\Lambda_0=+M~~(n=0),~~\pm\Lambda_{n} ~~(n=1,2,\cdots), \\
&K'~:~~-\Lambda_0=-M~~(n=0),~~\pm\Lambda_{n}~~(n=1,2,\cdots),
\end{align}\label{eigenvaluesKKdirac}
\end{subequations}
with $2(n+g)$ $(n=0, 1, 2, \cdots)$ degeneracy each.  
The corresponding eigenstates  are\footnote{
In the massless limit $M~\rightarrow ~0$, they are reduced to 
\be
\Psi_{\pm \Lambda_n, m;K}^{g}(M) ~
\longrightarrow~\Psi_{\pm \lambda_n, m;K}^{g},~~~\Psi_{\pm \Lambda_n, m;K'}^{g}(M) ~
\longrightarrow~\Psi_{\pm \lambda_n, m;K'}^{g}. 
\ee
}  
\begin{subequations}
\begin{align}
\Psi_{\pm \Lambda_n, m;K}^{g}(M)&= \sqrt{\frac{\Lambda_n+ \lambda_n}{2\Lambda_n}} 
~( \Psi_{\pm \lambda_n, m; K}^g \pm  \frac{M}{\Lambda_n +\lambda_n}\Psi_{\mp \lambda_n, m; K}^g)\nn\\
&=\frac{1}{2}\sqrt{\frac{\Lambda_n+\lambda_n}{\Lambda_n}} \begin{pmatrix}
(1\pm \frac{M}{\Lambda_n +\lambda_n})Y_{j=g-\frac{1}{2}+n, m}^{g-\frac{1}{2}} \\ 
\mp i(1\mp \frac{M}{\Lambda_n +\lambda_n})Y_{j=g+\frac{1}{2}+(n-1), m}^{g+\frac{1}{2}}
\end{pmatrix},    \\
\Psi_{\pm \Lambda_n, m;K'}^{g}(M)&=  \sqrt{\frac{\Lambda_n + \lambda_n}{2\Lambda_n}} ~( \Psi_{\pm \lambda_n, m; K'}^g \mp  \frac{M}{\Lambda_n +\lambda_n}\Psi_{\mp \lambda_n, m; K'}^g)\nn\\
&=\frac{1}{2}\sqrt{\frac{\Lambda_n+\lambda_n}{\Lambda_n}} \begin{pmatrix}
\mp i(1\pm \frac{M}{\Lambda_n +\lambda_n})Y_{j=g+\frac{1}{2}+(n-1), m}^{g+\frac{1}{2}}\\
(1\mp \frac{M}{\Lambda_n +\lambda_n})Y_{j=g-\frac{1}{2}+n, m}^{g-\frac{1}{2}} 
\end{pmatrix}.    
\end{align} \label{massdefeigenstatesgraphene}
\end{subequations}
The mass deformed graphene spectrum is given by 
\be
\pm \Lambda_n=\pm \sqrt{n(n+2g)+M^2}~~~~~(n=0, 1, 2, \cdots),
\ee
with degeneracy [Fig.\ref{graphenefig}]
\begin{align}
&\pm \Lambda_0=+M~~:~~2g~~~~~~~~~~~~~~(n=0),~~~~~~~~~\nn\\
&\pm \Lambda_n=-M~~:~~4(g+n)~~~~~~(n=1,2, \cdots).
\end{align}
The reflection symmetry of the spectra with respect to the zero energy still exists under the mass deformation, though either of the mass deformed Dirac-Landau operators at $K$ and $K'$ points does not respect the chiral symmetry. The reflection symmetry is guaranteed by 
\be
\{R~,~ (-i\fsl{\mathcal{D}}+M\sigma_z)_{K\oplus K'}\}=0, 
\label{diracchiralgraphene}
\ee
with 
\be
R=i\begin{pmatrix}
0 & \sigma_y \\
\sigma_y & 0 
\end{pmatrix}. 
\ee
Eq.(\ref{diracchiralgraphene}) can  readily be verified from the relation: 
\be
\sigma_y (-i\fsl{\mathcal{D}}_K+M\sigma_z) +   (-i\fsl{\mathcal{D}}_{K'}+M\sigma_z) \sigma_y=0. 
\ee
$R$ relates the eigenstates with opposite sign eigenvalues on $K$ and $K'$ points: 
\be
\Psi_{\pm\Lambda_n, m; K}^g (M)= i\sigma_y \Psi_{\mp \Lambda_n,m; K'}^g(M). 
\label{relpsikandkdash}
\ee

\begin{figure}[tbph]\center
\includegraphics*[width=140mm]{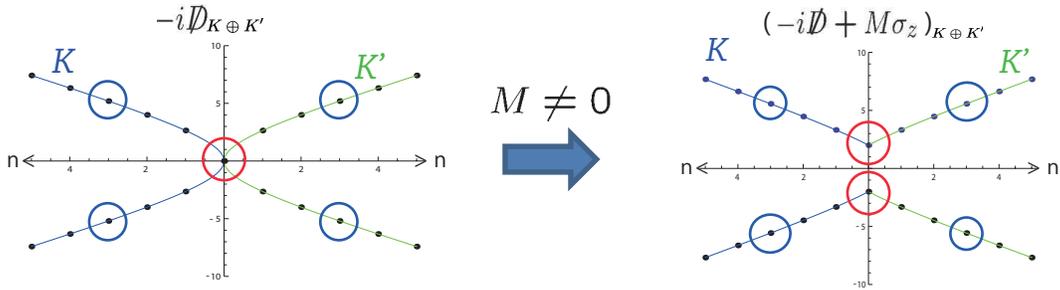}
\caption{ The circles represent the sizes of the fuzzy spheres on the corresponding Landau levels. ($g=3$ and $M=2$ are adopted in the figure.)  }
\label{graphenefuzzyspherefig}
\end{figure}

With use of the eigenstates  of $K$ and $K'$ valleys (\ref{massdefeigenstatesgraphene}),   valley fuzzy spheres are introduced as 
\begin{subequations}
\begin{align}
&\bs{X}^{K}_{\pm \Lambda_n}=\langle \Psi^{g}_{\pm \Lambda_n, ; K}(M)|\bs{x}|\Psi^{g}_{\pm \Lambda_n, ; K}(M)\rangle =-{\alpha'}^{(g)}_{\pm \Lambda_n}(M)\cdot \bs{S}_{s=n+g-\frac{1}{2}}, \\
&\bs{X}^{K'}_{\pm \Lambda_n}=\langle \Psi^{g}_{\pm \Lambda_n, ; K'}(M)|\bs{x}|\Psi^{g}_{\pm \Lambda_n, ; K'}(M)\rangle =-{\alpha'}^{(g)}_{\mp \Lambda_n}(M)\cdot \bs{S}_{s=n+g-\frac{1}{2}}, 
\end{align}
\end{subequations}
where (\ref{relpsikandkdash}) was used.  
Thus, we have 
\begin{subequations}
\begin{align}
&{\bs{X}^{K}_{ \Lambda_n}}\cdot{\bs{X}^{K}_{ \Lambda_n}} ={\bs{X}^{K'}_{ -\Lambda_n}}\cdot{\bs{X}^{K'}_{ -\Lambda_n}} ={{R'}^{(g)}_{ \Lambda_n}(M)}^2, \\
&{\bs{X}^{K}_{- \Lambda_n}}\cdot{\bs{X}^{K}_{- \Lambda_n}}={\bs{X}^{K'}_{ \Lambda_n}}\cdot{\bs{X}^{K'}_{ \Lambda_n}}  ={{R'}^{(g)}_{-\Lambda_n}(M)}^2, 
\end{align}
\end{subequations}
where ${R'}^{(g)}_{\pm\Lambda_n}(M)$ are given by (\ref{massdepradii}).  
As the mass parameter is turned on and increases, the four fuzzy spheres for  $n(\neq 0)$th Landau level change their sizes, two of which expand and the other two shrink, while  
the two fuzzy spheres for $n=0$ do not vary  their sizes [Fig.\ref{graphenefuzzyspherefig}]. 

\section{Summary}\label{sec:summarydiscussion}

 We gave a through study of the relativistic Landau models  and 
 derived  non-commutative geometry by applying the level projection method to the relativistic Landau models.    We obtained a concise expression of the   eigenstates of the Dirac-Landau operator on a sphere, which turned out to be related to non-relativistic Pauli-Sch\"odinger  eigenstates  
 by the  $SU(2)$ gauge transformation.  After the $SU(2)$ transformation,  the Dirac-Landau operator  acts as the boost operator of the Lorentz group. 
 We constructed the relativistic fuzzy spheres with use of the relativistic Landau level eigenstates and  found that  the fuzzy sphere of zero modes reduces its size while fuzzy spheres of non-zero Landau levels enhance their sizes compared to their non-relativistic counterparts.  
Under the mass deformation,   two fuzzy spheres of positive and negative relativistic Landau levels  vary their sizes keeping the sum of their radii constant, while the size of the  fuzzy sphere of zero modes does not vary.       
We also constructed super fuzzy spheres from a supersymmetric Landau model as the square of the Dirac-Landau operator, and discussed their behaviors with respect to the monopole charge.  
Finally, we investigated  graphene system.  Due to the valley degrees of freedom, each  Landau level is two-fold  degenerate compared to the single Dirac-Landau case, and there appear valley fuzzy spheres. We discussed the reflection symmetry of the graphene spectrum and clarified the particular properties of the valley spheres under the mass deformation.    

While we focused on the fuzzy geometry in the relativistic Landau models, the level projection itself is a versatile method to introduce fuzzy geometry from physical models.   
It may be interesting to apply the level projection to other manifolds to generate a variety of fuzzy geometry and investigate their geometrical behaviors  controlled by  physical parameters. 
We have not discussed many-body physics of the relativistic Landau system.  
The present analysis has an advantage for numerical calculations because of its rotational symmetry.   
 We will report 
applications of the present spherical formalism to  relativistic quantum Hall effect in a future publication \cite{Yonaga-Hasebe-Shibata-2015}. 

\section*{Acknowledgement}

 The author  is grateful to A. Furusaki, T. Morimoto, A. Sako and K. Shiozaki for fruitful discussions.  
 He would also like to thank  Furusaki Group, Riken for warm hospitality, where a part of this  work was performed.      This work was supported by JSPS KAKENHI Grant Number~16K05334 and 16K05138. 

\appendix

\section{Jacobi Polynomials}\label{sec:jacobi-poly}

Jacobi polynomials are defined by  
\be
P_n^{(\alpha, \beta)}(x)=\frac{(-1)^n}{2^n n!} (1-x)^{-\alpha}(1+x)^{-\beta} \frac{d^n}{dx^n} (1-x)^{n+\alpha}(1+x)^{n+\beta}, \label{jacobipolydef}
\ee
where $x\in [-1, 1]$. Normalization is  the following: 
\be
\int_{-1}^1 dx ~(1-x)^{-\alpha}(1+x)^{-\beta}{P_n^{(\alpha, \beta)}(x)}^* P_m^{(\alpha, \beta)}(x)= \frac{2^{\alpha+\beta+1}}{2n+\alpha+\beta+1}\frac{\Gamma(n+\alpha+a)\Gamma(n+\beta+1)}{n!~\Gamma(n+\alpha+\beta+1)}~\delta_{n,m}. 
\ee
The Jacobi polynomial is a solution of a second-order differential equation: 
\be
(1-x^2)\frac{d^2  P_n^{(\alpha, \beta)}(x)}{dx^2} -((\alpha+\beta+2)x +\alpha-\beta) \frac{d  P_n^{(\alpha, \beta)}(x)}{dx} +n(n+\alpha+\beta+1)P_n^{(\alpha, \beta)}(x)=0. 
\ee
For $n=0, 1, 2$, the Jacobi polynomials are given by  
\begin{align}
&P_{0}^{(\alpha,\beta)}=1, \nn\\
&P_1^{(\alpha, \beta)}(x)= \frac{1}{2}((\alpha+\beta+2) x +(\alpha-\beta)), \nn\\
&P_2^{(\alpha, \beta)}(x)=\frac{1}{8}(((\alpha+\beta)^2 + 7(\alpha+\beta)+12)x^2+2(\alpha+\beta+3)(\alpha-\beta)x +(\alpha-\beta)^2-(\alpha+\beta)-4). 
\end{align}

\section{From Three-sphere Point of View}\label{append:secmaurer}

The Dirac monopole set-up is mathematically equivalent to the 1st Hopf map [see Ref.\cite{Hasebe-2014-1} for instance]: 
\be
S^3 ~\overset{S^1}\rightarrow~S^2. 
\ee
The total manifold is $S^3$ and the algebras of the Landau problem on the two-sphere is naturally understood from the perspective of  $S^3$. 
The symmetry of the present system is the rotational symmetry of the total manifold $S^3$;  
\be
SO(4)~\simeq ~SU(2)_L\otimes SU(2)_R.
\ee


\subsection{$D$ functions}

With the Euler angle parametrization,  the $SU(2)$ element is expressed as  
\begin{align}
g(\chi, \theta, \phi)&=e^{i\frac{\chi}{2}\sigma_z} e^{i\frac{\theta}{2}\sigma_y} e^{i\frac{\phi}{2}\sigma_z}
=g^{\dagger}(-\phi,-\theta,-\chi)\nn\\
&~~~~(0\le \chi\le 4\pi,~0\le \theta\le \pi,~0\le \phi \le 2\pi).
\label{reldfuncmonopoeharm}
\end{align}
The Wigner's $D$ function, $D_{l,m,g}(\phi, \theta, \chi)$,  is introduced  as a generalization of (\ref{reldfuncmonopoeharm}) for arbitrary representation with $SU(2)$ Casimir index $l$: 
\begin{align}
D_{l, m, g}(\phi, \theta, \chi) &\equiv \langle l, m |e^{-i\phi S_z  } e^{-i\theta S_y} e^{-i\chi S_z}|l, g\rangle = D_{l, g, m}(-\chi,-\theta,-\phi)^* ~~~~~(m, g=l, l-1, \cdots, -l).
 \label{defofwignerDfunc}
\end{align}
For the fundamental representation, $D_{l=\frac{1}{2}, m, g}(-\chi, -\theta, -\phi)=g(\chi, \theta, \phi)_{m, g}$.  The $D$ function can be expressed as a simple product of functions of each angular coordinate: 
\be
D_{l, m, g}(\phi, \theta, \chi) =  e^{-i(m\phi +g\chi)   } d_{l, m, g}(\theta), 
\ee
with 
\be
d_{l, m, g}(\theta)\equiv  \langle l, m | e^{-i\theta S_y} |l, g\rangle, 
\ee
which has a symmetry under the interchange of two magnetic quantum numbers;   
\be
{d}_{l, m, g}(\theta)=(-1)^{m-g}{d}_{l, g, m }(\theta). 
\ee
The explicit form of  ${D}$ function is 
\be
D_{l, m, g}(\phi, \theta, \chi)=
\frac{(-1)^{m-g}}{2^m}
\sqrt{\frac{(l+m)!(l-m)!}{(l+g)!(l-g)!}}(1-x)^{\frac{m-g}{2}} (1+x)^{\frac{m+g}{2}} P_{l-m}^{(m-g, m+g)}(x)~\cdot ~
e^{-i(m\phi +g\chi)}, 
\ee
where $x=\cos\theta$ and  $P_{n}^{(\alpha, \beta)}$ denote the Jacobi polynomials (\ref{jacobipolydef}).  
The monopole harmonics (\ref{monopoleharmonicsjacobischwinger}) are related to the Wigner's $D$ functions as  \cite{Wu-Yang-1977}:  
\begin{align}
Y^g_{l,m} (\theta, \phi)&=   (-1)^{m+g}\sqrt{\frac{2l+1}{4\pi}}D_{l,-m, g}(\phi, \theta, 0)\nn\\
&=(-1)^{m+g}\sqrt{\frac{2l+1}{4\pi}}d_{l,-m, g}( \theta) e^{im\phi}
\label{schwingermonopoleharmonics}
\end{align}
or 
\be
D_{l,m, g}(\phi, \theta, \chi)=(-1)^{m-g}\sqrt{\frac{4\pi}{2l+1}}  ~Y_{l, -m}^{g}(\theta, \phi) e^{-ig\chi}. 
\ee

\subsection{Maurer-Cartan 1 form and left and right actions}

The $D$ function carries the $SU(2)$ Casimir index $l$ and two magnetic quantum numbers,  $m$ and $g$.  
We construct two independent sets of $SU(2)$ algebras whose simultaneous eigenstates to be $D$ function by using the Maurer-Cartan formulation.  

The left Maurer-Cartan 1 form is given by the formula 
\be
-ig^{\dagger}dg =e^i(L) \sigma_i, 
\ee
and from (\ref{reldfuncmonopoeharm})
\be
g(\chi, \theta, \phi)
=
\begin{pmatrix}
e^{i\frac{\chi}{2}} & 0 \\
 0 & e^{-i\frac{\chi}{2}}
\end{pmatrix}
\begin{pmatrix}
\cos\frac{\theta}{2} & \sin\frac{\theta}{2}\\
-\sin\frac{\theta}{2} & \cos\frac{\theta}{2}
\end{pmatrix}
\begin{pmatrix}
e^{i\frac{\phi}{2}} & 0 \\
 0 & e^{-i\frac{\phi}{2}}
\end{pmatrix}=
\begin{pmatrix}
e^{i\frac{1}{2}(\phi+\chi)} \cos\frac{\theta}{2} & e^{-i\frac{1}{2}(\phi-\chi)} \sin\frac{\theta}{2} \\
-e^{i\frac{1}{2}(\phi-\chi)} \sin\frac{\theta}{2} &  e^{-i\frac{1}{2}(\phi+\chi)} \cos\frac{\theta}{2} 
\end{pmatrix},
\ee
we have 
\begin{align}
&e^1(L)=\sin\theta\cos\phi d\chi-\sin\phi d\theta , \nn\\
&e^2(L)=\sin\theta\sin\phi d\chi+\cos\phi d\theta , \nn\\
&e^{3}(L)=\cos\theta d\chi +d\phi.  
\end{align}
Similarly, the right  Maurer-Cartan 1 form is given by   
\be
idg\cdot g^{\dagger} =e^i(R) \sigma_i, 
\ee
and 
\begin{align}
&e^1(R)=\sin\theta\cos\chi d\phi-\sin\chi d\theta , \nn\\
&e^2(R)=-\sin\theta\sin\chi d\phi-\cos\chi d\theta , \nn\\
&e^{3}(R)=-\cos\theta d\phi -d\chi.  
\end{align}
It is easy to check the $e^i(L)$ and $e^i(R)$ satisfy the Maurer-Cartan  equations: 
\be
de^i(L)-\frac{1}{2}\epsilon^{ijk}e^j(L)\wedge e^k(L)=0, ~~~~~~~~~
de^i(R)-\frac{1}{2}\epsilon^{ijk}e^j(R)\wedge e^k(R)=0. 
\label{maurercartanequss3}
\ee
The metric is read off from 
\be
ds^2=e^i(L) e^i(L)=e^i(R)e^i(R) = g_{\mu\nu}dx^{\mu} dx^{\nu}, ~~~~~~~~(x^{\mu}=\chi, \theta, \phi) 
\ee
as 
\be
g_{\mu\nu}=
\begin{pmatrix}
g_{\chi\chi} & g_{\chi\theta} & g_{\chi\phi} \\
g_{\theta\chi} & g_{\theta\theta} & g_{\theta\phi} \\
g_{\phi\chi} & g_{\phi\theta} & g_{\phi\phi} 
\end{pmatrix}
= 
\begin{pmatrix}
1 & 0 & \cos\theta  \\
0 & 1 & 0 \\
\cos\theta & 0  & 1
\end{pmatrix},  
\ee
and then 
\be
g^{\mu\nu}= \frac{1}{\sin^2\theta}
\begin{pmatrix}
1 & 0 & -\cos\theta  \\
0 & \sin^2\theta & 0 \\
-\cos\theta & 0  & 1
\end{pmatrix}. \label{metrics3}
\ee
$e^{i}_{~~\mu}$ are derived from $e^{i}=e^i_{~~\mu}dx^{\mu}$ and the dual Killing spinor  $e^{~~\mu}_{i}$ are introduced to satisfy $e^i_{~~\mu}e_i^{~~\nu}=\delta^{\mu}_{~~\nu}$. 
The Killing vectors dual to the left and right Maurer-Cartan 1 form are respectively given by 
\begin{align}
&L_i=-ie_i^{~~\mu}(L)\partial_{\mu}, \nn\\
&R_i=-ie_i^{~~\mu}(R)\partial_{\mu},
\end{align}
or 
\begin{align}
&L_x= -i  (-\sin\phi\partial_{\theta} -\cot\theta\cos\phi\partial_{\phi}  +\frac{\cos\phi}{\sin\theta}\partial_{\chi}), \nn\\
&L_y= -i (\cos\phi\partial_{\theta}-\cot\theta\sin\phi\partial_{\phi}  +\frac{\sin\phi}{\sin\theta}\partial_{\chi}), \nn\\
&L_z=-i\partial_{\phi}, 
\label{killingderivdual}
\end{align}
and 
\begin{align}
&R_x= i  (\sin\chi\partial_{\theta} +\cot\theta\cos\chi\partial_{\chi}  -\frac{\cos\chi}{\sin\theta}\partial_{\phi}), \nn\\
&R_y= i (\cos\chi\partial_{\theta}-\cot\theta\sin\chi\partial_{\chi}  +\frac{\sin\chi}{\sin\theta}\partial_{\phi}), \nn\\
&R_z=i\partial_{\chi}.  
\label{killingderivdualright}
\end{align}
(\ref{killingderivdual}) and (\ref{killingderivdualright}) are mutually transformed  by the interchange:   
\be
\phi ~\longleftrightarrow~-\chi, ~~~~~~\theta~\longleftrightarrow~-\theta.
\ee
They satisfy the two independent $SU(2)$ algebras: 
\be
[L_i, L_j]=i\epsilon_{ijk}L_k, ~~~[R_i,R_j]=i\epsilon_{ijk}R_k,~~~[L_i, R_j]=0. 
\label{twoindsu2s}
\ee
(\ref{twoindsu2s}) is a direct consequence of the Maurer-Cartan equation (\ref{maurercartanequss3}). 
The ladder operators for the two $SU(2)$ algebras are respectively constructed as 
\begin{align}
&L_+=L_x+iL_y=e^{i\phi}(\partial_{\theta}+i\cot\theta \partial_{\phi}-i\frac{1}{\sin\theta} \partial_{\chi}), \nn\\
&L_-=L_x-iL_y=-e^{-i\phi}(\partial_{\theta}-i\cot\theta \partial_{\phi}+i\frac{1}{\sin\theta} \partial_{\chi}), 
\label{ladderopessu2}
\end{align}
and 
\begin{align}
&R_+=R_x+iR_y=-e^{-i\chi}(\partial_{\theta}-i\cot\theta \partial_{\chi}+i\frac{1}{\sin\theta} \partial_{\phi}), \nn\\
&R_-=R_x-iR_y=e^{i\chi}(\partial_{\theta}+i\cot\theta \partial_{\chi}-i\frac{1}{\sin\theta} \partial_{\phi}).  
\label{ladderopessu2right}
\end{align}
They act to the $D$ functions as \cite{Dray1986}
\begin{align}
&L_+ D_{l, -m, g}(\theta,\phi,\chi)=-\sqrt{(l-m)(l+m+1)}~D_{l, -m-1, g}(\theta,\phi,\chi), \nn\\
&L_- D_{l, -m, g}(\theta,\phi,\chi)=-\sqrt{(l+m)(l-m+1)}~D_{l, -m+1, g}(\theta,\phi,\chi),  
\end{align}
and 
\begin{align}
&R_+ {D}_{l, -m, g}(\phi,\theta,\chi)=\sqrt{(l-g)(l+g+1)}~{D}_{l, -m, g+1}(\phi,\theta, \chi), \nn\\
&R_- {D}_{l, -m, g}(\phi,\theta,\chi)=\sqrt{(l+g)(l-g+1)}~{D}_{l, -m, g-1}(\phi,\theta, \chi). 
\end{align}
Thus, $\bs{L}$ and $\bs{R}$ are respectively  the left- and right-actions to $D$ functions.  
The $SU(2)$ Casimirs of $\bs{R}$ and $\bs{L}$ are equally given by  
\be
\bs{L}^2=\bs{R}^2=-\frac{1}{\sin\theta}\partial_{\theta}(\sin\theta\partial_{\theta})-\frac{1}{\sin^2\theta}({\partial_{\phi}}^2+{\partial_{\chi}}^2 -2{\cos\theta}\partial_{\phi}\partial_{\chi}). 
\ee
Therefore, $D_{l, -m, g}(\phi,\theta, \chi)$ is the simultaneous eigenstate of two independent $SU(2)$ algebras and the corresponding eigenvalues are  
\begin{align}
&\bs{L}^2=\bs{R}^2=l(l+1), \nn\\
&L_z=m,\nn\\
&R_z=g. 
\end{align}

By replacing $R_z=i\partial_{\chi}$ 
 with 
$g$, we obtain the
angular momentum operator $\bs{L}^{(g)}$ (\ref{explicitschwingerexsu2ooperators}) from the left Killing vector (\ref{killingderivdual}): 
\be
\bs{L}^{(g)}=\bs{L}|_{i\partial_\chi\rightarrow  g}, 
\ee
and the  edth operators (\ref{expliladdercharge}) from the right Killing vector (\ref{ladderopessu2right}): 
\begin{align}
&\eth_+^{(g)} =-R_{+}|_{i\partial_\chi\rightarrow  g},\nn\\
&\eth_-^{(g)} =R_{-}|_{i\partial_\chi\rightarrow  g}. 
\end{align}

\subsection{(2+2) spherical coordinate representation and $SO(4)$ spherical harmonics }

The above argument based on the Maurer-Cartan 1-form is mathematically elegant and the calculations are easy, but rather abstract. 
Here, we derive same results from a simple quantum mechanical argument. 
Calculations are rather laborious but straightforward  and familiar to any physicists.

From the $SU(2)$ group element $g$ ($g^{\dagger}g=1_2, ~\det g=1$),  $S^3$ coordinates $X_{\mu=1,2,3,4}$ ($\sum_{\mu=1}^4 X_{\mu}X_{\mu}=1$) are extracted as 
\be
g=\begin{pmatrix}
X_4 -iX_3 & -X_2-iX_1 \\
X_2-iX_1 & X_4+iX_3
\end{pmatrix}.
\ee
In the case of  (\ref{reldfuncmonopoeharm}), we have 
\begin{align}
&X_1=\sin\frac{\theta}{2}\sin\frac{1}{2}(\phi-\chi),\nn\\
&X_2=-\sin\frac{\theta}{2}\cos\frac{1}{2}(\phi-\chi),\nn\\
&X_3=-\cos\frac{\theta}{2}\sin\frac{1}{2}(\phi+\chi),\nn\\
&X_4=\cos\frac{\theta}{2}\cos\frac{1}{2}(\phi+\chi), 
\end{align}\label{2+2coordinates}
which is known as the (2+2) spherical coordinate representation.  
The metric on $S^3$ is derived as 
\be
\sum_{\mu=1}^4{dX_\mu}{dX_\mu}=\frac{1}{4}(d\chi^2+d\theta^2+d\phi^2+2\cos\theta d\chi d\phi),  
\ee
which is equal to (\ref{metrics3}) up to the unimportant proportional factor.  
The $SO(4)$ $\it{free}$ angular momentum operators are given by 
\be
L_{\mu\nu}=-iX_{\mu}\frac{\partial}{\partial X_{\nu}} +iX_{\nu}\frac{\partial}{\partial X_{\mu}},  
\ee
and  the corresponding $SU(2)_L\oplus SU(2)_R$ operators are constructed as  
\begin{subequations}
\begin{align}
&L_i=\frac{1}{4}\eta_{\mu\nu}^i L_{\mu\nu}=\frac{1}{4}\epsilon_{ijk}L_{ij}+\frac{1}{2}L_{i4}, \label{su2so4left}\\
&R_i=\frac{1}{4}\bar{\eta}_{\mu\nu}^i L_{\mu\nu}=\frac{1}{4}\epsilon_{ijk}L_{ij}-\frac{1}{2}L_{i4},   \label{su2so4right}
\end{align}
\end{subequations}
where $\eta_{\mu\nu}^i$ and $\bar{\eta}_{\mu\nu}^i$ are  the 'tHooft symbols:
\begin{align}
&\eta_{\mu\nu}^i=\epsilon_{\mu\nu i 4}+\delta_{\mu i}\delta_{\nu 4}- \delta_{\nu i}\delta_{\mu 4}, \nn\\
&\bar{\eta}_{\mu\nu}^i=\epsilon_{\mu\nu i 4}-\delta_{\mu i}\delta_{\nu 4}+ \delta_{\nu i}\delta_{\mu 4}. 
\end{align}
A bit of calculation   shows  that, in the (2+2) spherical coordinate representation,  (\ref{su2so4left}) and (\ref{su2so4right})  are exactly identical with the left and right dual Killing vectors, (\ref{killingderivdual}) and (\ref{killingderivdualright}). Therefore, the left, right dual Killing vectors are understood as the two independent $SU(2)$ sets of the free $SO(4)$ angular momentum.     
The $SO(4)$ Casimir is given as 
\be
\sum_{\mu <\nu =1}^4 {L_{\mu\nu}}^2=2(\bs{L}^2+\bs{R}^2)=k(k+2)~~~(k\equiv 2l=0, 1, 2, \cdots), 
\ee
and from the existence of two magnetic quantum numbers of the $D$ function, $m, s=-l, -l+1, \cdots, l$, the degeneracy of the irreducible representation of the $SO(4)$ Casimir index $k$ is\footnote{This result is consistent with the general formula of the $SO(D)$ spherical harmonics, whose Casimir eigenvalue is  $\sum_{\mu<\nu=1}^{D}{L_{\mu\nu}}^2=k(k+D-2)$ $(k=0, 1, 2, \cdots)$ with degeneracy $\frac{(k+D-3)!(2k+D-2)}{k!(D-2)!}$.} 
\be
(2l+1)^2=(k+1)^2.
\ee
The $D$ functions, $D_{l, -m, s}(\phi, \theta, \chi)$, which are the simultaneous irreducible representation of two $SU(2)$ algebras, constitutes the basis states of the $SO(4)$ spherical harmonics. In other words, the $D$ function is simply the $SO(4)$ spherical harmonics in the (2+2) spherical coordinate representation.

\subsection{Effective representation of the $SO(4)$ operators}

The $SU(2)$ group element (\ref{reldfuncmonopoeharm}) can be written as  
\be
g(\chi, \theta, \phi)=\begin{pmatrix}
u(\theta,\phi) ~e^{i\frac{1}{2}\chi} & v(\theta,\phi)~e^{i\frac{1}{2}\chi} \\
-v(\theta,\phi)^*~e^{-i\frac{1}{2}\chi} & u(\theta,\phi)^*~ e^{-i\frac{1}{2}\chi}
\end{pmatrix}
\ee
where $v$ and $u$ are 
  the components of the Hopf spinor, 
\be
\psi =\begin{pmatrix}
v  \\
u
\end{pmatrix}=\begin{pmatrix}
\sin\frac{\theta}{2} ~e^{-i\frac{1}{2}\phi} \\
\cos\frac{\theta}{2} ~e^{i\frac{1}{2}\phi} 
\end{pmatrix}.  
\ee
Since the monopole harmonics (\ref{monopoehamonicseffechopf}) are the homogeneous polynomials of the components of the Hopf spinor, the angular momentum and edth operators can effectively be expressed by the  Hopf spinor  in each of the Landau levels. 
The angular momentum and edth operators act to the Hopf spinor as 
\begin{subequations}
\begin{align}
&\bs{L}^{(\frac{1}{2})} \psi =-\frac{1}{2}\bs{\sigma}\psi,  ~~~~~~
\bs{L}^{(-\frac{1}{2})} {\psi}^* =\frac{1}{2}\bs{\sigma}^t{\psi}^*,  \\
&{\eth}_+^{(-\frac{1}{2})}{\psi} =-i\sigma_y {\psi}^*,~~~~\eth_-^{(\frac{1}{2})}\psi =i\sigma_y {\psi}^*, ~~~~\eth_+^{(\frac{1}{2})} \psi=\eth_-^{(-\frac{1}{2})}{\psi}^*=0,  
\end{align}
\end{subequations}
and are effectively expressed as 
\begin{subequations}
\begin{align}
&\bs{L} =-\frac{1}{2}\psi^t \bs{\sigma}^t \frac{\partial}{\partial \psi}+\frac{1}{2}\psi^{\dagger} \bs{\sigma} \frac{\partial}{\partial {\psi}^*}, \\
&\eth_+=\psi^t i\sigma_y \frac{\partial}{\partial \psi^*}, ~~~~\eth_-=\psi^{\dagger}i\sigma_y \frac{\partial}{\partial\psi}, 
\end{align} 
\end{subequations}
which satisfy 
\begin{align}
&[\eth_+, \eth_-]=2i\eth_z, \\
&[-i\eth_z, \eth_{\pm}]=\pm \eth_{\pm}.  
\end{align}
Here, the operator 
\be
-i\eth_z\equiv \frac{1}{2}(\psi^t \frac{\partial}{\partial \psi} -{\psi}^{\dagger}\frac{\partial}{\partial{\psi}^*}) 
\ee
 represents the monopole charge operator since its eigenvalue is  $g$ [see (\ref{monopoehamonicseffechopf})]. 
 Obviously, $\{{L}_x, L_y, L_z\}$ and $\{R_x, R_y, R_z\}\equiv \{-i\eth_{x}, -i\eth_{y}, -i\eth_z\}$ with
\begin{align}
&\eth_{x}\equiv \frac{1}{2}(\eth_+ +\eth_-)=i\frac{1}{2}\psi^t \sigma_y \frac{\partial}{\partial \psi^*}+i\frac{1}{2}\psi^{\dagger}\sigma_y \frac{\partial}{\partial\psi},\nn\\
&\eth_{y}\equiv -i\frac{1}{2}(\eth_+ -\eth_-)=\frac{1}{2}\psi^t \sigma_y \frac{\partial}{\partial \psi^*}-\frac{1}{2}\psi^{\dagger}\sigma_y \frac{\partial}{\partial\psi},
\end{align}
satisfy two independent $SU(2)$  algebras; 
\be
[L_i, L_j]=i\epsilon_{ijk}L_k,~~~~~[R_i, R_j]=i\epsilon_{ijk}R_k,~~~~~[L_i, R_j]=0. 
\ee
These results are consistent with  Ref.\cite{KarabaliNair2002}. 

\section{Geometric Quantities of Two-sphere}\label{sec:genforms2geo}

With the local coordinates,  
$\mu=\theta,\phi$, 
 $S^2$ metric  is expressed as 
\be
ds^2=d\theta^2+\sin^2\theta ~d\phi^2.   
\ee
From the formula  
\be
ds^2=\delta_{mn}e^m_{~\mu}e^n_{~\nu}dx^{\mu}dx^{\nu}, 
\ee
the zweibein of two-sphere is derived as\footnote{Choice of zweibein is not unique. For instance, we can adopt zweibein as 
\begin{align}
&e^1=\cos\phi d\theta -\sin\theta \sin\phi d\phi, \nn\\
&e^2=\sin\phi d\theta +\sin\theta \cos\phi d\phi,  
\label{zweibeindirac}
\end{align}
and consequently the  spin connection is 
\be
\omega_{12}= (1-\cos\theta)d\phi, 
\ee
which corresponds to the Dirac gauge (\ref{gaugediracfieldform}). 
}   
\be
e^m_{~~\mu}=\begin{pmatrix}
1 & 0 \\
0 & \sin\theta 
\end{pmatrix}~~~~~~~(m=1,2,~~~\mu=\theta, \phi)
\ee
and its  inverse that satisfy $e^m_{~~\mu}e_n^{~~\mu}=\delta_{m}^{~~n}$ and $e^m_{~~\mu}e_m^{~~\nu}=\delta^{\mu}_{~~\nu}$ is   
\be
e_{m}^{~~\mu}=\begin{pmatrix}
1 & 0 \\
0 & \sin^{-1}\theta
\end{pmatrix}. \label{zweischg}
\ee
Non-zero components of Christoffel symbol, $\Gamma^{\mu}_{~~\nu\rho}=\Gamma^{\mu}_{~~\rho\nu}$, are given by 
\be
\Gamma^{\theta}_{~~\phi\phi}=-\sin\theta\cos\theta, ~~~~\Gamma^{\phi}_{~~\theta\phi}=\Gamma^{\phi}_{~~\phi\theta}=\cot\theta,  
\ee
and from the formula,  
\be
\omega_{mn\mu} 
= -e_{n\nu} (\partial_{\mu}e_{m}^{~~\nu}+\Gamma_{\mu \rho}^{\nu}e_{m}^{~~\rho}),  
\label{compspinconnect}
\ee
we have 
\be
\omega_{12\mu}=(\omega_{ 12 \theta}, \omega_{12\phi})=(0, -\cos\theta).  
\ee
We adopt the $SO(2)$ gamma matrices 
$\gamma^1=\sigma_x,~~\gamma^2=\sigma_y$,   
to have 
\be
\sigma^{12}=-\sigma^{21}=-i\frac{1}{4}[\gamma^1, \gamma^2]=\frac{1}{2}\sigma_z, 
\ee
and then the spin connection, $\omega_{\mu}=\sum_{m<n =1,2}\omega_{m n\mu}\sigma^{mn}$, is constructed as 
\be
\omega_{\theta}=0,~~~~\omega_{\phi}=-\frac{1}{2}\cos\theta ~\sigma_z.  
\ee
Consequently, the Dirac operator, $-i\nabla_{\mu}=-i\partial_{\mu}+\omega_{\mu}$, is obtained as 
\be
-i\nabla_{\theta}=-i\partial_{\theta},~~~-i\nabla_{\phi}=-i\partial_{\phi}-\frac{1}{2}\cos\theta ~\sigma_z, 
\label{spineachderidiracop}
\ee
or 
\be
\fsl{\nabla}= \gamma^m e_m^{~\mu}\nabla_{\mu}=\sigma_x \nabla_{\theta}+\frac{1}{\sin\theta} \sigma_y \nabla_{\phi}=\sigma_x (\partial_{\theta}+\frac{1}{2}\cot\theta)+\frac{1}{\sin\theta}\sigma_y \partial_{\phi}.  
\ee
Square of the Dirac operator yields the Laplacian and the scalar curvature:    
\be
{\fsl{\nabla}}^2=\Delta-\frac{1}{4}{R}, 
\ee
where 
\be
\Delta=\frac{1}{\sqrt{g}}\nabla_{\mu}(g^{\mu\nu}\sqrt{g}\nabla_{\nu})=\frac{1}{\sin\theta} \partial_{\theta}(\sin\theta\partial_{\theta})+\frac{1}{\sin^2\theta}(\partial_{\phi}-i\frac{1}{2}\cos\theta \sigma_z)^2, 
\ee
and 
\be
{R}=-4i e_m^{~\mu}e_n^{~\nu}\sigma^{mn}[\nabla_{\mu},\nabla_{\nu}]=2.   
\ee
There are a number of works about the Dirac operator on a two-sphere \cite{AbrikosovJr2002,AbrikosovJr2001,CamporesiHiguchi1996,Tranutman1995,Tranutman1993}.   

\section{Dirac Gauge}\label{appnsecDiracform}

In the Dirac gauge, monopole gauge field is represented as  
\be
A=-g\frac{1}{{r(r+z)}}\epsilon_{ij3} x_j dx_i= g(1-\cos\theta)d\phi. 
\label{gaugediracfieldform}
\ee
The singularity lies on a semi-infinite line of the negative $z$ axis. 
The field strength is 
\be
F=dA=g\sin\theta d\theta\wedge d\phi. 
\ee
In the vector notation, the gauge field is given by  
\be
\bs{A}=\tan\frac{\theta}{2}\bs{e}_{\phi}. 
\ee
The covariant and total angular momentum operators are respectively expressed as 
\begin{align}
&\Lambda_x=L_x^{(0)}+g\cos\theta \tan\frac{\theta}{2}\cos\phi, \nn\\
&\Lambda_y=L_y^{(0)}+g\cos\theta \tan\frac{\theta}{2}\sin\phi, \nn\\
&\Lambda_z=L_z^{(0)}-g(1-\cos\theta), 
\end{align}
and 
\begin{align}
L_x^{(g)}&=\Lambda_x-g\frac{1}{r}x=i(\sin\phi\frac{\partial}{\partial\theta}+\cos\phi\cot\theta\frac{\partial}{\partial\phi})-g{\cos\phi}\tan\frac{\theta}{2},\nn\\ 
L_y^{(g)}&=\Lambda_y-g\frac{1}{r}y=-i(\cos\phi\frac{\partial}{\partial\theta}-\sin\phi\cot\theta\frac{\partial}{\partial\phi})-g{\sin\phi}\tan\frac{\theta}{2},\nn\\
L_z^{(g)}&=\Lambda_z-g\frac{1}{r}z=-i\frac{\partial}{\partial {\phi}}-g.  
\label{totalanwithmono2}
\end{align}
Square of $\bs{L}^{(g)}$ is 
\begin{align}
{({\bs{L}}^{(g)})}^2&=-\frac{1}{\sin\theta} \frac{\partial}{\partial\theta} (\sin\theta\frac{\partial}{\partial\theta})-\frac{1}{\sin^2\theta}(\frac{\partial}{\partial\phi} -ig (1-\cos\theta))^2+g^2\nn\\
&=-(1-x^2)\frac{\partial^2}{\partial x^2}+2x\frac{\partial}{\partial x} +\frac{1}{1-x^2} (i\frac{\partial}{\partial\phi} +g(1-x))^2+g^2, 
\end{align}
with $x=\cos\theta$.

The Dirac gauge is related  the Schwinger gauge  by $U(1)$ transformation:   
\be
A_{\text{S}}~\rightarrow~A_{\text{D}}=A_{\text{S}}+i(e^{ig\phi})d(e^{-ig\phi})=A_{\text{D}}+gd\phi,  
\label{gaugefieldsdiracform}
\ee
where $A_{\text{D}}$ denotes (\ref{gaugediracfieldform}) and $A_{\text{S}}$ represents (\ref{u1gauschfiel}),   
and then the monopole harmonics of the Dirac gauge are given by 
\be
\mathcal{Y}^g_{l,m} (\theta, \phi)=Y^g_{l,m} (\theta, \phi)\cdot e^{ig\phi}=(-1)^{m+g}\sqrt{\frac{2l+1}{4\pi}}D_{l,-m, g}(\phi, \theta, -\phi).  
\label{monodiracharmo}
\ee
where $Y^g_{l,m}$ represent the monopole harmonics in the Schwinger gauge (\ref{schwingermonopoleharmonics}). 
(\ref{monodiracharmo}) can be expressed as 
\be
\mathcal{Y}^g_{~l,m}(\theta, \phi)
= \sqrt{\frac{(2l+1)(l-m)!(l+m)!}{4\pi (l-g )!(l+g)!}} \biggl(\sin\frac{\theta}{2}\biggr)^{-(m+g)} \biggl(\cos\frac{\theta}{2}\biggr)^{-({m-g})}P_{l+m}^{(-m-g, -m+g)} (\cos\theta) \cdot e^{i(m+g)\phi}, 
\label{monoploleharmonicsjacobi}
\ee 
with $x=\cos\theta$, 
and are 
related to the ${D}$ functions as 
\be
\mathcal{Y}^{g}_{l,m}(\theta, \phi)=(-1)^{m+g}\sqrt{\frac{2l+1}{4\pi}} {D}_{l, -m , g} (\phi, \theta, -\phi). 
\ee
Due to the uniqueness of wavefunction, $m+g$ of the azimuthal angle part of  (\ref{monoploleharmonicsjacobi})  should be an integer 
\cite{Felsager1998}.

We can readily obtain the eigenstates  of the Dirac-Landau operator in the Dirac gauge by simply multiplying the phase factor $e^{ig\phi}$ to those of the Schwinger gauge: 
\begin{subequations}
\begin{align}
&n=0~~~~~~~~~:~~ 
{\Psi}^g_{\lambda_0=0, m}(\theta, \phi)=\begin{pmatrix}
{Y}^{g-\frac{1}{2}}_{j=g-\frac{1}{2}, m}(\theta, \phi)  
\\ 
0\end{pmatrix}
\cdot e^{ig\phi} =\begin{pmatrix}
\mathcal{Y}^{g-\frac{1}{2}}_{j=g-\frac{1}{2},m}(\theta, \phi)\cdot e^{i\frac{1}{2}\phi} \\
0
\end{pmatrix}, \\ 
&n=1,2,\cdots~:~ \Psi_{\pm \lambda_n, m}(\theta, \phi)=\frac{1}{\sqrt{2}}
\begin{pmatrix}
{Y}^{g-\frac{1}{2}}_{j,m}(\theta, \phi)\\
\mp i {Y}^{g+\frac{1}{2}}_{j,m}(\theta, \phi)
\end{pmatrix}\cdot e^{ig\phi}=\frac{1}{\sqrt{2}}
\begin{pmatrix}
\mathcal{Y}^{g-\frac{1}{2}}_{j,m}(\theta, \phi)\cdot e^{i\frac{1}{2}\phi} \\
\mp i \mathcal{Y}^{g+\frac{1}{2}}_{j,m}(\theta, \phi)\cdot e^{-i\frac{1}{2}\phi}
\end{pmatrix}, 
\end{align}
\end{subequations}
or 
\begin{subequations}
\begin{align}
&n=0~~~~~~~~~:~~ {\Psi}^g_{\lambda_0=0, m}(\theta, \phi)=(-1)^{g+m-\frac{1}{2}} 
 \sqrt{\frac{(2g)!}{4\pi (g+m-\frac{1}{2})! (g-m-\frac{1}{2})!}} \cdot e^{i(m+g)\phi}\nn\\
&~~~~~~~~~~~~~~~~~~\times 
\begin{pmatrix}
(\sin\frac{\theta}{2})^{(m+g-\frac{1}{2})}(\cos\frac{\theta}{2})^{(-m+g-\frac{1}{2})} \\
0
\end{pmatrix}  , \\ 
&n=1,2,\cdots~~:~~
\Psi_{\pm\lambda_n, m}(\theta, \phi)=
\frac{1}{2}\sqrt{\frac{(2l+1)(l-m)!(l+m)!}{2\pi}}\cdot e^{i(m+g)\phi}\cdot \nn\\ 
&~~~~~~~~~~~~~~~~~~\times \begin{pmatrix}
\frac{1}{\sqrt{(l-g+\frac{1}{2})! (l+g-\frac{1}{2})!}} (\sin\frac{\theta}{2})^{-(m+g-\frac{1}{2})}(\cos\frac{\theta}{2})^{-(m-g+\frac{1}{2})}\cdot P_{l+m}^{(-m-g+\frac{1}{2}, -m+g-\frac{1}{2})}(\cos\theta) \\
\mp i\frac{1}{\sqrt{(l-g-\frac{1}{2})! (l+g+\frac{1}{2})!}} (\sin\frac{\theta}{2})^{-(m+g+\frac{1}{2})}(\cos\frac{\theta}{2})^{-(m-g-\frac{1}{2})}\cdot P_{l+m}^{(-m-g-\frac{1}{2}, -m+g+\frac{1}{2})}(\cos\theta) 
\end{pmatrix}, \label{eigenstateofdiracopconcise}
\end{align}\label{eigenstateofdiracopconcisetot}
\end{subequations}
where $j=n+g-\frac{1}{2}$. The eigenvalues  are the same as of the Schwinger gauge: 
$\pm \lambda_n=\pm\sqrt{n(n+2g)}$ with $n=0,1,2,\cdots$.   
Notice when $g$ is an integer (half-integer), $j$ 
should be a half-integer (integer) and so $m$. 
 Consequently, $m+g$ of the azimuthal phase factor of (\ref{eigenstateofdiracopconcisetot}) is always a $\it{half\text{-}integer}$.

In the Dirac gauge, the edth operators and the ``boost'' operators  corresponding to  (\ref{expliladdercharge}) and (\ref{compboostop}) are respectively  represented as 
\begin{align}
&\eth_{+}^{(g)}
=e^{i\phi}(\partial_{\theta}  + g\tan\frac{\theta}{2} +i\frac{1}{\sin\theta} \partial_{\phi}), \nn\\
&\eth_{-}^{(g)}=e^{-i\phi}(\partial_{\theta} - g\tan\frac{\theta}{2}-i\frac{1}{\sin\theta} \partial_{\phi}), 
\label{edthdiracgauge}
\end{align}
and  
\begin{align}
&K_x^{(g)}\equiv -i(\cos\theta\cos\phi \frac{\partial}{\partial \theta} -\frac{1}{\sin\theta}\sin\phi\frac{\partial}{\partial\phi} -\sin\theta\cos\phi +ig \tan\frac{\theta}{2} \sin\phi), \nn\\
&K_y^{(g)}\equiv -i(\cos\theta\sin\phi \frac{\partial}{\partial \theta} +\frac{1}{\sin\theta}\cos\phi\frac{\partial}{\partial\phi} -\sin\theta\sin\phi -ig \tan\frac{\theta}{2} \cos\phi), \nn\\
&K_z^{(g)}\equiv i(\sin\theta\frac{\partial}{\partial\theta}+\cos\theta). 
\end{align}
Derivation of the edth operators may need some explanation. 
From (\ref{zweibeindirac}) the zweibein in the Dirac gauge is given by  
\be
e^m_{~~\mu}=
\begin{pmatrix}
\cos\phi & -\sin\theta\sin\phi \\
\sin\phi & \sin\theta\cos \phi
\end{pmatrix}~~~(m=x, y,~~\mu=\theta,\phi) \label{zweidiracg}
\ee
and its inverse that satisfies $e_m^{~~\mu}e^n_{~~\mu}=\delta_{m}^{~~n}$ and $e_m^{~~\mu}e^m_{~~\nu}=\delta^{\mu}_{~~\nu}$ is 
\be
e_{m}^{~~\mu}=\frac{1}{\sin\theta}
\begin{pmatrix}
{\sin\theta}\cos\phi & -{\sin\phi} \\
{\sin\theta}\sin\phi & {\cos\phi} 
\end{pmatrix}.
\ee
The edth operators
\be
\eth_m=e_m^{~~\mu}D_{\mu}
\ee
are given by 
\begin{align}
&\eth_x^{(g)}=\cos\phi D_{\theta}-\frac{\sin\phi}{\sin\theta}D_{\phi}, \nn\\
&\eth_y^{(g)}=\sin\phi D_{\theta}+\frac{\cos\phi}{\sin\theta}D_{\phi}, 
\end{align}
where $D_{\mu}$ are the covariant derivatives in the Dirac gauge:  
\be
D_{\theta}=\partial_{\theta},~~~D_{\phi}=\partial_{\phi}-ig(1-\cos\theta),  
\label{covderdirac}
\ee
and then we obtain 
\begin{align}
&\eth_{+}^{(g)}=\eth_x^{(g)}+ i\eth_y^{(g)}
=e^{i\phi}(D_{\theta} +i\frac{1}{\sin\theta} D_{\phi}), \nn\\
&\eth_{-}^{(g)}=\eth_x^{(g)}-i\eth_y^{(g)}=e^{-i\phi}(D_{\theta} -i\frac{1}{\sin\theta} D_{\phi}), 
\label{edthdiracgauge2}
\end{align}
which yield (\ref{edthdiracgauge}).  
The zweibeins in the Schwigner gauge 
$(e^S)_m^{~~\mu}=(\ref{zweischg})$ and the Dirac gauge $(e^D)^m_{~~\mu}=(\ref{zweidiracg})$ are related by the $SO(2)$ transformation, 
\be
R_{m}^{~~n}(\phi)\equiv (e^S)_m^{~~\mu}(e^D)^n_{~~\mu}=
\begin{pmatrix}
\cos\phi & \sin\phi \\
-\sin\phi & \cos\phi
\end{pmatrix}. 
\ee
Therefore, the edth operators in the Schwinger gauge $(\eth^S)_m=(e^S)_m^{~~\mu}(D^S)_{\mu}$ $((D^S)_{\mu}\equiv (\ref{effectofmonopolecov}))$ and  
the Dirac gauge $(\eth^D)_m=(e^D)_m^{~~\mu}(D^D)_{\mu}$ $((D^D)_{\mu}\equiv (\ref{covderdirac}))$ are related as  
\be
(\eth^S)_m^{(g)}=R_{m}^{~~n}(\theta) ~e^{-ig\phi}~(\eth^D)_{n}^{(g)} ~e^{ig\phi}~~~~(m,n=x,y), 
\ee
so $\eth_{\pm}^{(g)}=\eth_x^{(g)}\pm i\eth_y^{(g)}$ are 
\begin{align}
&(\eth^S)_{+}^{(g)}=e^{-i(g+1)\phi}~(\eth^D)_{+}^{(g)}~ e^{ig\phi}, \nn\\
&(\eth^S)_{-}^{(g)}=e^{-i(g-1)\phi}~(\eth^D)_{-}^{(g)}~ e^{ig\phi}, 
\end{align}
or 
\begin{align}
&(\eth^D)_{+}^{(g)}=e^{i(g+1)\phi}~(\eth^S)_{+}^{(g)}~ e^{-ig\phi}, \nn\\
&(\eth^D)_{-}^{(g)}=e^{i(g-1)\phi}~(\eth^S)_{-}^{(g)}~ e^{-ig\phi}. 
\label{relschdiracedthgauge}
\end{align}
(\ref{relschdiracedthgauge}) gives  (\ref{edthdiracgauge2}) through $(\eth^S)_{\pm}^{(g)}=(\ref{edthcovderi})$. Using (\ref{relschdiracedthgauge}), one may readily verify   the relations associated with the edth operators, such as  (\ref{commutationcovspheretot}) and (\ref{commutationrelaethls}), in the Dirac gauge.  

The Dirac operator  is constructed as  
\begin{align}
-i\fsl{D}&=-i\sigma^m \eth_m^{(g_s)}= 
-i\begin{pmatrix}
0 & \eth_-^{(g+\frac{1}{2})} \\
\eth_+^{(g-\frac{1}{2})} & 0 
\end{pmatrix} 
\nn\\
&=
-i\begin{pmatrix}
0 & e^{-i\phi}(D_{\theta}^{(g+\frac{1}{2})} -i\frac{1}{\sin\theta} D_{\phi}^{(g+\frac{1}{2})}) \\
e^{i\phi}(D_{\theta}^{(g-\frac{1}{2})} +i\frac{1}{\sin\theta} D_{\phi}^{(g-\frac{1}{2})})& 0 
\end{pmatrix} , 
\end{align}
where $D_{\mu}^{(g)}\equiv (\ref{covderdirac})$. 



\end{document}